\documentclass[trackchanges, twocolumn, astrosymb]{aastex701} % linenumbers, 

\graphicspath{{./}{figures/}}

\definecolor{amethyst}{rgb}{0.6, 0.4, 0.8}

\usepackage{graphicx}
\usepackage{amsmath}
\usepackage{amssymb}
\usepackage{makecell}
\usepackage{dirtytalk}
\usepackage{threeparttable}
\usepackage{multirow}
\usepackage{makecell}
\usepackage{enumitem}
\usepackage{pifont}

\usepackage{soul}

\usepackage{array}

\makeatletter
\def\switch@array{}
\makeatother

\usepackage{bm}
\usepackage[utf8]{inputenc}
\DeclareUnicodeCharacter{2212}{\bm{\text{--}}}

\begin{document}

\title{Discovery of the Lyman-\boldmath$\alpha$ Forest in a Star-Forming Galaxy at \boldmath$z \approx 7.57$: \\ Implications for the Timing and Topology of Cosmic Reionization}

\suppressAffiliations

\author[0000-0003-4337-6211]{Jakob M. Helton}
\affiliation{Department of Astronomy and Astrophysics, The Pennsylvania State University, University Park, PA 16802, USA}
\email{jakobhelton@psu.edu}

\author[0000-0002-7595-121X]{Joris Witstok}
\affiliation{Cosmic Dawn Center (DAWN), Copenhagen, Denmark}
\affiliation{Niels Bohr Institute, University of Copenhagen, Jagtvej 128, DK-2200, Copenhagen, Denmark}
\email{joris.witstok@nbi.ku.dk}

\author[0009-0002-5648-2670]{Marta Laska}
\affiliation{Department of Astronomy and Astrophysics, The Pennsylvania State University, University Park, PA 16802, USA}
\email{marta.laska@psu.edu}

\author[0000-0003-4565-8239]{Kevin N. Hainline}
\affiliation{Steward Observatory, University of Arizona, 933 N. Cherry Ave., Tucson, AZ 85721, USA}
\email{kevinhainline@arizona.edu}

\author[0000-0001-7151-009X]{Nikko J. Cleri}
\affiliation{Department of Astronomy and Astrophysics, The Pennsylvania State University, University Park, PA 16802, USA}
\affiliation{Institute for Computational and Data Sciences, The Pennsylvania State University, University Park, PA 16802, USA}
\affiliation{Institute for Gravitation and the Cosmos, The Pennsylvania State University, University Park, PA 16802, USA}
\email{cleri@psu.edu}

\author[0000-0003-2388-8172]{Francesco D'Eugenio}
\affiliation{Kavli Institute for Cosmology, University of Cambridge, Madingley Rd., Cambridge CB3 0HA, UK}
\affiliation{Cavendish Laboratory, University of Cambridge, 19 JJ Thomson Ave., Cambridge CB3 0HE, UK}
\email{francesco.deugenio@gmail.com}

\author[0009-0003-7423-8660]{Ignas Juod\v{z}balis}
\affiliation{Kavli Institute for Cosmology, University of Cambridge, Madingley Rd., Cambridge CB3 0HA, UK}
\affiliation{Cavendish Laboratory, University of Cambridge, 19 JJ Thomson Ave., Cambridge CB3 0HE, UK}
\affiliation{Max-Planck-Institut fur Astrophysik, Karl-Schwarzschild-Str 1, D-85748 Garching bei M\"{u}nchen, Germany}
\email{ij284@cam.ac.uk}

\author[0000-0002-5320-2568]{Nimisha Kumari}
\affiliation{AURA for the European Space Agency (ESA), Space Telescope Science Institute, 3700 San Martin Dr., Baltimore, MD 21218, USA}
\email{kumari@stsci.edu}

\author[0000-0001-6755-1315]{Joel Leja}
\affiliation{Department of Astronomy and Astrophysics, The Pennsylvania State University, University Park, PA 16802, USA}
\affiliation{Institute for Computational and Data Sciences, The Pennsylvania State University, University Park, PA 16802, USA}
\affiliation{Institute for Gravitation and the Cosmos, The Pennsylvania State University, University Park, PA 16802, USA}
\email{joel.leja@psu.edu}

\author[0000-0002-8224-4505]{Sandro Tacchella}
\affiliation{Kavli Institute for Cosmology, University of Cambridge, Madingley Rd., Cambridge CB3 0HA, UK}
\affiliation{Cavendish Laboratory, University of Cambridge, 19 JJ Thomson Ave., Cambridge CB3 0HE, UK}
\email{st578@cam.ac.uk}

\author[0000-0003-0883-2226]{Rachana Bhatawdekar}
\affiliation{European Space Agency (ESA), European Space Astronomy Centre (ESAC), Camino Bajo del Castillo s/n, 28692 Villanueva de la Ca\~{n}ada, Madrid, Spain}
\email{rachana.bhatawdekar@esa.int}

\author[0000-0002-8651-9879]{Andrew J. Bunker}
\affiliation{Department of Physics, University of Oxford, Denys Wilkinson Building, Keble Rd., Oxford OX1 3RH, UK}
\email{andy.bunker@physics.ox.ac.uk}

\author[0000-0002-2929-3121]{Daniel J. Eisenstein}
\affiliation{Center for Astrophysics $|$ Harvard and Smithsonian, 60 Garden St., Cambridge, MA 02138, USA}
\email{deisenstein@cfa.harvard.edu}

\author[0000-0001-7673-2257]{Zhiyuan Ji}
\affiliation{Steward Observatory, University of Arizona, 933 N. Cherry Ave., Tucson, AZ 85721, USA}
\email{zhiyuanji@arizona.edu}

\author[0000-0002-9280-7594]{Benjamin D. Johnson}
\affiliation{Center for Astrophysics $|$ Harvard and Smithsonian, 60 Garden St., Cambridge, MA 02138, USA}
\email{benjamin.johnson@cfa.harvard.edu}

\author[0000-0002-5104-8245]{Pierluigi Rinaldi}
\affiliation{Department of Astronomy, The University of Texas at Austin, Austin, TX 78712, USA}
\affiliation{Cosmic Frontier Center, The University of Texas at Austin, Austin, TX 78712, USA}
\email{prinaldi@utexas.edu}

\author[0000-0002-4271-0364]{Brant E. Robertson}
\affiliation{Department of Astronomy and Astrophysics, University of California, Santa Cruz, 1156 High St., Santa Cruz, CA 95064, USA}
\email{brant@ucsc.edu}

\author[0000-0001-9262-9997]{Christopher N. A. Willmer}
\affiliation{Steward Observatory, University of Arizona, 933 N. Cherry Ave., Tucson, AZ 85721, USA}
\email{cnaw@arizona.edu}

%% Use the \collaboration command to identify collaborations. This command takes an optional argument that is either a number or the word "all" which tells the compiler how many of the authors above the command to show. For example "\collaboration[all]{(DELVE Collaboration)}" wil include all the authors above this command. Mark off the abstract in the ``abstract'' environment.

\collaboration{all}{The JADES Collaboration}
\collaboration{all}{The complete list of authors and affiliations are provided at the end of the manuscript.}

% Adds information about the corresponding author

\correspondingauthor{Jakob M. Helton}
\email{jakobhelton@psu.edu}

% Defines properties used throughout

\newcommand{\ResultSpectroscopicRedshift}{$z \approx 7.57$}

\newcommand{\ResultEmissionLineEquivalentWidth}{$\mathrm{EW}_{\mathrm{H} \beta {+} [\mathrm{O\,\textsc{iii}}]} = 847 \pm 9\ \mathrm{\AA}$}

\newcommand{\ResultDustExtinction}{$E ( B-V ) = 0.047^{+0.067}_{-0.035}\ \mathrm{mag}$}

\newcommand{\ResultStellarMassLin}{$M_{\ast} \approx 10^{9.1}\ M_{\odot}$}

\newcommand{\ResultStarFormationRateLin}{$\mathrm{SFR}_{\mathrm{H}\alpha} = 31.1 \pm 2.2\ M_{\odot} / \mathrm{yr}$}

\newcommand{\ResultsIonizingPhotonProductionEfficiencyLin}{$\xi_{\mathrm{ion}} = 10^{25.62 \pm 0.08}\ \mathrm{Hz/erg}$}

\newcommand{\ResultsElectronTemperatureLin}{$T_{e} = 22600^{+1700}_{-2500}\ \mathrm{K}$}

\newcommand{\ResultsElectronDensityLinSelfConsistent}{$n_{e} = 1.5^{+1.5}_{-1.2} \times 10^{5}\ \mathrm{cm}^{-3}$}

\newcommand{\ResultsElectronTemperatureLinSelfConsistent}{$T_{e} = 18600^{+2300}_{-2300}\ \mathrm{K}$}

\newcommand{\ResultsGasPhaseOxygenAbundanceLog}{$12 + \mathrm{log}_{10} (\mathrm{O/H}) = 7.99^{+0.37}_{-0.41}$}

\newcommand{\ResultsGasPhaseOxygenAbundanceLogSolar}{$[\mathrm{O/H}] = -0.70^{+0.37}_{-0.41}$}

\newcommand{\ResultsCarbonToOxygenRatioLog}{$\mathrm{log}_{10} (\mathrm{C/O}) = -1.16^{+0.23}_{-0.20}$}

\newcommand{\ResultsCarbonToOxygenRatioLogSolar}{$[\mathrm{C/O}] = -0.93^{+0.23}_{-0.20}$}

\newcommand{\ResultsNitrogenToOxygenRatioLog}{$\mathrm{log}_{10} (\mathrm{N/O}) = -0.70^{+0.25}_{-0.20}$}

\newcommand{\ResultsNitrogenToOxygenRatioLogSolar}{$[\mathrm{N/O}] = +0.16^{+0.25}_{-0.22}$}

\newcommand{\ResultsNitrogenToCarbonRatioLog}{$\mathrm{log}_{10} (\mathrm{N/C}) = +0.47^{+0.10}_{-0.11}$}

\newcommand{\ResultsNitrogenToCarbonRatioLogSolar}{$[\mathrm{N/C}] = +1.10^{+0.10}_{-0.11}$}

\newcommand{\ResultsIonizationParameterLog}{$\log_{10} (U) = -1.966^{+0.067}_{-0.056}$}

\begin{abstract}

The escape of ionizing photons from galaxies during the Epoch of Reionization (EoR, at $z > 6$) is often inferred indirectly because the intervening neutral hydrogen in the intergalactic medium (IGM) is expected to absorb essentially all flux blueward of Lyman-$\alpha$ at these redshifts. However, in this work, we report the significant detection ($\approx 7 \sigma$) of flux blueward of Ly$\alpha$ in a massive (\ResultStellarMassLin) star-forming galaxy at \ResultSpectroscopicRedshift, \mbox{JADES-GS-z7-LAF}, using JWST/NIRSpec PRISM spectroscopy. We independently confirm this detection of the Ly$\alpha$ forest (LAF) using JWST/NIRCam F090W imaging ($\approx 10 \sigma$). The detected non-ionizing continuum in the LAF probes transmission through the intervening IGM, but does not directly constrain the galaxy’s ionizing escape fraction due to extreme differences between the absorption cross sections of Lyman-continuum and Lyman-series photons. The blueward flux in F090W is spatially coincident with the galaxy and emerges $\approx 0.4\ \mathrm{pkpc}$ from its UV centroid at $\lambda_{\mathrm{rest}} \approx 1500\ \mathrm{\AA}$. We find no evidence for a foreground interloper producing the observed flux, although we cannot formally reject this interpretation with the existing data. The transmitted flux implies an extraordinarily transparent sightline ($x_{\mathrm{H\,\textsc{i}}} \ll 0.1\%$) through the IGM at $z \approx 5.5-7.5$, a sightline that is spatially and kinematically coincident with numerous high-redshift galaxy overdensities, which possibly explains the emergence of this flux. We observe a smooth and gradual turnover around the Ly$\alpha$ break that is either consistent with an extreme damping-wing ($N_{\mathrm{H\,\textsc{i}}} \approx 10^{23.3}\ \mathrm{cm}^{-2}$) or a strong nebular continuum (contributing $\approx 60-70\%$ of the flux at $\lambda_{\mathrm{rest}} = 1500\ \mathrm{\AA}$). JADES-GS-z7-LAF is far more luminous ($M_{\mathrm{UV}} \approx -21.5$, similar to GN-z11) than the faint sources usually invoked as the drivers of cosmic reionization. If such systems are common in the early Universe, massive galaxies may contribute more to the EoR's ionizing photon budget than suggested by other indirect constraints.

% Paradoxically, the spectrum of this galaxy exhibits a strong Ly$\alpha$ damping wing consistent with a high column density of neutral hydrogen ($N_{\mathrm{H\,\textsc{i}}} \approx 10^{23.2}\ \mathrm{cm}^{-2}$), which we reconcile with a partial-covering geometry that allows ionizing photons to escape through a low column density channel offset from the bulk of the UV continuum.

\end{abstract}

%% Keywords should appear after the \end{abstract} command. The AAS Journals now uses Unified Astronomy Thesaurus (UAT) concepts:
%% You will be asked to selected these concepts during the submission process but this old "keyword" functionality is maintained in case authors want to include these concepts in their preprints.


%% From the front matter, we move on to the body of the paper. Sections are demarcated by \section and \subsection, respectively. Observe the use of the LaTeX \label command after the \subsection to give a symbolic KEY to the subsection for cross-referencing in a \ref command. You can use LaTeX's \ref and \label commands to keep track of cross-references to sections, equations, tables, and figures. That way, if you change the order of any elements, LaTeX will automatically renumber them.

%% Start of Section One.
\section{Introduction}
\label{SectionOne}

The reionization of neutral hydrogen permeating the intergalactic medium (IGM) represents the Universe's last major phase transition \citep[for a recent review, see][]{Robertson:2022}. Our current understanding of cosmic reionization is that it began around $z \approx 15-20$ when the first stars and galaxies began to form in dark matter halos \citep[e.g.,][]{Barkana:2001, Abel:2002, Bromm:2004} and ended around $z \approx 5-6$ \citep[e.g.,][]{Becker:2001, Fan:2006, McGreer:2015, Kulkarni:2019, Keating:2020, Bosman:2022, Zhu:2022}. Reionization's midpoint is less constrained, but found to be around $z \approx 7-8$ by measuring Thomson scattering optical depths from the cosmic microwave background (CMB) radiation \citep[e.g.,][]{Planck:2020} and by measurements of Lyman-$\alpha$ (Ly$\alpha$) damping wings toward quasars \citep[for a review, see][]{Fan:2023} and galaxies \citep[e.g.,][]{Umeda:2024, Huberty:2025, Mason:2026, Umeda:2026}. Star-forming galaxies are often considered the dominant sources of ionizing photons in the early Universe, with active galactic nuclei (AGN) providing a negligible contribution. However, there is an ongoing debate about the sources of reionization, and some have even argued for a photon budget crisis where there are \textit{too many} ionizing photons \citep[e.g.,][]{Munoz:2024}, which would be inconsistent with observations of CMB radiation and Ly$\alpha$ forests.

For these reasons, understanding the ionizing photon budget is crucial. The ionizing photon budget per unit time and comoving volume ($\dot{n}_{\mathrm{ion}}$) is dependent on the luminosity density of galaxies ($\rho_{\mathrm{UV}}$), the ionizing photon production efficiency ($\xi_{\mathrm{ion}}$), and the ionizing photon escape fraction ($f_{\mathrm{esc}}$). Assuming the canonical value for ionizing photon production efficiencies, relatively large average escape fractions ($f_{\mathrm{esc}} \approx 10-20\%$) are needed for galaxies to be the dominant source of reionization \citep[e.g.,][]{Ouchi:2009, Robertson:2013, Robertson:2015, Finkelstein:2019, Naidu:2020}. However, \citet{Papovich:2026} recently found that the inferred average escape fractions for $z \approx 6-8$ galaxies are relatively small ($f_{\mathrm{esc}} \approx 1-4\%$) with little redshift evolution. But a similar analysis from \citet{Giovinazzo:2026a} found slightly higher values ($f_{\mathrm{esc}} \approx 10-15\%$). We should note that for these works, ionizing photon escape fractions are inferred indirectly since it remains difficult to make direct measurements of $f_{\mathrm{esc}}$ during the Epoch of Reionization (EoR, at $z > 6$).

Within this discussion of the ionizing photon budget, luminous and massive galaxies occupy an uncertain position. Theoretical arguments predict that the dense gas reservoirs fueling star formation in these systems will suppress the escape of ionizing photons \citep[e.g.,][]{Ferrara:2013, Xu:2016}. Observationally motivated reionization models will often assign the photon budget to numerous and faint dwarf galaxies \citep[e.g.,][]{Finkelstein:2019, Atek:2024}, while some other models instead assign this budget to rare and luminous massive galaxies with relatively high escape fractions \citep[e.g.,][]{Naidu:2020, Summerfield:2026}. Yet populations of bright Lyman continuum emitters (LCEs) are known at $z \approx 3-4$ \citep[e.g.,][]{Steidel:2018, Fletcher:2019, Ji:2020, Marques-Chaves:2022} where escape appears to proceed through ionized channels in an otherwise opaque neutral medium \citep[the so-called ``picket-fence'' geometry;][]{Heckman:2011}. Together, these results demonstrate that high masses and high column densities do not always prevent ionizing photon escape along favorable sightlines. However, it is unclear whether such favorable sightlines exist during the EoR, and even if these sightlines did exist, the radiation must cross a foreground IGM that is predominantly neutral. % For reference, the highest redshift LCE candidate was recently discovered at $z \approx 4.4$ \citep[][]{Goovaerts:2026, Zhu:2026}.

It is extremely unlikely for such a sightline to exist at $z > 6$ since the IGM saturates to Lyman-series photons once the neutral hydrogen fraction in the IGM reaches $x_{\mathrm{H\,\textsc{i}}} \approx 10^{-4}$ \citep[][]{Gunn:1965, Fan:2006}. This means that complete Gunn-Peterson absorption occurs quickly and easily during the EoR, as evidenced by the complete absorption troughs extending across $\approx 100\ \mathrm{Mpc}/h$ that have been observed at $z \approx 6$ \citep[e.g.,][]{Becker:2015, Zhu:2021}. Even among the most luminous high-redshift quasars known, Lyman-$\alpha$ (Ly$\alpha$) forest transmission has not yet been observed at $z > 7$. For reference, \citet{Barnett:2017} provides one of the highest redshift cases of transmission along the sightline of the quasar ULAS J1120+0641 at $z \approx 7.1$. Galaxy spectroscopy reflects the same physics with JWST/NIRSpec observations of galaxies at $z > 7$ ubiquitously showing complete absorption blueward of Ly$\alpha$ \citep[e.g.,][]{Meyer:2025, Umeda:2026}, and often showing softened Ly$\alpha$ breaks shaped by IGM damping wings in addition to proximate damped Ly$\alpha$ absorbers \citep[DLAs; e.g.,][]{DEugenio:2024, Hainline:2024, Heintz:2024, Heintz:2025}. With these results in mind, detecting flux blueward of Ly$\alpha$ (corresponding to $\lambda_{\mathrm{rest}} = 912-1216\ \mathrm{\AA}$) within a normal star-forming galaxy at $z > 7$ would be a surprising discovery that would require an extraordinarily transparent sightline through gas that is, on average, several orders of magnitude too neutral to transmit.

Such a discovery would deliver three incredibly important measurements unavailable by any other means at this epoch. First, the transmitted flux blueward of Ly$\alpha$ would directly constrain the neutral fraction along the sightline as a function of redshift, complementing statistical analyses of damping wings and Ly$\alpha$ visibility \citep[e.g.,][]{Mason:2018, Umeda:2024}. Second, together with a proximate DLA at the systemic redshift, it would directly constrain the geometry of escape, since transmitted flux alongside high-column density absorption would be an expected signature of the picket-fence geometry \citep[][]{Heckman:2011}. Finally, it would tell us something about the sources of reionization by connecting an individual galaxy to the large ionized bubbles traced using high-redshift Ly$\alpha$ emitters out to $z \approx 13$ \citep[LAEs; e.g.,][]{Saxena:2023, Saxena:2024, Witstok:2024, Witstok:2025a, Witstok:2025b, Napolitano:2026}.

In this paper, we report the significant detection of transmitted flux blueward of Ly$\alpha$ in a relatively massive (\ResultStellarMassLin) star-forming galaxy at \ResultSpectroscopicRedshift, which we refer to as JADES-GS-z7-LAF. This detection was originally discovered with JWST/NIRSpec spectroscopy and is independently confirmed with JWST/NIRCam imaging.  We note a companion paper that reports a similar discovery using the same public data from JWST for the same galaxy \citep[][]{Yue:2026}. Our paper proceeds as follows. In Section~\ref{SectionTwo}, we describe the data and observations used in our analysis. In Section~\ref{SectionThree}, we present the analysis and key results. In Section~\ref{SectionFour}, we discuss the scientific interpretation of these results and place them in the context of understanding cosmic reionization. Finally, in Section~\ref{SectionFive}, we summarize our findings and conclude with broader implications for galaxy formation and evolution in the early Universe. Throughout this work, we provide magnitudes in the AB system \citep[][]{Oke:1983} and report wavelengths in air. We assume the standard flat $\Lambda$CDM cosmology from Planck18 with $H_{0} = 67.4\ \mathrm{km/s/Mpc}$ and $\Omega_{m} = 0.315$ \citep[][]{Planck:2020}. We adopt the solar abundance pattern from \citet{Asplund:2021} with $12 + \mathrm{log}_{10}(\mathrm{O/H}) = 8.69$, $\mathrm{log}_{10}(\mathrm{C/O}) = -0.23$, and $\mathrm{log}_{10}(\mathrm{N/O}) = -0.86$. We quote uncertainties as $68\%$ confidence intervals, unless otherwise stated. 

%% Start of Section Two.
\section{Data \& Observations}
\label{SectionTwo}

The primary set of observations used in this work were acquired by the JWST Advanced Deep Extragalactic Survey \citep[JADES;][]{Eisenstein:2026}\footnote[1]{\href{https://archive.stsci.edu/hlsp/jades}{https://archive.stsci.edu/hlsp/jades} \\ \href{https://jades.herts.ac.uk/search/}{https://jades.herts.ac.uk/search/} \\ \href{https://jades-survey.github.io/scientists/data.html}{https://jades-survey.github.io/scientists/data.html}}, which is an ambitious observing program containing infrared imaging and spectroscopy in the Great \mbox{Observatories} Origins Deep Survey (GOODS) extragalactic deep fields \citep[][]{Giavalisco:2004} in both the equatorial north (GOODS-N) and south (GOODS-S).

In particular, we use JWST/NIRSpec multi-object spectroscopy (MOS) obtained with the micro-shutter assembly (MSA) from the JADES Data Release 4 \citep[DR4;][]{Curtis-Lake:2026, Scholtz:2026}. Our target was observed by program ID 1286 in the JADES medium tier program (\texttt{goods-s-mediumjwst}) with five different grating-filter combinations. One of these is low-resolution with $R \approx 30-300$ (PRISM/CLEAR), three are medium-resolution with $R \approx 1000$ (G140M/F070LP, G235M/F140LP, and G395M/F290LP), and one is high-resolution with $R \approx 2700$ (G395H/F290LP). Each of the grating-filter combinations were observed for a total exposure time of $t_{\mathrm{obs}} \approx 2.4\ \mathrm{hr}$. Both the low-dispersion prism and the medium-resolution gratings provide a continuous wavelength coverage of $\lambda_{\mathrm{obs}} \approx 0.6-5.5\ \mu\mathrm{m}$.

We additionally make use of JWST/NIRCam imaging products \citep[][]{Johnson:2026} and photometric catalogs \citep[][]{Robertson:2026} from the JADES Data Release 5 (DR5). Our target (NIRCam ID $217704$) is located in the JADES deep region, which encompasses the Hubble Ultra Deep Field \citep[HUDF;][]{Beckwith:2006} in GOODS-S. In this region of the sky, there are NIRCam observations in $11$ photometric bandpasses: F090W, F115W, F150W, F182M, F200W, F210M, F277W, F335M, F356W, F410M, and F444W. Aside from F182M and F210M\footnote{We note that FRESCO also includes shallow observations of F444W with $t_{\mathrm{obs}} \approx 0.3\ \mathrm{hr}$.}, which were acquired by the First Reionization Epoch Spectroscopically Complete Observations \citep[FRESCO;][]{Oesch:2023} survey with exposure times of $t_{\mathrm{obs}} \approx 0.7-1.0\ \mathrm{hr}$, all other NIRCam filters mentioned here were acquired by JADES with exposure times of $t_{\mathrm{obs}} \approx 6.9-17.8\ \mathrm{hr}$ and provide continuous coverage of $\lambda_{\mathrm{obs}} \approx 0.8-5.0\ \mu\mathrm{m}$.

JADES DR5 also contains the JWST/MIRI imaging products from the Systematic MIRI Legacy Extragalactic Survey \citep[SMILES;][]{Alberts:2024, Rieke:2024}. SMILES is an eight-band imaging survey that takes full advantage of MIRI's continuous wavelength coverage of $\lambda_{\mathrm{obs}} \approx 5.0-26.7\ \mu\mathrm{m}$. We make use of self-consistent, forced circular-aperture photometry that was measured from these imaging products, which are provided in \citet{Robertson:2026}. Together with the other imaging products, there are a total of $19$ JWST filters with available data for our target.

We note the use of \texttt{FitsMap} \citep[][]{Hausen:2022} to visually inspect the available imaging products and photometric catalogs from JADES\footnote{\href{https://jades-survey.github.io/viewer/}{https://jades-survey.github.io/viewer/}}. \texttt{FitsMap} is a simple and lightweight tool for displaying interactive astronomical image and catalog data.

We searched for ALMA observations using the ECOology for Galaxies using ALMA archive and Legacy (ECOGAL) surveys \citep[][]{Lee:2025}. The ECOGAL survey is an ALMA data-mining effort that uniformly reduces archival data, creates science-ready ALMA images, and links them to JWST/HST legacy datasets in well-studied survey fields. We accessed the ECOGAL data products using the DAWN JWST Archive (DJA)\footnote{\href{https://dawn-cph.github.io/dja/index.html}{https://dawn-cph.github.io/dja/index.html}}. We found zero ECOGAL footprints at our target's location and thus no available ALMA observations. We hoped there were observations of JADES-GS-z7-LAF's $[\mathrm{O\,\textsc{iii}}] \lambda 88\,\mu\mathrm{m}$ line to improve our inference of its nebular properties, since this line is sensitive to, e.g., density, temperature, and gas-phase oxygen abundance.

%% Start of Section Three.
\section{Analysis \& Key Results}
\label{SectionThree}

\begin{figure*}
    \centering
    \includegraphics[width=0.9\linewidth]{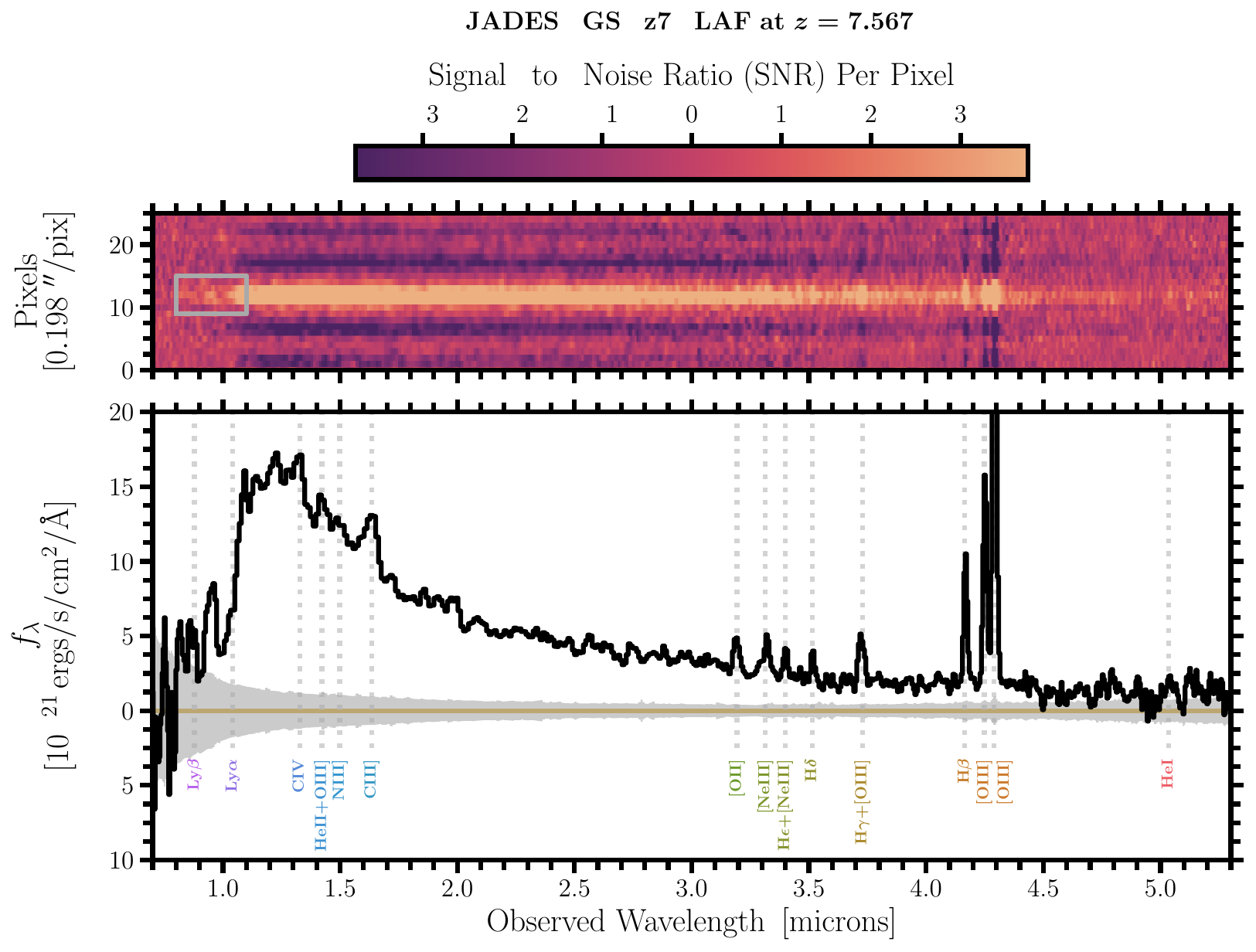}
    \caption{\textbf{The observed spectrum of JADES-GS-z7-LAF using the PRISM/CLEAR configuration of NIRSpec.} \textit{Top panel:} The final 2D spectrum is provided. The corresponding color bar for the measured signal-to-noise ratio per pixel is also shown. JWST/NIRSpec's detector plate scale is $0.198\ \mathrm{arcsec/pixel}$, as shown by the $y$-axis label. The gray bounding box indicates the location of $\lambda_{\mathrm{rest}} \approx 950-1250\ \mathrm{\AA}$, which is where we observe emission from the Ly$\alpha$ forest (LAF). \textit{Bottom panel:} The final 1D spectrum is provided. Gray vertical lines indicate the locations of the strongest emission lines in the rest-frame UV and optical. These lines allow us to robustly determine the spectroscopic redshift of this star-forming galaxy ($z = 7.567$). \label{fig:Full_PRISM_Spectrum}}
\end{figure*}

We initially identified JADES-GS-z7-LAF by searching for galaxies with significant detections ($> 3 \sigma$) of $[\mathrm{O\,\textsc{iii}}] \lambda 4363$ in the JADES DR4, which contains the full spectroscopic sample of galaxies from JADES. Upon visual inspection of this source's NIRSpec/PRISM spectrum (see Figure~\ref{fig:Full_PRISM_Spectrum}), we observed obvious flux appearing blueward of the Ly$\alpha$ break at $\lambda_{\mathrm{rest}} = 1216\ \mathrm{\AA}$ in both the two-dimensional (2D; top panel) and one-dimensional (1D; bottom panel) spectra. Flux is detected in the low-resolution ($R \approx 100$) spectrum at $\lambda_{\mathrm{obs}} \approx 0.92-1.04\ \mu\mathrm{m}$, which corresponds to redshifts of $z \approx 6.56-7.57$ and probes a line-of-sight physical size of $d_{\mathrm{LOS}} \approx 43.2\ \mathrm{pMpc}$ (or $d_{\mathrm{LOS}} \approx 349\ \mathrm{cMpc}$ in comoving units). This feature has only ever been observed among the most luminous quasars known at $z \approx 7$ \citep[e.g.,][]{Barnett:2017}. Given the large neutral fractions at $z > 6$, and the correspondingly high opacity of the IGM, this was an unexpected finding that warranted further investigation.

JADES-GS-z7-LAF was first selected by \citet{Yan:2012} as the second brightest galaxy candidate at $z \approx 8$ (i.e., a dropout in WFC3/F105W) using optical imaging from the Advanced Camera for Surveys (ACS) and near-infrared imaging from the Wide Field Camera 3 (WFC3), both on HST. This galaxy was known as both ISO-$164$ and AUTO-$100$ in \citet{Yan:2012}. It later appears as DEEP\_$33109$ in the compilation at  $z \approx 4-8$ from \citet{Finkelstein:2012}, who report a photometric redshift of $z_{\mathrm{phot}} = 7.61_{-0.30}^{+0.95}$, rest-frame UV absolute magnitude of $M_{\mathrm{UV}} = -21.27_{-0.29}^{+0.13}$, rest-frame UV continuum slope of $\beta_{\mathrm{UV}} = -2.10_{-0.39}^{+0.60}$, and inferred stellar mass of $\mathrm{log}_{10} (M_{\odot} / M_{\odot}) = 9.9_{-0.7}^{+0.4}$. This galaxy also appears in many later compilations, e.g., \citet{Bouwens:2014} and \citet{Harikane:2016}.

JADES-GS-z7-LAF was included in the JADES DR4 \citep[][]{Curtis-Lake:2026, Scholtz:2026}. It was originally allocated a slit as an ``oddball'' high-priority target rather than through the standard selection. A spectroscopic redshift of $z_{\mathrm{spec}} = 7.5668$ is reported in the JADES DR4 redshift catalog \citet{Curtis-Lake:2026}, as measured from the medium-resolution ($R \approx 1000$) data. This is formally $\approx 2 \sigma$ discrepant with a similar value of $z_{\mathrm{spec}} = 7.5670 \pm 0.0001$ that we measure in Section~\ref{SectionThreeOne} using $\mathrm{H}\beta + [\mathrm{O\,\textsc{iii}}]$ (see also Table~\ref{tab:PhysicalProperties}).

JADES-GS-z7-LAF was recently identified as a bright $\mathrm{H}\beta + [\mathrm{O\,\textsc{iii}}]$ emitter in the wide field slitless spectroscopy (WFSS) observed by FRESCO at $\lambda_{\mathrm{obs}} \approx 3.9-5.0\ \mu\mathrm{m}$. This galaxy is the brightest $\mathrm{H}\beta + [\mathrm{O\,\textsc{iii}}]$ emitter in the spectroscopic catalog from \citet{Helton:2024b} and is associated with an extreme galaxy overdensity at $z \approx 7.6$ in GOODS-S, JADES-GS-OD-$7.561$, with an inferred halo mass of $M_{\mathrm{halo}} \approx 10^{11.9}\ M_{\odot}$; this is the most massive protocluster candidate at $z > 6$ in the catalog from \citet{Helton:2024b}. Beyond being the brightest line emitter, JADES-GS-z7-LAF was notable as being one of the only galaxies where the inferred photometric redshift was significantly smaller than the measured spectroscopic redshift because of its blueward-of-Ly$\alpha$ flux in NIRCam/F090W (for more about this, see Section~\ref{SectionThreeTwo} and Figure~\ref{fig:RGB_Thumbnail_Overlay}). This galaxy was identified in \citet{Meyer:2024} as FRESCO-GS-$03081$.

\begin{table}[]
    \centering
    \caption{A summary of the empirical and inferred physical properties of \mbox{JADES-GS-z7-LAF}. For reference, all chemical abundances and abundance patterns are quoted relative to the solar values from \citet[][]{Asplund:2021}.}
    \label{tab:PhysicalProperties}
    \begin{threeparttable}
        \begin{tabular}{>{\raggedright\arraybackslash\hangindent=1.5em}p{0.66\columnwidth} >{\raggedleft\arraybackslash\hangindent=1.5em}p{0.25\columnwidth}}
            \hline
            \hline
            \multicolumn{2}{c}{Empirical Properties} \\
            \hline
            Identification Number & $217704$ \\
		      R.A. [degrees (ICRS)] & $+53.17259$ \\
		      Decl. [degrees (ICRS)] & $-27.74393$ \\
            Photometric Redshift ($z_{\mathrm{phot}}$) & $7.213 \pm 0.017$ \\
            Spectroscopic Redshift ($z_{\mathrm{spec}}$) & $7.5670 \pm 0.0001$ \\
            Ultraviolet Luminosity ($M_{\mathrm{UV}}$) & $-21.55 \pm 0.02$ \\
            Ultraviolet Continuum Slope ($\beta_{\mathrm{UV}}$) & $-2.23 \pm 0.06$ \\
            Apparent Magnitude ($m_{\mathrm{F150W}} / \mathrm{mag}$) & $25.59 \pm 0.01$ \\
            Apparent Magnitude ($m_{\mathrm{F277W}} / \mathrm{mag}$) & $25.52 \pm 0.01$ \\
            Apparent Magnitude ($m_{\mathrm{F444W}} / \mathrm{mag}$) & $24.97 \pm 0.01$ \\
            \noalign{\vskip 1pt}
            \hline
            \hline
            \multicolumn{2}{c}{Inferred Properties from Emission Line Fitting\,\tnote{a}} \\
            \hline
            $\mathrm{EW}_{\mathrm{H} \beta {+} [\mathrm{O\,\textsc{iii}}]} / \mathrm{\AA}$ & $846.6^{+9.1}_{-9.1}$ \\
            $R_{3}$ & $4.61^{+0.40}_{-0.35}$ \\
            $R_{23}$ & $6.50^{+0.57}_{-0.49}$ \\
            $O_{32}$ & $17.58^{+3.81}_{-2.64}$ \\
            $[\mathrm{O\,\textsc{iii}}] \,\lambda\, 4363 / \mathrm{H} \gamma$ & $0.383^{+0.080}_{-0.077}$ \\
            $[\mathrm{O\,\textsc{iii}}] \,\lambda\, 4363 / [\mathrm{O\,\textsc{iii}}] \,\lambda\, 5007$ & $0.047^{+0.009}_{-0.009}$ \\
            \noalign{\vskip 1pt}
            \hline
            Extinction ($E ( B-V ) / \mathrm{mag}$) & $0.047^{+0.067}_{-0.035}$ \\
            Electron Density ($\log_{10} (n_{e} / \mathrm{cm}^{-3})$) & $5.18^{+0.30}_{-0.64}$ \\
            Electron Temperature ($T_{e} / \mathrm{K}$) & $18600^{+2300}_{-2300}$ \\
            Ionization Parameter ($\log_{10} (U)$) & $-1.966^{+0.067}_{-0.056}$ \\
            Ionizing Photon Production Efficiency ($\log_{10} (\xi_{\mathrm{ion}} / \{\mathrm{Hz / erg}\})$) & $25.620^{+0.100}_{-0.058}$ \\
            Gas-Phase Oxygen Abundance ($[\mathrm{O/H}]$) & $-0.70^{+0.37}_{-0.41}$ \\
            Nitrogen-to-Oxygen Ratio ($[\mathrm{N/O}]$) & $+0.16^{+0.25}_{-0.22}$ \\
            Carbon-to-Oxygen Ratio ($[\mathrm{C/O}]$) & $-0.93^{+0.23}_{-0.20}$ \\
            \noalign{\vskip 1pt}
            \hline
            \hline
            \multicolumn{2}{c}{Inferred Properties from Panchromatic SED Fitting\,\tnote{b}} \\
            \noalign{\vskip 1pt}
            \hline
            Stellar Mass Formed ($\mathrm{log}_{10} (M_{\ast}/M_{\odot})$) & $9.07^{+0.10}_{-0.20}$ \\
            Mass-Weighted Stellar Age ($t_{\ast} / \mathrm{Myr}$) & $27.1^{+26.5}_{-9.4}$ \\
            Recent Star-Formation Rate Ratio ($\mathrm{SFR}_{3} / \mathrm{SFR}_{10}$) & $0.42^{+1.74}_{-0.36}$ \\
            Star-Formation Rate in Last $10\ \mathrm{Myr}$ ($\mathrm{SFR}_{10} / (M_{\odot}/\mathrm{yr})$) & $21.4^{+4.2}_{-3.8}$ \\
            Star-Formation Rate in Last $100\ \mathrm{Myr}$ ($\mathrm{SFR}_{100} / (M_{\odot}/\mathrm{yr})$) & $10.9^{+2.8}_{-4.4}$ \\
            Specific Star-Formation Rate in Last $10\ \mathrm{Myr}$ ($\mathrm{log}_{10} (\mathrm{sSFR}_{10} / \mathrm{yr}^{-1})$) & $-7.74^{+0.25}_{-0.15}$ \\
            Specific Star-Formation Rate in Last $100\ \mathrm{Myr}$ ($\mathrm{log}_{10} (\mathrm{sSFR}_{100} / \mathrm{yr}^{-1})$) & $> -8.23\ (3 \sigma)$ \\
            Lookback Time of the First Stars ($t_{\mathrm{birth}} / \mathrm{Myr}$) & $511^{+67}_{-58}$ \\
            Redshift of the First Stars ($z_{\mathrm{birth}}$) & $20.4^{+8.3}_{-3.7}$ \\
            Dust Attenuation ($A_{V} / \mathrm{mag}$) & $0.235^{+0.055}_{-0.091}$ \\
            \noalign{\vskip 1pt}
            \hline
	\end{tabular}
	\begin{tablenotes}
	    \footnotesize
            \item \textbf{Notes.}
            \item[a] For details about emission line fitting, see Section~\ref{SectionThreeOne}.
            \item[b] For details about panchromatic SED fitting, see Section~\ref{SectionThreeEight}.
    \end{tablenotes}
    \end{threeparttable}
\end{table}

In this section, we first perform emission line fitting to robustly determine the spectroscopic redshift and inferred nebular properties (Section~\ref{SectionThreeOne}). We then measure flux blueward of Ly$\alpha$ using the NIRSpec/PRISM spectrum and determine the spatial profile of this emission (Section~\ref{SectionThreeTwo}). We perform multi-component morphological fitting to better understand the location of the blueward-of-Ly$\alpha$ flux and how it corresponds spatially with the rest of the galaxy (Section~\ref{SectionThreeThree}). We test for any instrumental systematics that could be responsible for this blueward-of-Ly$\alpha$ flux and determine that the observed emission is real (Section~\ref{SectionThreeFour}) and then test for any interloping foreground galaxies that could be responsible for this blueward-of-Ly$\alpha$ flux and determine that this is unlikely (Section~\ref{SectionThreeFive}). Since the observed emission is real, we carefully model the rest-frame UV continuum properties (Section~\ref{SectionThreeSix}) while fitting the Ly$\alpha$ damping wing absorption (Section~\ref{SectionThreeSeven}). We model the panchromatic spectral energy distribution (SED) to self-consistently determine the properties of the stellar populations, nebular gas, and dust (Section~\ref{SectionThreeEight}). We finally correlate the blueward-of-Ly$\alpha$ flux with the spatial and kinematic locations of foreground galaxies to see if these overlap with one another (Section~\ref{SectionThreeNine}).

\subsection{Emission Line Fitting, Spectroscopic Redshift Determination, \& Inferred Nebular Properties}
\label{SectionThreeOne}

In Figure~\ref{fig:Full_PRISM_Spectrum}, we show the low-resolution spectrum of \mbox{JADES-GS-z7-LAF} at observed-frame wavelengths of $\lambda_{\mathrm{obs}} \approx 0.7-5.3\ \mu\mathrm{m}$. These data were obtained using JWST/NIRSpec's PRISM/CLEAR configuration, and this particular reduction is from JADES DR4. We provide the 2D and 1D spectra in the upper and lower panels, respectively. Visually, there are numerous spectroscopic features that are clearly detected above the noise level (see the gray shaded region in the lower panel). These features include a continuum detected with high significance\footnote{The continuum is detected with a median per-pixel significance of $\gtrsim 10 \sigma$ at $\lambda_{\mathrm{obs}} \approx 1-3\ \mu\mathrm{m}$ in the NIRSpec/PRISM spectrum.} across almost the full spectral range and a prominent discontinuity from the Ly$\alpha$ break around $\lambda_{\mathrm{obs}} \approx 1.0\ \mu\mathrm{m}$. Surprisingly, we observe emission blueward of the Ly$\alpha$ break from the Ly$\alpha$ forest (LAF) in both the 2D and 1D spectra (see the gray bounding box at rest-frame wavelengths of $\lambda_{\mathrm{rest}} \approx 950-1250\ \mathrm{\AA}$ in the upper panel), which we return to in Section~\ref{SectionThreeTwo}.

As shown in the lower panel of Figure~\ref{fig:Full_PRISM_Spectrum} with the gray, vertical lines, there are roughly ten emission lines in the rest-frame UV and optical that are detected with high significance. These occasionally include individual lines (e.g., $\mathrm{H}\beta$ and the deblended $\mathrm{[O\,\textsc{iii}]} \lambda\lambda 4959{,}5007$ doublet), but often are blended doublets (e.g., $\mathrm{C\,\textsc{iii}]} \lambda\lambda 1907{,}1909$ and $\mathrm{[O\,\textsc{ii}]} \lambda\lambda 3727{,}3729$) or blended nearby lines from different atomic species that have similar wavelengths (e.g., $\mathrm{H}\gamma+\mathrm{[O\,\textsc{iii}]} \lambda 4363$). As described below, using these lines, we can robustly determine the spectroscopic redshift and nebular properties of JADES-GS-z7-LAF.

For all of the subsequent spectroscopic analyses, we apply slit loss corrections by comparing the observed JWST/NIRCam photometry described in Section~\ref{SectionTwo} with synthetic JWST/NIRSpec photometry. These slit-loss corrections are in addition to the default correction from the pipeline, which assumed JADES-GS-z7-LAF was a point source. We found that $\approx 58\%$ ($\approx 53\%$) of the total flux is lost at $\approx 1\ \mu\mathrm{m}$ ($\approx 5\ \mu\mathrm{m}$) in the JWST/NIRSpec data. We used a first-order polynomial of the functional form $a_{0} + a_{1} \lambda_{\mathrm{obs}}$ to fit the multiplicative slit-loss correction as a function of wavelength $\lambda_{\mathrm{obs}}$ with units of $\mu \mathrm{m}$. We fit all available photometric filters redward-of-Ly$\alpha$ using a weighted least-squares approach implemented by the \texttt{scipy.optimize.curve\_fit} routine. We determine polynomial coefficients of $a_{0} = 2.441 \pm 0.062$ and $a_{1} = -0.065 \pm 0.019\ \mu\mathrm{m}^{-1}$, although we note that the reduced chi-square (${\chi_{\nu}}^{2} = \chi^{2} / \mathrm{dof}$ with eight degrees of freedom) is larger than one and therefore suggests a real, non-negligible source of uncertainty not being included in the quoted values for the polynomial coefficients.

To measure the emission-line properties, we separately fit the PRISM and R1000 grating spectroscopy provided by JWST/NIRSpec using a two-stage Bayesian inference framework built around the Python implementation of the \texttt{MultiNest} multimodal nested sampling algorithm \citep[][]{Feroz:2009}, \texttt{PyMultiNest} \citep[][]{Perrin:2014}. We closely follow the methodology described in \citet{Witstok:2025a} using publicly available emission line fitting code from GitHub\footnote{\href{https://github.com/joriswitstok/emission_line_fitting}{https://github.com/joriswitstok/emission\_line\_fitting}}. We adopt a first-order polynomial for the continuum and a Gaussian profile for each of the emission lines throughout.

We first determine the spectroscopic redshift ($z_{\mathrm{spec}}$) and velocity dispersion ($\sigma_{v}$) of \mbox{JADES-GS-z7-LAF} \mbox{using} $\mathrm{H}\beta$ and the $\mathrm{[O\,\textsc{iii}]} \lambda\lambda 4959{,}5007$ doublet, which are the strongest emission lines available in our data set. These three lines are separately fit with Gaussian line profiles using a shared redshift and velocity dispersion. Since the $\mathrm{[O\,\textsc{iii}]} \lambda\lambda 4959{,}5007$ doublet ratio is determined by atomic physics, we fix the line ratio to $f_{5007}/f_{4959} = 2.98$, as calculated with \texttt{PyNeb} \citep[][]{Luridiana:2015}. From the low-resolution PRISM/CLEAR spectrum, we measure $z_{\mathrm{spec}} = 7.5711 \pm 0.0003$ and $\sigma_{v} = 19^{+19}_{-12}\ \mathrm{km/s}$, although we note that these lines are entirely unresolved when compared to the instrumental resolution ($R \approx 200$ and $\sigma_{\mathrm{instr}} \approx 600\ \mathrm{km/s}$ at $\lambda_{\mathrm{obs}} \approx 4.3\ \mu\mathrm{m}$). Alternatively, from the medium-resolution G395M/F290LP spectrum, we measure $z_{\mathrm{spec}} = 7.5670 \pm 0.0001$ and $\sigma_{v} = 79.1^{+4.3}_{-4.2}\ \mathrm{km/s}$. We note that these two redshifts formally disagree with one another at high significance ($\approx 15 \sigma$). However, the quoted confidence intervals assume perfect data quality and do not account for any systematic sources of uncertainty, such as imperfect wavelength calibrations and line spread functions (LSFs). \citet{Scholtz:2026} reports a redshift-independent offset of $\Delta v = 72 \pm 8\ \mathrm{km/s}$ between the PRISM and R1000 grating spectroscopy for the JADES data products, which accounts for roughly half of the observed offset ($\Delta v = 141 \pm 9\ \mathrm{km/s}$).

\begin{table*}[]
    \centering
    \caption{A summary of the emission-line properties of \mbox{JADES-GS-z7-LAF}. As described in Section~\ref{SectionThreeOne}, we perform emission line modeling using a two-stage Bayesian inference framework. Prior to the analysis, we apply slit loss corrections by comparing the observed JWST/NIRCam photometry with synthetic JWST/NIRSpec photometry. Line fluxes and rest-frame equivalent widths are provided separately for the low-resolution (PRISM/CLEAR) and medium-resolution (G140M/F070LP or G395M/F290LP) spectroscopy. For undetected lines, we provide $3 \sigma$ upper limits. We adopt medium-resolution results as fiducial throughout. From the medium-resolution spectrum, we measure $z_{\mathrm{spec}} = 7.5670 \pm 0.0001$ and $\sigma_{v} = 79.1^{+4.3}_{-4.2}\ \mathrm{km/s}$ by fitting $\mathrm{H} \beta {+} \mathrm{[O\,\textsc{iii}]} \lambda\lambda 4959{,}5007$.}
    \label{tab:EmissionLineProperties}
    \begin{threeparttable}
        \begin{tabular}{lcccc}
            \hline
            \hline
            & \multicolumn{2}{c}{PRISM/CLEAR} & \multicolumn{2}{c}{G140M/F070LP or G395M/F290LP} \\
            Emission Line & $\mathrm{Flux}\ [10^{-20}\ \mathrm{erg/s/cm^{2}}]$ & $\mathrm{Equivalent\ Width}\ [\mathrm{\AA}]$ & $\mathrm{Flux}\ [10^{-20}\ \mathrm{erg/s/cm^{2}}]$ & $\mathrm{Equivalent\ Width}\ [\mathrm{\AA}]$ \\
            \noalign{\vskip 1pt}
            \hline
            $\mathrm{N\,\textsc{iv}]}\,\lambda\lambda\,1483{,}1486$ & $< 118.2\ (3\sigma)$ & $< 8.6\ (3\sigma)$ & $< 158.4\ (3\sigma)$ & $< 3.9\ (3\sigma)$ \\
            $\mathrm{C\,\textsc{iv}}\,\lambda\lambda\,1548{,}1551$ & $255.37 \pm 78.30$ & $9.35 \pm 2.85$ & $238.20 \pm 48.19$ & $6.63 \pm 1.34$ \\
            $\mathrm{He\,\textsc{ii}}\,\lambda\,1640$ & $< 170.4\ (3\sigma)$ & $< 6.3\ (3\sigma)$ & $145.58 \pm 30.19$ & $4.90 \pm 1.02$ \\
            $\mathrm{O\,\textsc{iii}]}\,\lambda\lambda\,1661{,}1666$ & $< 249.0\ (3\sigma)$ & $< 9.2\ (3\sigma)$ & $191.71 \pm 51.65$ & $6.82 \pm 1.84$ \\
            $\mathrm{N\,\textsc{iii}]}\,\lambda\lambda\,1747{-}1754$ & $< 143.4\ (3\sigma)$ & $< 5.3\ (3\sigma)$ & $307.30 \pm 60.05$ & $13.83 \pm 2.70$ \\
            $\mathrm{C\,\textsc{iii}]}\,\lambda\lambda\,1907{,}1909$ & $405.06 \pm 64.02$ & $20.48 \pm 3.20$ & $270.20 \pm 45.71$ & $8.54 \pm 1.44$ \\
            $[\mathrm{Ne\,\textsc{v}}]\,\lambda\,3346$ & $< 20.2\ (3\sigma)$ & $< 3.0\ (3\sigma)$ & \nodata & \nodata \\
            $[\mathrm{Ne\,\textsc{v}}]\,\lambda\,3426$ & $< 4.1\ (3\sigma)$ & $< 0.6\ (3\sigma)$ & \nodata & \nodata \\
            $[\mathrm{O\,\textsc{ii}}]\,\lambda\lambda\,3727{,}3729$ & $133.66 \pm 22.36$ & $25.47 \pm 4.14$ & \nodata & \nodata \\
            $[\mathrm{Ne\,\textsc{iii}}]\,\lambda\,3869$\tnote{a} & $146.36 \pm 18.50$ & $30.01 \pm 3.61$ & $132.79 \pm 15.81$ & $21.75 \pm 2.45$ \\
            $\mathrm{H} \epsilon\,{+}\,[\mathrm{Ne\,\textsc{iii}}]\,\lambda\,3968$\tnote{a} & $63.69 \pm 17.84$ & $15.05 \pm 4.21$ & $< 67.8\ (3\sigma)$ & $< 13.6\ (3\sigma)$ \\
            $\mathrm{H} \delta$ & $97.02 \pm 16.41$ & $25.50 \pm 4.19$ & $117.10 \pm 15.13$ & $16.70 \pm 2.06$ \\
            $\mathrm{H} \gamma$\tnote{c} & $175.86 \pm 17.34$ & $44.75 \pm 4.00$ & $213.12 \pm 18.20$ & $53.21 \pm 3.96$ \\
            $\mathrm{[O\,\textsc{iii}]}\,\lambda\,4363$\tnote{c} & $< 21.3\ (3\sigma)$ & $< 5.4\ (3\sigma)$ & $81.48 \pm 15.47$ & $26.58 \pm 4.93$ \\
            $\mathrm{He\,\textsc{ii}}\,\lambda\,4686$ & $< 20.3\ (3\sigma)$ & $< 5.4\ (3\sigma)$ & $< 29.9\ (3\sigma)$ & $< 9.5\ (3\sigma)$ \\
            $\mathrm{H} \beta$ & $386.04 \pm 26.37$ & $115.49 \pm 5.75$ & $376.21 \pm 23.96$ & $70.12 \pm 3.08$ \\
            $[\mathrm{O\,\textsc{iii}}]\,\lambda\,4959$\tnote{b} & $592.22 \pm 29.30$ & $161.01 \pm 2.12$ & $581.40 \pm 28.84$ & $175.86 \pm 2.41$ \\
            $[\mathrm{O\,\textsc{iii}}]\,\lambda\,5007$\tnote{b} & $1767.16 \pm 87.44$ & $457.34 \pm 6.03$ & $1734.89 \pm 86.05$ & $600.64 \pm 8.23$ \\
            $\mathrm{He\,\textsc{i}}\,\lambda\,5876$ & $< 58.7\ (3\sigma)$ & $< 32.7\ (3\sigma)$ & $< 77.8\ (3\sigma)$ & $< 21.6\ (3\sigma)$ \\
            $[\mathrm{O\,\textsc{i}}]\,\lambda\,6300$ & $< 42.9\ (3\sigma)$ & $< 25.1\ (3\sigma)$ & $< 3.2\ (3\sigma)$ & $< 2.2\ (3\sigma)$ \\
            \noalign{\vskip 1pt}
            \hline
	\end{tabular}
	\begin{tablenotes}
	    \footnotesize
            \item \textbf{Notes.}
            \item[a] Since the $[\mathrm{Ne\,\textsc{iii}}] \lambda\lambda 3869{,}3968$ doublet ratio is determined by atomic physics, we fix the line ratio to $f_{3869}/f_{3968} = 3.31$.
            \item[b] Since the $\mathrm{[O\,\textsc{iii}]} \lambda\lambda 4959{,}5007$ doublet ratio is determined by atomic physics, we fix the line ratio to $f_{5007}/f_{4959} = 2.98$.
            \item[c] $\mathrm{H} \gamma$ and $\mathrm{[O\,\textsc{iii}]} \lambda 4363$ are unresolved and blended together in the low-resolution PRISM/CLEAR spectroscopy.
    \end{tablenotes}
    \end{threeparttable}
\end{table*}

We adopt the redshift and velocity dispersion that was measured from the medium-resolution G395M/F290LP spectrum as fiducial throughout and then use these measurements as Gaussian priors to determine fluxes and equivalent widths for the other emission lines available in our data set. The full list of fitted lines is provided in Table~\ref{tab:EmissionLineProperties}, which includes a summary of the emission-line properties of JADES-GS-z7-LAF. Line fluxes and rest-frame equivalent widths are provided separately for the low- and medium-resolution spectroscopy. Uncertainties are quoted as $68\%$ confidence intervals and we provide $3 \sigma$ upper limits for undetected lines. We note that $\mathrm{H} \gamma$ and $\mathrm{[O\,\textsc{iii}]} \lambda 4363$ are unresolved and blended together in the low-resolution PRISM/CLEAR spectroscopy. Since the $[\mathrm{Ne\,\textsc{iii}}] \lambda\lambda 3869{,}3968$ ratio is determined by atomic physics, we fix the line ratio to $f_{3869}/f_{3968} = 3.31$ as calculated with \texttt{PyNeb} \citep[][]{Luridiana:2015}.

To quickly summarize the results that are presented in Table~\ref{tab:EmissionLineProperties}, we detect $10$ ($12$) emission lines and multiplets in the low-resolution (medium-resolution) spectroscopy with $\mathrm{S/N} > 3$. We find overall agreement between the fluxes and equivalent widths measured by the different spectra. However, there are a few notable discrepancies (e.g., $\mathrm{N\,\textsc{iii}]} \lambda\lambda 1747{-}1754$ and $\mathrm{C\,\textsc{iii}]} \lambda\lambda 1907{,}1909$), which we attribute to either an uncertain underlying continuum at those wavelengths or spectrally unresolved lines in the low-resolution data. To confirm the validity of our emission line measurements and their quoted uncertainties, we performed synthetic injection-recovery tests. For the remainder of this work, we adopt the fluxes and equivalent widths derived from the medium-resolution data as fiducial, unless otherwise stated. In the next few paragraphs, we will discuss more specific line fluxes, flux ratios, and equivalent widths, then use those to infer the nebular properties of JADES-GS-z7-LAF.

We measure a rest-frame optical equivalent width of \ResultEmissionLineEquivalentWidth, which is roughly consistent with, but larger than, the typical values observed for luminous ($M_{\mathrm{UV}} \approx -20$) galaxies at $z \approx 8$ from JADES \citep[][]{Endsley:2024}, as measured using JWST/NIRCam photometry. Together, the optical equivalent width and UV luminosity of JADES-GS-z7-LAF suggest that it is a relatively massive ($M_{\ast} \approx 10^{9}\ M_{\odot}$), young ($t_{\ast} \approx 50\ \mathrm{Myr}$) star-forming galaxy experiencing a strong recent upturn in its star-formation history over the last $\approx 50\ \mathrm{Myr}$. We will return to discussing the detailed stellar population properties of this galaxy in Section~\ref{SectionThreeEight} (see Table~\ref{tab:PhysicalProperties}).

We measure Balmer decrements (i.e., ratios of Balmer lines) of $\mathrm{H} \gamma / \mathrm{H} \beta = 0.559 \pm 0.048$ and $\mathrm{H} \delta / \mathrm{H} \beta = 0.305 \pm 0.040$, which should be compared with intrinsic values of $0.473$ and $0.263$ when assuming Case B recombination. Both of these measured ratios are above the intrinsic Case B values, suggesting little to no dust in this galaxy. We infer the dust extinction by jointly fitting the $\mathrm{H} \beta$, $\mathrm{H} \gamma$ and $\mathrm{H} \delta$ fluxes using a wavelength-dependent dust attenuation factor $10^{{-0.4 \, k(\lambda) \, E(B-V)}}$ and assuming the \citet{Cardelli:1989} extinction law for $k(\lambda)$ with $R_{V} = 3.1$. With this methodology, we simultaneously fit all three Balmer line fluxes rather than treating each ratio as independent. Using this method, we infer a dust extinction of \ResultDustExtinction, consistent with JADES-GS-z7-LAF having zero dust.

After accounting for the luminosity distance at the observed redshift, we measure an $\mathrm{H} \beta$ line luminosity of $L_{\mathrm{H} \beta} = 2.74 \pm 0.19 \times 10^{42}\ \mathrm{erg/s}$ from the flux reported in Table~\ref{tab:EmissionLineProperties}. We convert $L_{\mathrm{H} \beta}$ to an $\mathrm{H} \alpha$ line luminosity by assuming the intrinsic Case B recombination value for the Balmer decrement, then use this measurement to infer a star formation rate of \ResultStarFormationRateLin\ by adopting an updated calibration from \citet{Kramarenko:2026} to convert $\mathrm{H} \alpha$ line luminosities into star formation rates. We further estimate a star formation rate surface density of $\Sigma_{\mathrm{SFR}} = 9.04 \pm 0.65\ M_{\odot} / \mathrm{yr} / \mathrm{kpc}^{2}$ by using the measured rest-frame UV half-light radius from the JADES DR5 ($r_{\mathrm{UV}} = 0.1451^{\prime\prime} = 0.739\ \mathrm{pkpc}$) and the definition $\Sigma_{\mathrm{SFR}} = \mathrm{SFR} / \{ 2 \pi ( r_{\mathrm{UV}} )^{2} \}$.

We can further use this measurement of the $\mathrm{H} \alpha$ line luminosity, along with the monochromatic luminosity of the UV ($L_{\mathrm{UV}}$) as measured at $\lambda_{\mathrm{rest}} = 1500\ \mathrm{\AA}$, to infer an ionizing photon production efficiency of \ResultsIonizingPhotonProductionEfficiencyLin\ by using the standard calibration from \citet{Osterbrock:2006} to convert $\mathrm{H} \alpha$ line luminosities into the number density of ionizing photons per unit time. The inferred ionizing photon production efficiency is larger than the canonical value of $\xi_{\mathrm{ion}} \approx 10^{25.2}\ \mathrm{Hz/erg}$ but far below the stellar population maximum of $\xi_{\mathrm{ion}} \approx 10^{26.0}\ \mathrm{Hz/erg}$. Such a high value is possibly related to the burstiness of the star-formation history, which is an idea we return to in Section~\ref{SectionThreeEight}.

For simplicity, we have implicitly assumed an escape fraction of zero for our inference of the star formation rates and ionizing photon production efficiency. We note that a non-zero escape fraction would increase the inferred values for both of these physical properties.

As previously mentioned, we detect the faint auroral $\mathrm{[O\,\textsc{iii}]} \lambda 4363$ line at $\approx 5 \sigma$. The strength of this transition (i.e., the $2s^{2} \, 2p^{2} \, ^{1}S_{0} \rightarrow 2s^{2} \, 2p^{2} \, ^{1}D_{2}$ electronic transition of doubly ionized oxygen) is strongly dependent on collisional excitation that populates the upper level. This means that the line ratio $\mathrm{[O\,\textsc{iii}]} \lambda 4363 / \mathrm{[O\,\textsc{iii}]} \lambda 5007$ (i.e., $2s^{2} \, 2p^{2} \, ^{1}D_{2} \rightarrow 2s^{2} \, 2p^{2} \, ^{3}P_{2}$) is strongly correlated with the electron temperature of the nebular gas. Using the measured $\mathrm{[O\,\textsc{iii}]} \lambda 4363 / \mathrm{[O\,\textsc{iii}]} \lambda 5007$ ratio from Table~\ref{tab:PhysicalProperties}, we infer an electron temperature of \ResultsElectronTemperatureLin\ as calculated with \texttt{PyNeb} \citep[][]{Luridiana:2015}. This is among the highest $\mathrm{[O\,\textsc{iii}]}$ temperatures ever observed at $z > 2$ \citep[e.g.,][]{Sanders:2024, Sanders:2026}, possibly telling us something interesting about the nebular and stellar population properties of JADES-GS-z7-LAF.

\begin{figure}
    \centering
    \includegraphics[width=1.0\linewidth]{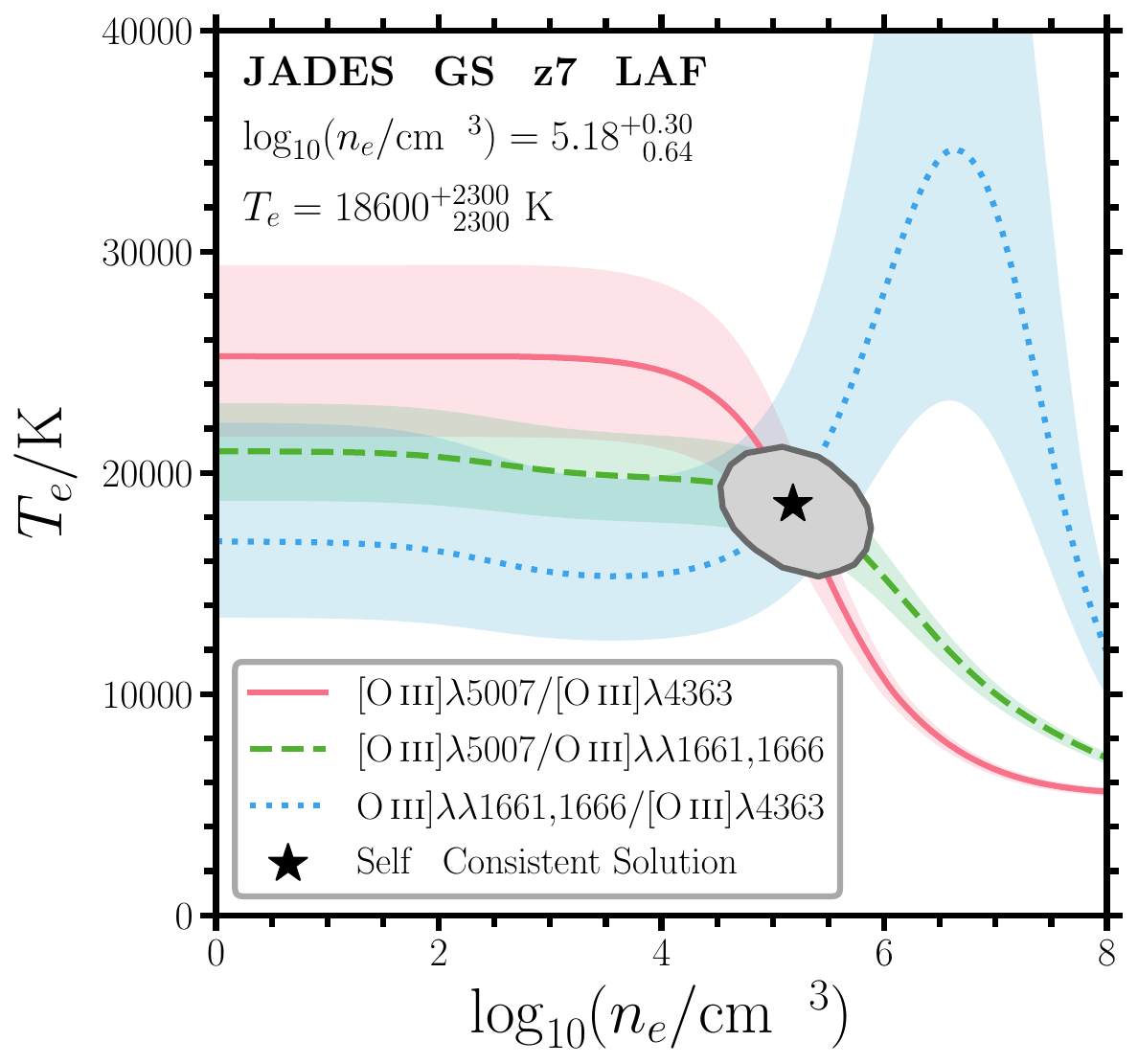}
    \caption{\textbf{Simultaneous constraints on \boldmath$n_{e}$ and \boldmath$T_{e}$.} We self-consistently infer the electron density and electron temperature using the line ratios $[\mathrm{O\,\textsc{iii}}] \lambda 5007 / [\mathrm{O\,\textsc{iii}}] \lambda 4363$ (red solid line), $[\mathrm{O\,\textsc{iii}}] \lambda 5007 / \mathrm{O\,\textsc{iii}}] \lambda 1666$ (green dashed line), and $\mathrm{O\,\textsc{iii}}] \lambda 1666 / [\mathrm{O\,\textsc{iii}}] \lambda 4363$ (blue dotted line). Shaded regions represent $68\%$ confidence intervals. This figure demonstrates how combining three lines from the same ionic species allows $n_{e}$ and $T_{e}$ to be constrained simultaneously, which is crucial for accurately determining gas-phase oxygen abundances. \label{fig:Electron_Density_And_Temperature}}
\end{figure}

We also detect the semi-forbidden $\mathrm{[O\,\textsc{iii}]} \lambda\lambda 1661{,}1666$ doublet at $\approx 4 \sigma$ in the medium-resolution ($R \approx 1000$) G140M/F070LP spectroscopy. This transition (i.e., the $2s \, 2p^{3} \, ^{5}S_{2}^{\,\circ} \rightarrow 2s^{2} \, 2p^{2} \, ^{3}P_{1{,}2}$ intercombination electronic transition of doubly ionized oxygen) is populated from a much higher excited state than either of the $\mathrm{[O\,\textsc{iii}]}$ optical transitions discussed above. The result is that the $\mathrm{[O\,\textsc{iii}]} \lambda\lambda 1661{,}1666$ doublet has orders of magnitude higher critical densities and different temperature sensitivities than $\mathrm{[O\,\textsc{iii}]} \lambda 4363$ and $\mathrm{[O\,\textsc{iii}]} \lambda 5007$. This distinction is important because the $\mathrm{[O\,\textsc{iii}]} \lambda 4363 / \mathrm{[O\,\textsc{iii}]} \lambda 5007$ ratio cannot constrain both electron temperature and density at $n_{e} \gtrsim 10^{4}\ \mathrm{cm}^{-3}$. Such densities have now been observed in several high-redshift galaxies \citep[e.g.,][]{Topping:2025, Hsiao:2026}.

For the reasons outlined above, we self-consistently infer the electron density and temperature for $\mathrm{[O\,\textsc{iii}]}$ by adopting the methodology described in \citet{Hsiao:2026}, which was based on the methods of \citet{Berg:2025_GISM} and \citet{ArellanoCordova:2026}. To accomplish this, we use \texttt{PyNeb} \citep[][]{Luridiana:2015} to construct temperature-density curves for the three line ratios $[\mathrm{O\,\textsc{iii}}] \lambda 5007 / [\mathrm{O\,\textsc{iii}}] \lambda 4363$, $[\mathrm{O\,\textsc{iii}}] \lambda 5007 / \mathrm{O\,\textsc{iii}}] \lambda 1666$, and $\mathrm{O\,\textsc{iii}}] \lambda 1666 / [\mathrm{O\,\textsc{iii}}] \lambda 4363$. To construct these curves, we needed to use the $\mathrm{[O\,\textsc{iii}]}$ transition and electron impact excitation collision rates from \citet{Tayal:2017} within \texttt{PyNeb} since its default dataset does not include the upper level needed to compute emissivities for $\mathrm{[O\,\textsc{iii}]} \lambda\lambda 1661{,}1666$. The curves for $T_{e} (n_{e})$ are provided in Figure~\ref{fig:Electron_Density_And_Temperature}, which demonstrates how combining three lines from the same ionic species allows us to simultaneously constrain $n_{e}$ and $T_{e}$ for certain combinations of these parameters. We infer \ResultsElectronDensityLinSelfConsistent\ and \ResultsElectronTemperatureLinSelfConsistent. Such a high inferred density lowers the inferred temperature at fixed $\mathrm{[O\,\textsc{iii}]} \lambda 4363 / \mathrm{[O\,\textsc{iii}]} \lambda 5007$ line ratio. The inferred density is much larger than typical conditions seen in the local Universe ($z \approx 0$) and provides further evidence for a redshift evolution toward increasing electron density at higher redshifts \citep[e.g.,][]{Isobe:2023}, reaching extreme densities that are comparable to the critical densities of optical $\mathrm{[O\,\textsc{iii}]}$ lines. This redshift evolution is possibly caused by smaller physical sizes and thus larger surface densities at higher redshifts.

Although the temperature-density relations converge to a self-consistent solution for JADES-GS-z7-LAF in Figure~\ref{fig:Electron_Density_And_Temperature}, we caution the reader that the uncertainties in the line ratios still allow for a low-density solution of $n_{e} \ll 10^{5}\ \mathrm{cm}^{-3}$. This is similar to the results presented in \citet{Hsiao:2026} for a galaxy at $z \approx 5$, SPURS-A2744-544. The large inferred electron density has implications for two-photon nebular emission, as we discuss in Section~\ref{SectionThreeSeven} in the context of the smooth and gradual turnover in JADES-GS-z7-LAF's SED around the Ly$\alpha$ break. This is because the inferred gas density is nearly an order of magnitude larger than the critical density of two-photon emission \citep[$n_{e{,}\mathrm{crit}} \approx 2 \times 10^{4}\ \mathrm{cm}^{-3}$; e.g.,][]{Osterbrock:2006}.

We use the inferred electron temperature to determine the gas-phase oxygen abundance of JADES-GS-z7-LAF using the so-called direct ($T_{e}$) method and by assuming $\mathrm{O/H} \approx \mathrm{O^{++}/H} + \mathrm{O^{+}/H}$. This is a valid approximation since the vast majority ($\gtrsim 95\%$) of oxygen should be either singly or doubly ionized for high-redshift galaxies with low metallicities and high ionization parameters \citep[e.g.,][]{Berg:2019}. We calculate the ionic abundances of $\mathrm{O^{++}/H}$ and $\mathrm{O^{+}/H}$ with \texttt{PyNeb} \citep[][]{Luridiana:2015}, adopting $T_{e}(\mathrm{[O\,\textsc{iii}]})$ for the high-ionization zone and $T_{e}(\mathrm{[O\,\textsc{ii}]})$ for the low-ionization zone. $T_{e}(\mathrm{[O\,\textsc{ii}]})$ is indirectly inferred from $T_{e}(\mathrm{[O\,\textsc{iii}]})$ using an empirical relation\footnote{We indirectly infer $T_{e}(\mathrm{[O\,\textsc{ii}]}) = 14800^{+2300}_{-2300}\ \mathrm{K}$, where the quoted uncertainty includes both the measurement uncertainty on $T_{e}(\mathrm{[O\,\textsc{iii}]})$ and the empirical relation's intrinsic scatter.} from the DEep Spectra of Ionized REgions Database Extended (DESIRED-E) project \citep[][]{OrteGarcia:2026}. Because $\mathrm{[O\,\textsc{ii}]}$ is not significantly detected in our medium-resolution spectrum, we instead infer its grating-equivalent flux by rescaling the grating $\mathrm{[O\,\textsc{iii}]}$ flux with the $O_{32}$ ratio measured from the low-resolution spectrum. Summing the two ionic species, we infer a total gas-phase oxygen abundance of \ResultsGasPhaseOxygenAbundanceLog\ ($\approx 20\%\ Z_{\odot}$), which confirms that JADES-GS-z7-LAF is a metal-poor galaxy.

The large inferred electron density has implications for the gas-phase oxygen abundance since at extremely high densities ($n_{e} \approx 10^{5-6}\ \mathrm{cm}^{-3}$), the $[\mathrm{O\,\textsc{iii}}] \lambda 5007$ line is suppressed once collisional de-excitation dominates over spontaneous de-excitation, which occurs when the gas density is larger than the line's critical density. If we instead use \ResultsElectronTemperatureLin\ as measured from the $\mathrm{[O\,\textsc{iii}]} \lambda 4363 / \mathrm{[O\,\textsc{iii}]} \lambda 5007$ ratio and further assume $n_{e} = 300\ \mathrm{cm}^{-3}$, we would infer a total gas-phase oxygen abundance of $12 + \mathrm{log}_{10} (\mathrm{O/H}) = 7.36^{+0.10}_{-0.07}$ ($\approx 5\%\ Z_{\odot}$), which is $\approx 0.6-0.7\ \mathrm{dex}$ smaller than our fiducial value. Thus, a failure to account for high densities can lead to a systematic misclassification of extremely metal-poor galaxies (EMPGs) along with biased conclusions about chemical enrichment in the very early Universe. \citet{Hsiao:2026} arrived at a similar conclusion using a sample of four EMPG candidates at $z \approx 5-6$.

We additionally constrain the carbon and nitrogen abundance patterns relative to oxygen by leveraging our detections of $\mathrm{C\,\textsc{iv}} \lambda\lambda 1548{,}1551$, $\mathrm{N\,\textsc{iii}]} \lambda\lambda 1747{-}1754$, and $\mathrm{C\,\textsc{iii}]} \lambda\lambda 1907{,}1909$. We assume $\mathrm{C} \approx \mathrm{C^{3+}} + \mathrm{C^{++}}$ and $\mathrm{N} \approx \mathrm{N^{++}}$ since the vast majority ($\gtrsim 95\%$) of carbon and nitrogen should be either doubly or triply ionized for galaxies with low metallicities and high ionization parameters \citep[e.g.,][]{Berg:2019}. We separately calculate ionic abundances for carbon and nitrogen with \texttt{PyNeb} \citep[][]{Luridiana:2015}, inferring \ResultsCarbonToOxygenRatioLog\ and \ResultsNitrogenToOxygenRatioLog, which corresponds to \ResultsCarbonToOxygenRatioLogSolar\ and \ResultsNitrogenToOxygenRatioLogSolar\ relative to solar. These results correspond to a sub-solar carbon-to-oxygen ratio but a super-solar nitrogen-to-oxygen ratio. We additionally infer \ResultsNitrogenToCarbonRatioLog, or \ResultsNitrogenToCarbonRatioLogSolar, reflecting the abundance patterns described above. The elevated nitrogen-to-carbon ratio is unusual for a galaxy with this gas-phase metallicity, which might point toward an unusual mode of chemical enrichment or nitrogen production in JADES-GS-z7-LAF. As discussed in \citet{Naidu:2026} for MoM-z14 at $z = 14.44$ \citep[see also, e.g.,][]{Cameron:2023, Ji:2026}, these abundance patterns are routinely observed in local globular clusters and the oldest metal-poor stars in the Milky Way \citep[e.g.,][]{Belokurov:2023, Belokurov:2024}. We note that these results depend on a line flux measurement for the blended five-component $\mathrm{N\,\textsc{iii}]}$ multiplet, motivating deeper and higher-resolution follow-up spectroscopy to confirm the inferred chemical abundance pattern.

Finally, we estimate the ionization parameter using the $O_{32} = \mathrm{[O\,\textsc{iii}]} \lambda\lambda 4959{,}5007 / \mathrm{[O\,\textsc{ii}]} \lambda\lambda 3727{,}3729$ ratio as measured from the low-resolution spectrum. After adopting an empirical relation from the Keck Baryonic Structure Survey \citep[KBSS;][]{Strom:2018}, we infer \ResultsIonizationParameterLog\ for JADES-GS-z7-LAF, consistent with high-ionization and high-excitation nebular conditions that are typical of low-metallicity high-redshift galaxies \citep[e.g.,][]{Cleri:2026}.

\subsection{Measurement of Flux Blueward of Lyman-$\alpha$}
\label{SectionThreeTwo}

Now that we have fit the emission lines, determined the spectroscopic redshift, and inferred some of the most important nebular properties, we return to discussing the blueward-of-Ly$\alpha$ flux. Qualitatively, the top panel of Figure~\ref{fig:Full_PRISM_Spectrum} demonstrates that the observed 2D cross-dispersion profile appears constant across the complete spectral range, including on either side of the Ly$\alpha$ break at $\lambda_{\mathrm{rest}} = 1216\ \mathrm{\AA}$. To quantify this behavior, we directly measure the light profile from the NIRSpec/PRISM 2D spectrum as a function of wavelength.

\begin{figure*}
    \centering
    \includegraphics[width=0.9\linewidth]{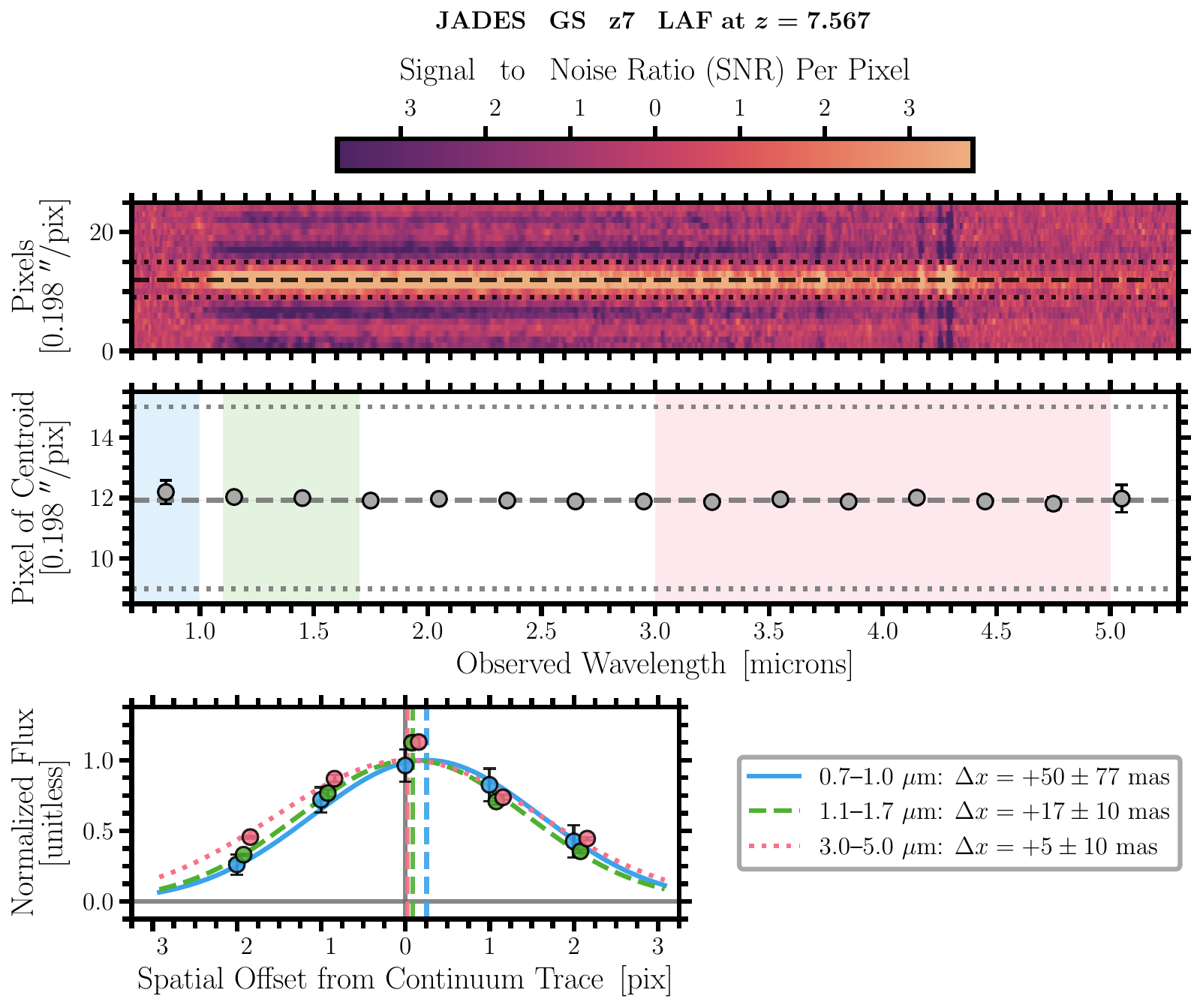}
    \caption{\textbf{The observed cross-dispersion profile and inferred spatial offsets from the NIRSpec/PRISM spectrum.} \textit{Top panel:} Similar to Figure~\ref{fig:Full_PRISM_Spectrum}, the final 2D spectrum is provided. The corresponding color bar for the measured signal-to-noise ratio per pixel is also shown. The dashed horizontal line represents the fitted centroid of the bright continuum while the two dotted horizontal lines indicate the narrow spatial window used to fit the centroid at every wavelength. \textit{Middle panel:} Filled circles indicate the best-fit spatial centroid as measured in wavelength bins with $\Delta \lambda_{\mathrm{obs}} = 0.3\ \mu\mathrm{m}$. The blue, green, and red shaded regions represent the three wavelength regions compared below. \textit{Bottom panels:} The wavelength-collapsed spatial profile in each of the three wavelength regions is shown as a function of pixel offset from the reference centroid with best-fit Gaussian models overlaid. Each wavelength region's fitted spatial offset from the bright continuum ($\Delta x$) is annotated in the legend with units of milliarcseconds. The measured centroids indicate the faint continuum blueward of the Ly$\alpha$ break is spatially coincident with the rest-frame UV and optical emission, suggesting a single source rather than chance alignment with an interloper. \label{fig:Cross_Dispersion_Profile}}
\end{figure*}

In Figure~\ref{fig:Cross_Dispersion_Profile}, we provide the observed spatial (or cross-dispersion) profile from the NIRSpec/PRISM spectrum. Similar to Figure~\ref{fig:Full_PRISM_Spectrum}, the final 2D spectrum is provided in the top panel. We fit a centroid to the bright continuum across $\lambda_{\mathrm{obs}} = 1.5-3.5\ \mu\mathrm{m}$ using a Gaussian profile, as shown by the dashed horizontal lines in the top two panels and the solid vertical line in the bottom panel. This centroid defines the reference continuum trace, which was determined from a narrow spatial window of five JWST/NIRSpec pixels, as indicated by the dotted horizontal lines in the top two panels. Such a narrow spatial window was used to avoid the nodded negative traces. We then fit for centroids across $\lambda_{\mathrm{obs}} = 0.7-5.2\ \mu\mathrm{m}$ using wavelength bins of $\Delta \lambda_{\mathrm{obs}} = 0.3\ \mu\mathrm{m}$, as shown by the grey data points in the middle panel. Finally, we fit three broad wavelength regions covering the continuum blueward of the Ly$\alpha$ break (blue shaded region in the middle panel and blue lines in the lower panel), the rest-frame UV (green shaded region and lines), and the rest-frame optical (red shaded region and lines). Each wavelength region's offset from the bright continuum is annotated in the legend with units of milliarcseconds.

\begin{figure*}
    \centering
    \includegraphics[width=0.9\linewidth]{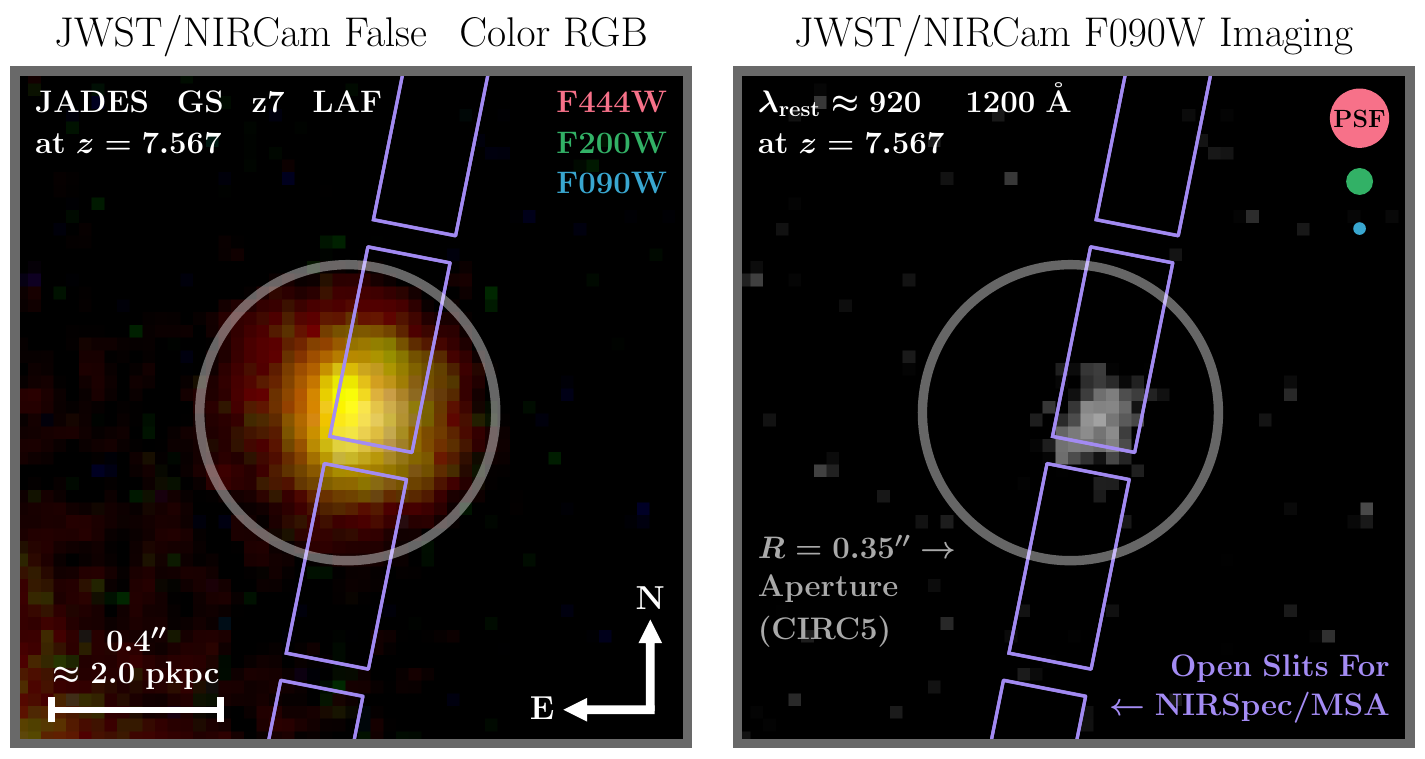}
    \caption{\textbf{Comparing the spatial structure of emission from JADES-GS-z7-LAF across different wavelengths with JWST/NIRCam.} \textit{Left panel:} We show a false-color RGB using imaging in the F444W (red; rest-frame optical), F200W (green; rest-frame UV), and F090W (blue; blueward-of-Ly$\alpha$) filters. A scalebar and compass are provided in the lower left and lower right, respectively. \textit{Right panel:} We show the F090W imaging, which probes rest-frame wavelengths of $\lambda_{\mathrm{rest}} \approx 920-1200\ \mathrm{\AA}$. Color-coded circles represent the FWHM of each filter's PSF in the upper right corner. \textit{Both panels:} The grey circle represents a circular aperture with radius $R = 0.35\ \mathrm{arcsec}$ (i.e., CIRC5 from the JADES DR5). The purple rectangles represent the assigned slitlets for JWST/NIRSpec's MSA. The blueward-of-Ly$\alpha$ flux measured by F090W emerges $\approx 0.4\ \mathrm{pkpc}$ to the west (right) of JADES-GS-z7-LAF's UV centroid, but is roughly coincident with the disk component of the galaxy (see Section~\ref{SectionThreeThree}). \label{fig:RGB_Thumbnail_Overlay}}
\end{figure*}

If the blueward-of-Ly$\alpha$ flux originated from a spatially distinct source from JADES-GS-z7-LAF (e.g., an interloping foreground galaxy), then we would expect its light to trace a different position along the shutter's spatial direction (i.e., the cross-dispersion profile). However, the measured centroids shown in Figure~\ref{fig:Cross_Dispersion_Profile} indicate that the faint continuum blueward of the Ly$\alpha$ break, the rest-frame UV emission, and the rest-frame optical emission are all spatially coincident with one another, suggesting a single source is producing the flux at all wavelengths rather than a chance alignment with an interloper.

As an independent test, we additionally searched the available NIRCam/F090W imaging data for photometric evidence of any emission blueward of the Ly$\alpha$ break. At the spectroscopic redshift of our galaxy, $z = 7.567$, this filter should probe emission at $\lambda_{\mathrm{rest}} \approx 920-1200\ \mathrm{\AA}$. The left panel of Figure~\ref{fig:RGB_Thumbnail_Overlay} shows a false-color RGB using imaging in the F444W (red; rest-frame optical), F200W (green; rest-frame UV), and F090W (blue; blueward-of-Ly$\alpha$) filters. The right panel of Figure~\ref{fig:RGB_Thumbnail_Overlay} shows the F090W imaging data. JADES DR5 reports a $\approx 10 \sigma$ detection in F090W at the location of JADES-GS-z7-LAF. The blueward-of-Ly$\alpha$ flux measured by F090W emerges $\approx 0.4\ \mathrm{pkpc}$ to the west (right) of JADES-GS-z7-LAF's UV centroid, but is roughly coincident with the disk component of the galaxy (see also Section~\ref{SectionThreeThree} for details about the multi-component morphological fitting).

To quickly summarize, we find both spectroscopic ($\approx 7 \sigma$) and photometric ($\approx 10 \sigma$) evidence of flux blueward of Ly$\alpha$ in JADES-GS-z7-LAF using NIRSpec/PRISM spectroscopy and NIRCam/F090W imaging. These two data sets are fully independent of one another. In the NIRSpec/PRISM spectrum, the peak centroid for the blueward-of-Ly$\alpha$ flux is at $\lambda_{\mathrm{obs}} = 0.9540 \pm 0.0021\ \mu\mathrm{m}$, which corresponds to $\lambda_{\mathrm{rest}} = 1114 \pm 3\ \mathrm{\AA}$ at \ResultSpectroscopicRedshift. When compared to the expected wavelength of Ly$\alpha$, this flux has a velocity offset of $\Delta v = -25200 \pm 600\ \mathrm{km/s}$, which is at least an order of magnitude larger than any physically plausible bulk velocity shift for Ly$\alpha$.

\subsection{Multi-Component Morphological Fitting}
\label{SectionThreeThree}

We model the available JWST/NIRCam imaging of JADES-GS-z7-LAF across nine photometric filters to explicitly test whether the source's blueward-of-Ly$\alpha$ flux traces the same spatial structure as its rest-frame UV and optical emission. To accomplish this, we utilize the Python package \texttt{pysersic} \citep[][]{Pasha:2023} for determining this galaxy's structural properties through Bayesian inference. We sample posteriors using the gradient-based No U-Turn Sampler \citep[\texttt{NUTS};][]{Hoffman:2011} by assuming four chains with 1,000 warm-up samples and 10,000 sampling draws for each chain; we also assume a target acceptance probability of $90\%$ to ensure convergence. Our morphological modeling assumptions are described below.

We provide results for \mbox{JADES-GS-z7-LAF} from the multi-component morphological fitting in Table~\ref{tab:DerivedPhotometry}, which includes the measured photometry, and Table~\ref{tab:DerivedMorphology}, which includes the inferred structural parameters. To quickly summarize, the measured CIRC5 ($0.35\ \mathrm{arcsec}$ in radius; for reference, see the circular grey apertures shown in Figure~\ref{fig:RGB_Thumbnail_Overlay}) fluxes from the JADES DR5 \citep[][]{Robertson:2026} are consistent with the total measured fluxes from the morphological fitting, although we note that the bulge component appears bluer in the rest-frame UV than the total value, while the disk component appears redder. We also note that the centroids of F090W and the disk are offset from one another by $\Delta \theta = 1.267 \pm 0.230$ pixels ($0.0380 \pm 0.0069\ \mathrm{arcsec}$ or $0.193 \pm 0.035\ \mathrm{kpc}$), which we reconcile with a partial-covering geometry.

We jointly fit a two-component morphological model using eight photometric filters redward-of-Ly$\alpha$ (F115W, F150W, F200W, F277W, F335M, F356W, F410M, and F444W). The two components include a bulge modeled by a S\'{e}rsic profile with free S\'{e}rsic index $n$ and a disk modeled by an exponential profile with fixed S\'{e}rsic index $n = 1$. For the bulge component, we fit for six structural parameters ($x$- and $y$-positions of the centroid, effective radius $r_{\mathrm{eff}}$, S\'{e}rsic index $n$, ellipticity $b/a$, and position angle $\theta$) that are shared jointly across all eight filters; we independently fit for the flux $f_{i}$ in each of the filters. For the disk component, we fit for the same structural parameters as the bulge, except for the S\'{e}rsic index, which is fixed at $n = 1$. We also fit a single-component morphological model to the blueward-of-Ly$\alpha$ flux traced by F090W using the same methodology as the bulge.

In an attempt to avoid degeneracy between the bulge and disk in the two-component model, we parametrize the disk's effective radius as $r_{\mathrm{eff,\,disk}} = r_{\mathrm{eff,\,bulge}} + \Delta r_{\mathrm{eff}}$ with $\Delta r_{\mathrm{eff}} \geq 0$; this ensures that the disk's effective radius is strictly equal to or larger than the bulge's radius.

As previously mentioned, all three components (bulge, disk, and F090W) have their own independent position and orientation in the image, rather than being forced to share one spatial location. This required going around the fitting software's standard multi-component tool (\texttt{FitMulti}), which fixes every morphological component to a common center, and instead assembling the model from its more basic building blocks (\texttt{render\_sersic} and \texttt{render\_exp}) by hand. In practice, each filter sums their models with \texttt{combine\_scene} and compares the fitted model to the observations using \texttt{gaussian\_loss}, where the loss terms are fed into the same \texttt{numpyro} model.

\begin{figure*}
    \centering
    \includegraphics[width=1.0\linewidth]{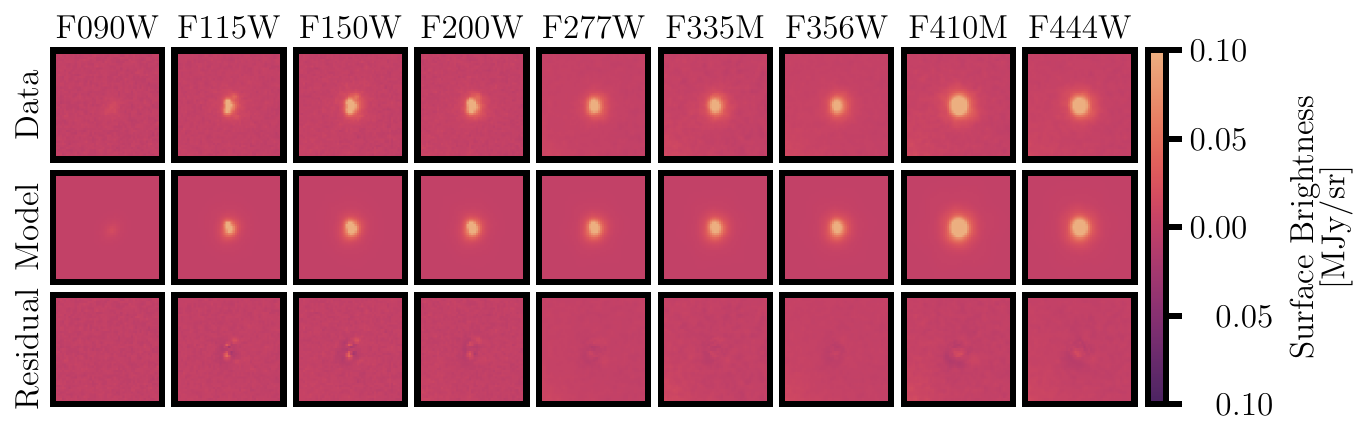}
    \caption{\textbf{Two-component morphological modeling of JADES-GS-z7-LAF using imaging from JWST/NIRCam.} We model nine photometric filters to determine this galaxy's structural properties using \texttt{pysersic} \citep[][]{Pasha:2023}, as described in Section~\ref{SectionThreeThree}. Each panel represents a $1.5^{\prime\prime} \times 1.5^{\prime\prime}$ thumbnail with data shown in the top panels, best-fit models shown in the middle panels, and residuals (data minus best-fit models) shown in the bottom panels. A color bar for the measured per-pixel surface brightness is shown on the right; this color bar is shared among each of the panels. Results from this morphological fitting are provided in Table~\ref{tab:DerivedPhotometry}, which includes the measured photometry, and Table~\ref{tab:DerivedMorphology}, which includes the inferred structural parameters. After accounting for spatially correlated noise, the reduced chi-squares are ${\chi_{\nu}}^{2} \lesssim 1$ for each filter, which demonstrates the quality of our fits. However, we note a coherent residual structure near JADES-GS-z7-LAF's UV centroid. \label{fig:Morphological_Fitting_Results}}
\end{figure*}

\begin{table*}[]
    \centering
    \caption{The measured photometry for JADES-GS-z7-LAF. As described in Section~\ref{SectionThreeThree}, we perform multi-component morphological fitting on the available, well-detected images from JWST/NIRCam using \texttt{pysersic} \citep[][]{Pasha:2023}. F090W is fit independently with a single S\'{e}rsic component while the remaining eight filters are simultaneously fit with two S\'{e}rsic components (i.e., a bulge and a disk). We also report the measured CIRC5 ($0.35\ \mathrm{arcsec}$ in radius) photometry from the JADES DR5 \citep[][]{Robertson:2026} and find overall consistency in the measured photometry from these different methods.}
    \label{tab:DerivedPhotometry}
    \begin{threeparttable}
        \begin{tabular}{c|c|cccc|c}
            \hline
            \hline
            Filter & CIRC5 Flux [nJy] & Total Flux [nJy] & F090W Flux [nJy] & Bulge Flux [nJy] & Disk Flux [nJy] & Total/CIRC5 \\
            \noalign{\vskip 1pt}
            \hline
            F090W & $23.3 \pm 2.3$ & $22.0 \pm 3.2$ & $22.0 \pm 3.2$ & n/a & n/a & $0.94$ \\
            \noalign{\vskip 1pt}
            \hline
            F115W & $151.3 \pm 2.1$ & $141.6 \pm 1.0$ & n/a & $69.0 \pm 1.7$ & $72.6 \pm 1.7$ & $0.94$ \\
            F150W & $211.5 \pm 2.5$ & $198.2 \pm 1.3$ & n/a & $100.3 \pm 2.4$ & $97.8 \pm 2.3$ & $0.94$ \\
            F200W & $198.5 \pm 2.5$ & $186.7 \pm 1.2$ & n/a & $88.2 \pm 2.3$ & $98.6 \pm 2.2$ & $0.94$ \\
            F277W & $224.9 \pm 2.0$ & $196.0 \pm 1.6$ & n/a & $83.5 \pm 3.2$ & $112.4 \pm 2.9$ & $0.87$ \\
            F335M & $240.9 \pm 3.3$ & $208.9 \pm 1.7$ & n/a & $96.1 \pm 3.5$ & $112.8 \pm 3.2$ & $0.87$ \\
            F356W & $244.7 \pm 2.2$ & $210.3 \pm 1.2$ & n/a & $95.6 \pm 2.9$ & $114.7 \pm 2.7$ & $0.86$ \\
            F410M & $462.1 \pm 2.8$ & $416.4 \pm 1.9$ & n/a & $215.6 \pm 5.0$ & $200.7 \pm 4.5$ & $0.90$ \\
            F444W & $375.0 \pm 2.7$ & $336.7 \pm 1.5$ & n/a & $175.9 \pm 4.0$ & $160.7 \pm 3.7$ & $0.90$ \\
            \noalign{\vskip 1pt}
            \hline
	\end{tabular}
    \end{threeparttable}
\end{table*}

\begin{table*}[]
    \centering
    \caption{The measured morphological parameters for JADES-GS-z7-LAF. As described in Section~\ref{SectionThreeThree} and Table~\ref{tab:DerivedPhotometry}, we perform multi-component morphological fitting on JWST/NIRCam images using \texttt{pysersic} \citep[][]{Pasha:2023}. From left to right, we report effective half-light radii (in pixels, angular, and physical units), S\'{e}rsic indices, ellipticities, and position angles. The blueward-of-Ly$\alpha$ flux measured by F090W has morphological properties that are fully consistent with those of the disk, suggesting a similar physical origin. However, we note that the centroids of F090W and the disk are offset from one another by $\Delta \theta = 1.267 \pm 0.230$ pixels ($0.0380 \pm 0.0069\ \mathrm{arcsec}$ or $0.193 \pm 0.035\ \mathrm{kpc}$), which we reconcile with a partial-covering geometry.}
    \label{tab:DerivedMorphology}
    \begin{threeparttable}
        \begin{tabular}{c|ccccccc}
            \hline
            \hline
            Component & $r_{\mathrm{eff}}\ \mathrm{\left[ pix \right]}$ & $r_{\mathrm{eff}}\ \mathrm{\left[ arcsec \right]}$ & $r_{\mathrm{eff}}\ \mathrm{\left[ kpc \right]}$ & $n$ & $b/a$ & $\theta\ \mathrm{\left[ deg \right]}$ \\
            \noalign{\vskip 1pt}
            \hline
            F090W & $4.33 \pm 0.84$ & $0.1297 \pm 0.0251$ & $0.660 \pm 0.128$ & $1.01 \pm 0.34$ & $0.275 \pm 0.110$ & $143.5 \pm 17.4$ \\
            \noalign{\vskip 1pt}
            \hline
            Bulge & $3.552 \pm 0.031$ & $0.1065 \pm 0.0009$ & $0.542 \pm 0.005$ & $4.42 \pm 0.14$ & $0.285 \pm 0.022$ & $9.9 \pm 1.7$ \\
            Disk & $3.557 \pm 0.031$ & $0.1067 \pm 0.0009$ & $0.543 \pm 0.005$ & $1\ \mathrm{\left( fixed \right)}$ & $0.272 \pm 0.011$ & $136.9 \pm 1.5$ \\
            \noalign{\vskip 1pt}
            \hline
	\end{tabular}
    \end{threeparttable}
\end{table*}

For each of the nine photometric filters, we fit cutouts with sizes of $99 \times 99\ \mathrm{pix}$ ($\approx 3.0 \times 3.0\ \mathrm{arcsec}$). We mask nearby sources using the JADES DR5 segmentation map so that only sky ($\texttt{segid} = 0$) or target ($\texttt{segid} = 217704$) pixels contribute to the likelihood. Within the adopted box size, $\approx 79\%$ of the pixels belong to masked neighbors and $\approx 21\%$ of the pixels belong to JADES-GS-z7-LAF; only a handful of pixels belong to the sky and are genuinely target-free. We fit each cutout by convolving the morphological model with each filter's point spread function (PSF) using STScI's STPSF \citep[formerly known as WebbPSF;][]{Perrin:2012, Perrin:2014}. PSFs are generated by first oversampling, then resampling to the mosaic's pixel scale, and finally rotating to the mosaic frame using each filter's aperture position angle.

We rescale the per-pixel errors by an empirically measured noise-inflation term to account for spatially correlated noise in the drizzled mosaics. This error rescaling is especially important for the long-wavelength filters (F277W, F335M, F356W, F410M, and F444W), since those mosaics are oversampled by roughly a factor of two relative to their native pixel scale. The noise-inflation terms are determined separately for each filter by measuring the variance in the reduced chi-square statistics (${\chi_{\nu}}^{2} = {\chi}^{2}/\mathrm{dof}$, where $\mathrm{dof}$ represents number of degrees of freedom) across many nearby regions that are free of sources based on the JADES DR5 segmentation map.

After accounting for spatially correlated noise with the aforementioned empirically measured noise-inflation terms, the reduced chi-squares are ${\chi_{\nu}}^{2} \lesssim 1$ for each filter, which demonstrates the quality of our fits. However, we note that there is a coherent residual structure near the UV centroid of JADES-GS-z7-LAF. This residual is most obvious at shorter-wavelengths in the F115W, F150W, and F200W filters, but it is also obvious at longer-wavelength in the F410M and F444W filters.

To quantify the residual structure, we measured the lag-1 spatial autocorrelation as a function of wavelength by shifting the residual image in each filter by one pixel in the $x$-direction and computing the Pearson correlation coefficient between each pixel and its immediate right-hand neighbor. If the noise model were correct and the source model captured all real structure, the residuals would be independent and the correlation coefficient would be centered around zero. A large, positive value means that neighboring residual pixels tend to have the same sign and similar magnitude. After accounting for correlated noise, we measure spatial autocorrelation coefficients much larger than zero ($\approx 0.3-0.5$ at shorter wavelengths and $\approx 0.7-0.9$ at longer wavelengths).

Because of the coherent residual structure near the UV centroid of JADES-GS-z7-LAF, we fit some other morphological models with additional structural freedom to see if they were preferred by the data. In particular, we tried a two-component model with independent per-filter S\'{e}rsic parameters, a two-component model with the disk having a free S\'{e}rsic index $n$, a two-component model with a linear background, and a three-component model with a second disk (i.e., with free S\'{e}rsic index $n$). We found that none of these other morphological models with additional structural freedom were able to significantly reduce the reduced chi-squares nor autocorrelation coefficients. Instead, the additional components often had their parameters pinned at the prior \mbox{boundaries} (e.g., with effective radii against the upper bound). Thus, while we are unable to fit the coherent residual structure near the UV centroid of \mbox{JADES-GS-z7-LAF}, it should be considered a real, unmodeled small-scale feature rather than noise or an instrumental artifact.

\subsection{Testing for Instrumental Systematics}
\label{SectionThreeFour}

It is possible that morphological broadening of the line spread function (LSF) could be responsible for the observed blueward-of-Ly$\alpha$ flux, since the NIRSpec/PRISM LSF has extended wings, and especially at $\lambda_{\mathrm{obs}} \lesssim 1\ \mu\mathrm{m}$. This means that light from JADES-GS-z7-LAF's bright UV continuum could in principle leak into the opaque region blueward of the Ly$\alpha$ break at $\lambda_{\mathrm{rest}} = 1216\ \mathrm{\AA}$, thereby mimicking genuine flux that is blueward-of-Ly$\alpha$. We test for this using the following methodology and determine that the likelihood of the observed emission coming from an instrumental systematic is negligible.

We forward modeled JADES-GS-z7-LAF's bright UV continuum through JWST/NIRSpec's MSA optics using \texttt{msafit} \citep[][]{deGraaff:2024}\footnote{\href{https://github.com/annadeg/jwst-msafit}{https://github.com/annadeg/jwst-msafit}}. To accomplish this, we construct three effective wavelength-dependent LSF variants that bracket the true resolving power at our target's specific position within the shutter\footnote{We note that \texttt{msafit} only provides LSFs for JWST/NIRSpec's third quadrant (Q3), but JADES-GS-z7-LAF's shutters are in the second quadrant (Q2). We adopt the Q3 LSFs and assume there is no quadrant-dependent instrumental effects.}. These three variants include an idealized point source, a best-fit morphology derived from modeling a S\'{e}rsic profile to the observed F150W image (see also Section~\ref{SectionThreeThree} for a detailed discussion of the morphological fitting), and unrealistic S\'{e}rsic profile that is $20\%$ larger than the best-fit morphology. For each of the three LSFs, we construct three simple models for the spectral energy distribution (SED) with each described by zero intrinsic flux blueward of Ly$\alpha$ and a power-law continuum extending redward of Ly$\alpha$ into the rest-frame UV. These three models for the SED include power-law continuum slopes of $\beta_{\mathrm{UV}} = \{ -2.4, \, -2.2, \, -2.0 \}$, which correspond to the measured photometric value from Table~\ref{tab:PhysicalProperties} ($\beta_{\mathrm{UV}} \approx -2.2$), along with significantly bluer and redder slopes than what we measure directly from the data. Finally, we apply IGM attenuation using the patchy-reionization model provided by \citet{Mason:2020}.

For each of these nine physical models (three LSF variants times three power-law continuum slopes), we convolve the model with its corresponding wavelength-dependent Gaussian LSF. We then add $1{,}000$ independent noise realizations from the observed spectrum's pixel covariance and re-measure the blueward-of-Ly$\alpha$ flux by adopting the same methodology used for the observed data. Without noise, the maximum blueward-of-Ly$\alpha$ flux arising purely from instrumental LSF leakage is $f \approx 1.6 \times 10^{-18}\ \mathrm{erg/s/cm^{2}}$ when integrated over $\lambda_{\mathrm{rest}} = 912-1216\ \mathrm{\AA}$ for the most pessimistically broadened LSF and the bluest SED. With noise, the maximum blueward-of-Ly$\alpha$ flux is $f \approx 1.4 \times 10^{-17}\ \mathrm{erg/s/cm^{2}}$ when integrating over the same window. This corresponds to the maximum across all nine physical models and $1{,}000$ independent noise realizations. These values are roughly two times smaller than our measured flux value of $f \approx 2.80 \pm 0.42 \times 10^{-17}\ \mathrm{erg/s/cm^{2}}$ from Section~\ref{SectionThreeTwo}.

Notably, none of these models include damping-wing absorption or nebular continuum emission. We note that the presence of a DLA or a strong nebular continuum would make it even harder to fit the observed emission. We conclude that an instrumental systematic from the LSF cannot explain the observed blueward-of-Ly$\alpha$ flux, even when considering $\approx 3 \sigma$ outliers under the most pessimistic modeling assumptions.

% After speaking with experts on the JWST/NIRSpec instrument science team, we determined that this was the only instrumental systematic that could potentially explain the observed blueward-of-Ly$\alpha$ flux. However, one other possibility is contamination from a foreground interloper, which is what we consider in Section~\ref{SectionThreeFive}.

\subsection{Testing for Foreground Interlopers}
\label{SectionThreeFive}

In Section~\ref{SectionThreeFour}, we explored the possibility of morphological  broadening of the LSF being responsible for the observed blueward-of-Ly$\alpha$ flux and determined that the likelihood of the observed emission coming from an instrumental systematic is negligible. However, it is also possible that an interloping foreground galaxy could be responsible for the observed emission. We test for this using the following methodology and determine that the likelihood of the observed flux coming from a foreground interloper is negligible, although we cannot formally reject this interpretation with the existing data.

For the time being, let's assume that the blueward-of-Ly$\alpha$ flux in the NIRSpec/PRISM spectroscopy and NIRCam/F090W imaging was produced by an interloping foreground galaxy. Since there is no emission at wavelengths shorter than $\lambda_{\mathrm{obs}} \approx 0.78\ \mu\mathrm{m}$, we further assume that the foreground galaxy has its Ly$\alpha$ break at this wavelength, which corresponds to a redshift of $z \approx 5.4$. This is roughly the same redshift as an extreme high-redshift galaxy overdensity in GOODS-S \citep[][]{Helton:2024a, Helton:2024b}. Based on these assumptions, and assuming $\beta_{\mathrm{UV}} = -2$, we would infer $M_{\mathrm{UV}} = -18.59 \pm 0.11$ using the measured F090W flux ($f_{\mathrm{F090W}} = 23.3 \pm 2.3\ \mathrm{nJy}$ from the CIRC5 photometry reported in Table~\ref{tab:DerivedPhotometry}).

At $z \approx 5.4$, $\mathrm{[O\,\textsc{iii}]} \lambda 5007$ would be at $\lambda_{\mathrm{obs}} \approx 3.2\ \mu\mathrm{m}$ and $\mathrm{H}\alpha$ at $\lambda_{\mathrm{obs}} \approx 4.2\ \mu\mathrm{m}$ for the hypothetical foreground galaxy. As the two strongest rest-frame optical emission lines, we would expect to observe these lines for most star-forming Ly$\alpha$ break galaxies at this redshift. Using the conversion factors from \citet{Kennicutt:2012}, we convert the UV luminosity into a star-formation rate ($\mathrm{SFR}_{\mathrm{UV}} \approx 1\ M_{\odot} / \mathrm{yr}$), then into an $\mathrm{H}\alpha$ luminosity ($L_{\mathrm{H}\alpha} \approx 2.0 \times 10^{41}\ \mathrm{erg/s}$), and finally into a predicted $\mathrm{H}\alpha$ flux ($f_{\mathrm{H}\alpha} \approx 60.3 \times 10^{-20}\ \mathrm{erg/s/cm^{2}}$) based on the luminosity distance at $z \approx 5.4$. The predicted $\mathrm{H}\alpha$ flux is far above the noise floor at its expected location. For reference, we measure a $3 \sigma$ upper limit of $\approx 22.8 \times 10^{-20}\ \mathrm{erg/s/cm^{2}}$ at $\lambda_{\mathrm{obs}} \approx 4.2\ \mu\mathrm{m}$ from the medium-resolution ($R \approx 1000$) NIRSpec/G395M spectrum.

If we instead assume typical high-redshift conversion factors for the star-formation rates of low-metallicity galaxies \citep[see also the discussions in][]{Shapley:2023, Helton:2026a}, the predicted $\mathrm{H}\alpha$ flux would be larger by a factor of $2-3 \times$, thus worsening the discrepancy between our prediction and the upper limit on the line flux at the expected location of $\mathrm{H}\alpha$.

Finally, we perform a probabilistic test to determine the likelihood of a coincident galaxy appearing within a certain radius of JADES-GS-z7-LAF by integrating the UV luminosity function (LF) at $z = 5-7$ down to $M_{\mathrm{UV}} = -18.5$, which roughly corresponds to the UV luminosity that we would measure from the F090W flux if we assume the F090W flux arises from an interloping foreground galaxy at those redshifts. We adopt the Schechter function parameters from \citet{Bouwens:2021} and linearly interpolate those parameters between their redshift bins of $z \approx 5$, $6$, and $7$. We infer an incredibly low probability of finding at least one galaxy within $0.1^{\prime\prime}$ of JADES-GS-z7-LAF, $P (N \geq 1) \approx 0.014\%$, by assuming Poissonian statistics. The inferred probability increases by $\approx 100 \times$ if we instead require at least one galaxy within $1^{\prime\prime}$, $P (N \geq 1) \approx 1.4\%$. However, we note that a separation of $1^{\prime\prime}$ is inconsistent with the quoted offset ($\Delta \theta \lesssim 0.1^{\prime\prime}$) from Sections~\ref{SectionThreeTwo} and \ref{SectionThreeThree}.

\subsection{Inferring Rest-Frame UV Continuum Properties}
\label{SectionThreeSix}

We fit for the rest-frame UV continuum properties of JADES-GS-z7-LAF using a similar methodology as the one described in \citet{Topping:2024} and \citet{Helton:2026b}. We use the CIRC5 photometry from the JADES DR5 \citep[][]{Robertson:2026} to infer rest-frame UV absolute magnitudes ($M_{\mathrm{UV}}$) and continuum slopes ($\beta_{\mathrm{UV}}$). The continuum slopes are inferred by fitting a simple power law to the observed broad-band photometry at $\lambda_{\mathrm{rest}} \approx 1200-2600\ \mathrm{\AA}$ \citep[$f_{\lambda} \propto \lambda^{\beta_{\mathrm{UV}}}$;][]{Calzetti:1994} while the continuum luminosities are inferred by measuring $\nu L_{\nu}$ at $\lambda_{\mathrm{rest}}$ using the same power-law fit as used for the continuum slopes. We obtain $M_{\mathrm{UV}} = -21.55 \pm 0.02$ and $\beta_{\mathrm{UV}} = -2.23 \pm 0.06$ for JADES-GS-z7-LAF using the F150W, F200W, and F277W filters from JWST/NIRCam. We note that the F150W filter contains strong nebular emission lines (e.g., $\mathrm{[C\,\textsc{iv}]} \lambda\lambda 1548{,}1551$ and $\mathrm{[C\,\textsc{iii}]} \lambda\lambda 1907{,}1909$) that will bias the continuum luminosities bright and continuum slopes blue when compared to a pure continuum.

JADES-GS-z7-LAF's UV absolute magnitude places it as one of the brightest galaxies at $z > 3$ with spectroscopic confirmation in JADES \citep[e.g.,][]{Helton:2026a} while its continuum slope is typical for galaxies of this brightness at $z \approx 7-8$ \citep[e.g.,][]{Topping:2024, Saxena:2026} and is considered somewhat reddened. This reddening is potentially caused by an evolved stellar population, moderate dust attenuation, relatively high stellar metallicities, or contributions from a nebular continuum (i.e., two-photon emission). For reference, JADES-GS-z7-LAF's UV absolute magnitude is roughly identical to that of GN-z11 at $z \approx 10.6$. We argue that JADES-GS-z7-LAF's brightness is further evidence against an interloper and thus supports the Ly$\alpha$ forest interpretation, since it would be incredibly unlikely for a faint foreground interloper to almost perfectly overlap with one of the brightest galaxies at $z > 3$ with spectroscopic confirmation in JADES.

We then separately infer rest-frame UV continuum properties for the bulge and disk components using the same methodology to see if there are any differences in their properties. We find that the bulge is described by $M_{\mathrm{UV}} = -20.783 \pm 0.032$ and $\beta_{\mathrm{UV}} = -2.460 \pm 0.074$ while the disk is described by $M_{\mathrm{UV}} = -20.674 \pm 0.032$ and $\beta_{\mathrm{UV}} = -1.979 \pm 0.067$. Thus, the bulge is brighter and bluer than the disk in the rest-frame UV continuum. However, the disk is brighter and redder than the bulge in the rest-frame optical continuum as evidenced by the elevated fluxes in the F277W, F335M, and F356W filters while the bulge has stronger rest-frame optical emission lines when compared to the disk as evidenced by the elevated fluxes in F410M relative to F356W. Altogether, these results suggest that the bulge is younger and more actively star-forming than the disk component.

\subsection{Damping-Wing \& Nebular Continuum Fitting}
\label{SectionThreeSeven}

\begin{figure*}
    \centering
    \includegraphics[width=1.0\linewidth]{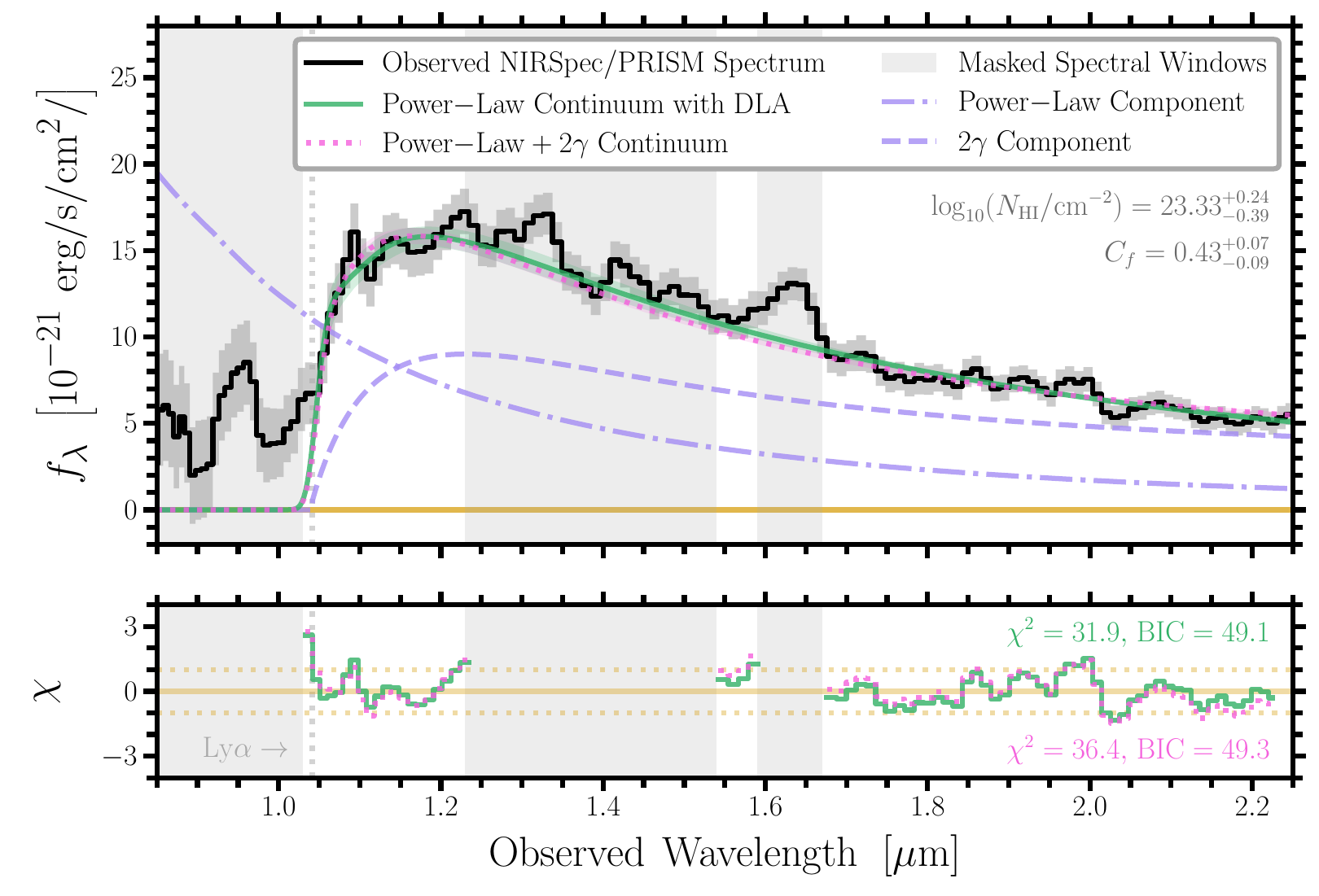}
    \caption{\textbf{Fitting the smooth and gradual turnover in JADES-GS-z7-LAF's spectrum around the Ly\boldmath$\alpha$ break.} We model the observed spectral feature in the NIRSpec/PRISM data using the methods described in \citet{Witstok:2025b}. The first physical model assumes this feature is associated with the damping-wing absorption that arises from large quantities of neutral hydrogen in dense intervening absorbing systems (solid black line). The second model assumes this feature is associated with the $2 \gamma$ nebular continuum emission that arises from $2s \rightarrow 1s$ forbidden transitions in $\mathrm{H\,\textsc{ii}}$ regions (dashed blue line). Based on the inferred BICs, we find that the two models are equally preferred ($\Delta \mathrm{BIC} \approx 0$). Our preferred interpretation suggests that $\approx 60-70\%$ of the observed flux at $\lambda_{\mathrm{rest}} = 1500\ \mathrm{\AA}$ is associated with nebular emission. This interpretation is supported by the detection ($\approx 3.4 \sigma$) of a Balmer jump in the observed NIRSpec/PRISM spectrum (see discussion in Section~\ref{SectionThreeSeven} and Figure~\ref{fig:Balmer_Jump_Fit}). \label{fig:Damping_Wing_Fitting_Results}}
\end{figure*}

Qualitatively, there is a smooth and gradual turnover in JADES-GS-z7-LAF's SED around the Ly$\alpha$ break, as we observe in the NIRSpec/PRISM spectrum (Figure~\ref{fig:Full_PRISM_Spectrum}). This spectral feature is commonly associated with the damping-wing absorption that sometimes arises from large quantities of neutral hydrogen in dense intervening absorbing systems \citep[e.g.,][]{Heintz:2024, Heintz:2025}. Alternatively, this feature is sometimes associated with the two-photon ($2 \gamma$) nebular continuum that arises from $2s \rightarrow 1s$ forbidden transitions in $\mathrm{H\,\textsc{ii}}$ regions \citep[e.g.,][]{Cameron:2024, Katz:2025}. For the $2 \gamma$ emission, this occurs when neutral hydrogen atoms are populated by recombination cascades, thus producing a continuum from photon pairs summing to $10.2\ \mathrm{eV}$. This nebular continuum can mimic strong damping-wing absorption with $N_{\mathrm{H\,\textsc{i}}} \gtrsim 10^{22}\ \mathrm{cm^{-2}}$ since $2 \gamma$ emission peaks around $\lambda_{\mathrm{rest}} \approx 1500\ \mathrm{\AA}$ and drops off toward Ly$\alpha$. In addition to $\mathrm{H\,\textsc{ii}}$ regions, we note that $2 \gamma$ emission can also arise from warm cooling flows that may be common in the high-redshift Universe \citep[e.g.,][]{Dijkstra:2009}.

To model the smooth and gradual turnover around the Ly$\alpha$ break, we follow the methodology described in \citet{Witstok:2025b} using publicly available fitting code from GitHub\footnote{\href{https://github.com/joriswitstok/lymana_absorption}{https://github.com/joriswitstok/lymana\_absorption}} \citep[for reference, see also][]{DEugenio:2024, Hainline:2024}. This code utilizes a Bayesian inference framework built around the Python implementation of the \texttt{MultiNest} multimodal nested sampling algorithm \citep[][]{Feroz:2009}, \texttt{PyMultiNest} \citep[][]{Perrin:2014}. We describe the fitting below.

For damping-wing model, the absorbing column is described by two physically distinct components acting upon a power-law continuum ($f_{\lambda} \propto \lambda^{\beta_{\mathrm{UV}}}$) at rest-frame UV wavelengths. By using the power-law continuum, we remain agnostic about the nature of the ionizing source.

One component of our absorbing medium represents the DLA and is parametrized by a column density of neutral hydrogen $N_{\mathrm{H\,\textsc{i}}}$ and a covering fraction $C_{f}$. The covering fraction accounts for a partial, non-uniform covering of the source, which could be caused by an inhomogeneous interstellar medium (ISM) or an ionization cone that allows Ly$\alpha$ photons to escape unattenuated. The absorption cross-section is based on the Voigt profile approximation provided by \citet{Tasitsiomi:2006}, but using an updated approximation with a quantum mechanical correction from \citet{Bach:2015}. The second component represents the IGM using the patchy-reionization model provided by \citet{Mason:2020}. We assume the mean cosmic density for gas in the DLA and IGM ($\overline{n_{\mathrm{H}}} \approx 1.20 \times 10^{-4}\ \mathrm{cm}^{-3}$ at \ResultSpectroscopicRedshift). We further assume this gas is at rest with respect to our source. The damping-wing model has four free parameters: $M_{\mathrm{UV}}$, $\beta_{\mathrm{UV}}$, $\log_{10} (N_{\mathrm{H\,\textsc{i}}} / \mathrm{cm^{-2}})$, and $C_{f}$.

For our fiducial $2 \gamma$ model, we consider a combination consisting of a power-law continuum (using the same parametrization as above) and a $2 \gamma$ nebular continuum \citep[e.g.,][]{Katz:2025}. $2 \gamma$ emission has a fixed shape uniquely determined by the gas density and temperature \citep[e.g.,][]{Spitzer:1951} and thus only requires one free parameter, the normalization $O$. The shape of $2 \gamma$ emission is fixed because it reflects the uniform probability distribution of dividing the energy of Ly$\alpha$ between two photons.  We adopt the self-consistent $\mathrm{[O\,\textsc{iii}]}$ solution presented in Table~\ref{tab:PhysicalProperties} and Figure~\ref{fig:Electron_Density_And_Temperature} for the electron density and temperature. However, we caution that the inferred density for $\mathrm{[O\,\textsc{iii}]}$ is nearly an order of magnitude larger than the critical density of $2 \gamma$ emission \citep[$n_{e{,}\mathrm{crit}} \approx 2 \times 10^{4}\ \mathrm{cm}^{-3}$; e.g.,][]{Osterbrock:2006}. We can reconcile these potentially conflicting solutions if the $2 \gamma$ emission arises from lower density gas than the $\mathrm{[O\,\textsc{iii}]}$ emission. Lower density gas certainly exists, as evidenced by the presence of $\mathrm{[O\,\textsc{ii}]}$ with critical densities of $n_{e{,}\mathrm{crit}} \approx 2-3 \times 10^{3}\ \mathrm{cm}^{-3}$. The $2 \gamma$ model has three free parameters: $O$, $M_{\mathrm{UV}}$, and $\beta_{\mathrm{UV}}$.

We fit the low-resolution NIRSpec/PRISM spectrum over a spectral window covering rest-frame wavelengths $\lambda_{\mathrm{rest}} = 1200-2700\ \mathrm{\AA}$ (at \ResultSpectroscopicRedshift, this corresponds to observed-frame wavelengths $\lambda_{\mathrm{obs}} \approx 1.03-2.31\ \mu\mathrm{m}$). We mask prominent strong nebular emission lines, such as $\mathrm{[C\,\textsc{iv}]} \lambda\lambda 1548{,}1551$ and $\mathrm{[C\,\textsc{iii}]} \lambda\lambda 1907{,}1909$).

Figure~\ref{fig:Damping_Wing_Fitting_Results} illustrates our results from fitting the smooth and gradual turnover in JADES-GS-z7-LAF's spectrum. We use the information criterion (BIC) to evaluate fits from different physical models since it considers differences in the number of free parameters between models by using a complexity penalty. We find that the fiducial $2 \gamma$ model containing three free parameters is equally preferred ($\Delta \mathrm{BIC} \approx 0$) when compared to the fiducial DLA model containing four free parameters, which would require an extreme DLA with $N_{\mathrm{H\,\textsc{i}}} \approx 10^{23.3}\ \mathrm{cm}^{-2}$. Such a high column density of neutral gas would be surprising to discover in a galaxy that also has blueward-of-Ly$\alpha$ flux and would be one of the most extreme DLAs ever discovered \citep[e.g.,][]{Heintz:2024, Heintz:2025}. Adopting a similar methodology to the one used in \citet{Heintz:2025_z14}, we assume spherical symmetry and that the neutral gas has the same half-light radius as the rest-frame UV light to estimate a total neutral gas mass of $\mathrm{log}_{10} ( M_{\mathrm{H\,\textsc{i}}} / M_{\odot} ) \approx 9.0_{-0.4}^{+0.2}$ from the inferred column density. If we instead assume the neutral gas has a radius $\approx 3 \times$ larger than the rest-frame UV light, as was assumed in \citet{Heintz:2025_z14}, we would estimate a total neutral gas mass of $\mathrm{log}_{10} ( M_{\mathrm{H\,\textsc{i}}} / M_{\odot} ) \approx 10.0_{-0.4}^{+0.2}$. We can reconcile these two competing observations (i.e., blueward-of-Ly$\alpha$ flux alongside an enormous DLA) by invoking a partial-covering geometry that allows Lyman-series photons to escape through a low column density channel offset from the bulk of the UV continuum. This requires a contrived geometry and is not our preferred interpretation of the spectroscopic data.

Our preferred interpretation of the data suggests that $\approx 60-70\%$ of the observed flux at $\lambda_{\mathrm{rest}} = 1500\ \mathrm{\AA}$ is associated with strong nebular continuum emission, which would imply that the ionizing photon production efficiency inferred in Section~\ref{SectionThreeOne} is underestimated by $\Delta \xi_{\mathrm{ion}} \approx 0.4-0.5\ \mathrm{dex}$. In this proposed scenario, the intrinsic value would be $\xi_{\mathrm{ion}} \approx 10^{26.0-26.1}\ \mathrm{Hz/erg}$, which is an extreme value consistent with the stellar population maximum \citep[e.g.,][]{Schaerer:2025}.

\begin{figure}
    \centering
    \includegraphics[width=1.0\linewidth]{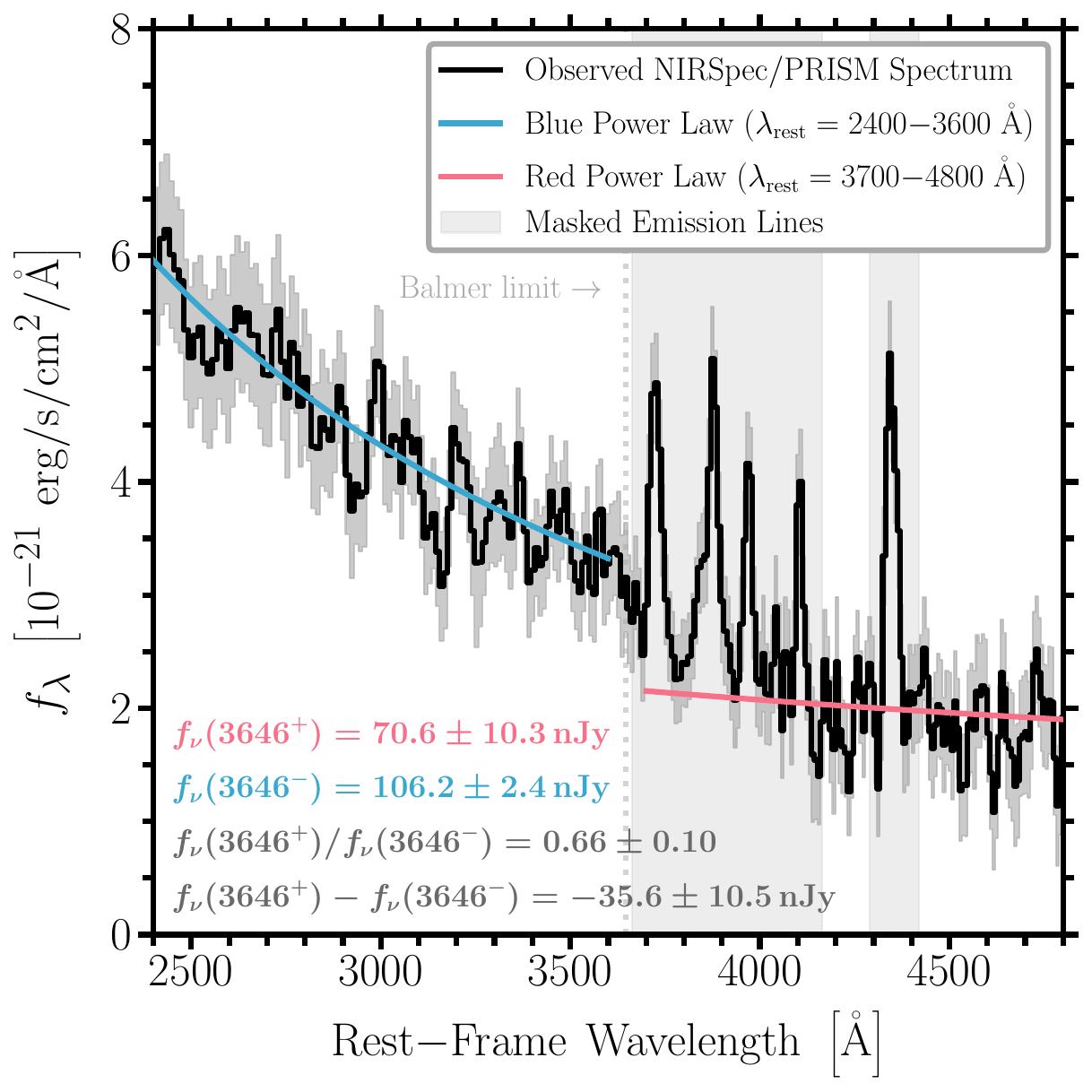}
    \caption{\textbf{Fitting the rest-frame optical continuum.} We fit a power-law continuum ($f_{\lambda} \propto \lambda^{\beta_{\mathrm{opt}}}$) on either side of the Balmer limit to measure the strength of the Balmer jump by comparing the predicted fluxes at $\lambda_{\mathrm{rest}} = 3645\ \mathrm{\AA}$. We mask emission lines with significant detections above $3 \sigma$, as shown by the shaded regions. The blue continuum is larger than the red at $\lambda_{\mathrm{rest}} = 3646\ \mathrm{\AA}$ by $\Delta f_{\nu} = 35.6 \pm 10.5\ \mathrm{nJy}$, which confirms a $\approx 3.4 \sigma$ detection of the Balmer jump. This result is consistent with the interpretation of a strong nebular continuum within JADES-GS-z7-LAF (see Section~\ref{SectionThreeSeven}). \label{fig:Balmer_Jump_Fit}}
\end{figure}

If the $2 \gamma$ nebular continuum emission is dominating the rest-frame UV luminosity of JADES-GS-z7-LAF, as suggested by Figure~\ref{fig:Damping_Wing_Fitting_Results}, it is predicted that the free-bound nebular continuum should also be quite strong. Free-bound emission increases toward longer wavelengths up until the Balmer continuum limit at $\lambda_{\mathrm{rest}} = 3645\ \mathrm{\AA}$. At this wavelength, a Balmer jump occurs in the free-bound continuum \citep[see also, e.g.,][]{Katz:2025}. To measure the Balmer jump, we fit power-law continua in $f_{\lambda}$ on either side of the Balmer continuum limit and compare the predicted continuum fluxes at $\lambda_{\mathrm{rest}} = 3645\ \mathrm{\AA}$ from the two fits. We mask prominent emission lines to avoid biasing the continuum measurements. Figure~\ref{fig:Balmer_Jump_Fit} illustrates our fitting results. On the blue side, we measure $f_{\nu{,}\mathrm{blue}} = 106.2 \pm 2.4\ \mathrm{nJy}$, while on the red side, we measure $f_{\nu{,}\mathrm{red}} = 70.6 \pm 10.3\ \mathrm{nJy}$. These measurements correspond to $f_{\nu{,}\mathrm{red}} / f_{\nu{,}\mathrm{blue}} = 0.66 \pm 0.100$ and $\Delta f_{\nu} = -35.6 \pm 10.5\ \mathrm{nJy}$. This represents a $\approx 3.4 \sigma$ detection of the Balmer jump and provides further support for the interpretation of a strong nebular continuum.

\subsection{Panchromatic Spectral Energy Distribution Fitting}
\label{SectionThreeEight}

We self-consistently determine the properties of the stellar populations, nebular gas, and dust by modeling the panchromatic SED for JADES-GS-z7-LAF using the Bayesian SED fitting code \texttt{Prospector} \citep[][]{Johnson:2021} and \texttt{Cue} \citep[][]{Li:2024, Li:2025}, the fast and flexible neural network emulator trained on the \texttt{Cloudy} photoionization modeling code \citep[][]{Ferland:1998, Chatzikos:2023}. We adopt stellar population synthesis models from the Flexible Stellar Population Synthesis (\texttt{FSPS}) code \citep[][]{Conroy:2009, Conroy:2010}. We adopt synthetic stellar libraries from C3K \citep[][]{Conroy:2019} in addition to stellar evolutionary tracks and isochrones from MIST \citep[][]{Choi:2016, Dotter:2016}. Finally, we choose to samples posterior distributions with the dynamic nested sampling code \texttt{dynesty} \citep[][]{Speagle:2020}.

\begin{figure*}
    \centering
    \includegraphics[width=1.0\linewidth]{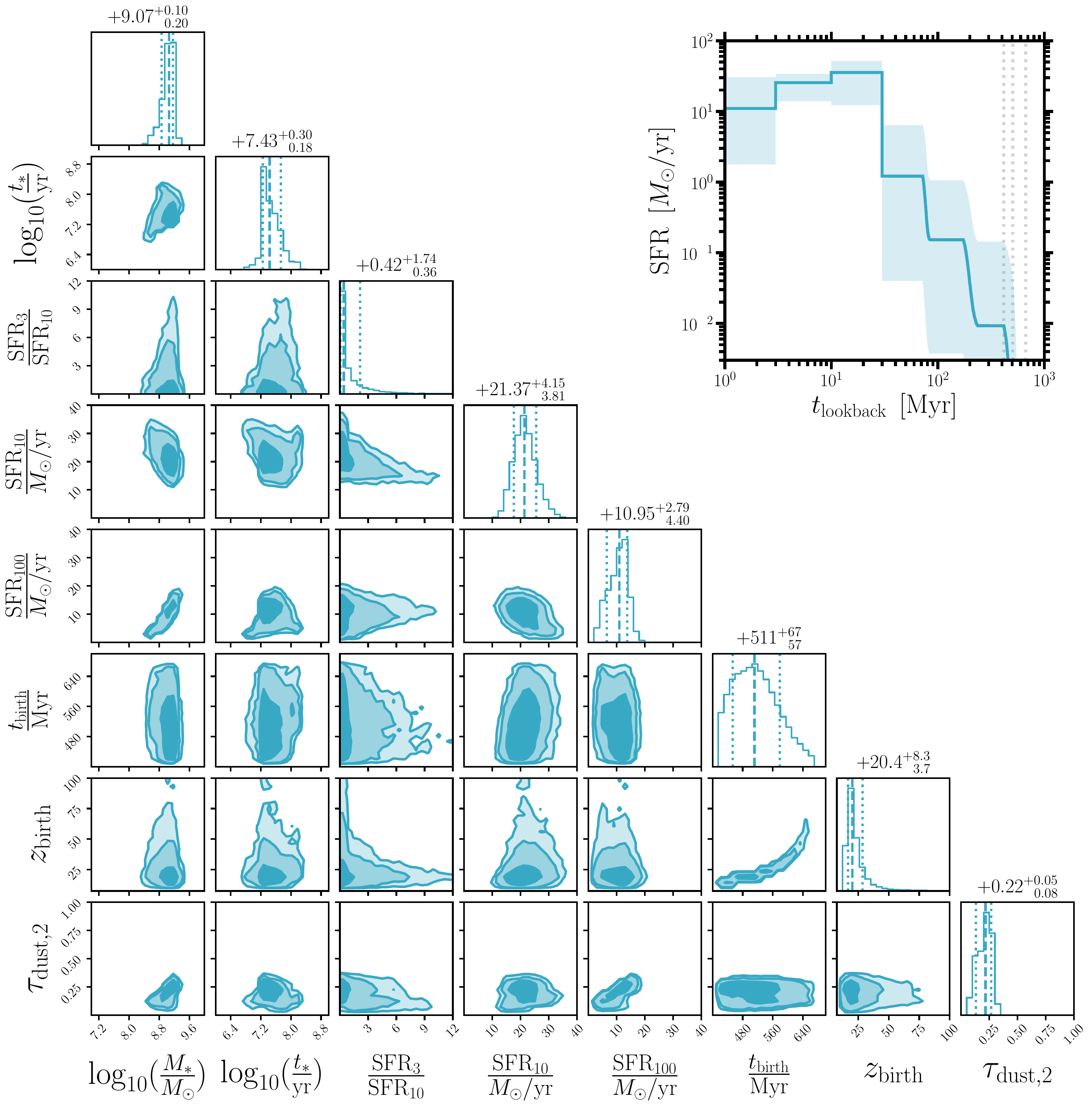}
    \caption{\textbf{Constraints on physical properties by simultaneously modeling the properties of the stellar populations, nebular gas, and dust.} We utilize the Bayesian SED fitting code \texttt{Prospector} \citep[][]{Johnson:2021} and \texttt{Cue} \citep[][]{Li:2024, Li:2025}, the fast and flexible neural network emulator trained on the \texttt{Cloudy} photoionization modeling code, to model the panchromatic SED of JADES-GS-z7-LAF. \textit{Lower left:} The joint posterior distributions are shown for stellar mass, mass-weighted stellar age, ratio of recent star-formation rates, star-formation rates averaged over the previous $10$ and $100\ \mathrm{Myr}$, the lookback time and redshift corresponding to when the first stars formed, and the $V$-band optical depth. The contours in the off-diagonal panels represent the $68\%$, $95\%$, and $99\%$ confidence intervals of the modeling. The dashed and dotted lines in the diagonal panels represent the  medians and $68\%$ confidence intervals. \textit{Upper right:} The solid line represents the median for the inferred star-formation history while the shaded region represents the $68\%$ confidence interval. \textit{All panels:} JADES-GS-z7-LAF is a relatively massive ($M_{\ast} \approx 10^{9.1}\,M_{\odot}$) and young ($t_{\ast} \approx 27\,\mathrm{Myr}$) star-forming galaxy with minimal dust attenuation that formed most of its stars in a recent burst ($\approx 10-30\,\mathrm{Myr}$ ago) and has experienced a declining star-formation history ever since. \label{fig:Prospector_Corner_Plot}}
\end{figure*}

To our knowledge, this is the first time someone has used the combination of \texttt{Prospector} and \texttt{Cue} to model the SED of a high-redshift galaxy. \texttt{Cue} provides more flexibility than the pre-computed \texttt{Cloudy} photoionization models that are typically used. \texttt{Cue} will typically approximate the intrinsic ionizing spectrum with a four-part piecewise power law. However, we choose for \texttt{Cue}'s intrinsic ionizing spectrum to be fixed by \texttt{Prospector}'s stellar population predictions. The assumed physical model will be described in detail in a forthcoming paper (M.~Laska et~al., in preparation). We describe the most important details of the physical model below.

We adopt a nonparametric SFH for our fiducial model, which is characterized by the logarithmic ratios of star-formation rates between six adjacent time bins that are spaced logarithmically in lookback time. We assume a rising prior on the star-formation ratios that roughly follows the redshift evolution of dark matter halo accretion rates \citep[e.g.,][]{Tacchella:2018}, as was first used in \citet[][]{Turner:2025}. Previous implementations of this nonparametric SFH have typically assumed no star formation occurs before $z = 20$. However, as will be discussed in a forthcoming paper (J.~M.~Helton et~al., in preparation), we parametrize the redshift corresponding to when the first stars form, which we call the ``birth'' redshift, or $z_{\mathrm{birth}}$. This allows us to try and constrain the onset of star formation within JADES-GS-z7-LAF.

Our fiducial model has $16$ free parameters of physical interest: systemic redshift, stellar metallicity, total stellar mass formed, $V$-band optical depth, the lookback time corresponding to when the first stars formed, five ratios of star-formation rates in adjacent time bins, five parameters for the nebular gas (gas-phase oxygen abundance, carbon-to-oxygen ratio, nitrogen-to-oxygen ratio, ionization parameter, and electron density), and the column density of neutral hydrogen in a DLA. Our model also has two free nuisance parameters for each emission line detected at $> 3 \sigma$ in the NIRSpec/PRISM spectrum (see Table~\ref{tab:EmissionLineProperties}), for a total of $20$ more nuisance parameters. We include two final nuisance parameters for the calibration model, one for an outlier fraction and the other as a noise inflation term. We fit the low-resolution NIRSpec/PRISM spectrum at all wavelengths with $\lambda_{\mathrm{rest}} > 1216\ \mathrm{\AA}$ together with the available JWST/NIRCam and JWST/MIRI photometry as measured using the CIRC5 ($R = 0.35^{\prime\prime}$) circular aperture from the JADES DR5. Except for MIRI/F560W, all other mid-infrared photometric points are upper limits.

Figure~\ref{fig:Prospector_Corner_Plot} provides constraints on the joint posterior distributions for some of the most important physical properties of JADES-GS-z7-LAF. In the lower left, from left to right, these properties include total stellar mass formed, the mass-weighted stellar age ($t_{\ast}$), a ratio of recent star-formation rates ($\mathrm{SFR}_{3} / \mathrm{SFR}_{10}$), star-formation rates averaged over the previous $10$ and $100\ \mathrm{Myr}$ ($\mathrm{SFR}_{10}$ and $\mathrm{SFR}_{100}$), the lookback time and redshift that correspond to when the first stars formed ($t_{\mathrm{birth}}$ and $z_{\mathrm{birth}}$), and the $V$-band optical depth ($\tau_{\mathrm{dust}{,}2}$). In the upper right, we show the inferred star-formation history. For all of the panels, dashed lines represent the medians of the \texttt{Prospector} predictions while dotted lines or shaded regions represent the $68\%$ confidence intervals.

The results provided in Figure~\ref{fig:Prospector_Corner_Plot} and Table~\ref{tab:PhysicalProperties} demonstrate that JADES-GS-z7-LAF is a relatively massive ($M_{\ast} \approx 10^{9.1}\ M_{\odot}$) and young ($t_{\ast} \approx 27\ \mathrm{Myr}$) star-forming galaxy that formed most of its stars in a recent burst of star formation ($\approx 10-30\ \mathrm{Myr}$ ago), but has experienced a declining star-formation history ever since. It has a small amount of dust attenuation at rest-frame optical wavelengths. The time corresponding to when the first stars formed, $t_{\mathrm{birth}}$ is almost entirely unconstrained.

When looking at the residuals $\chi$ between observations and \texttt{Prospector} model predictions, we found overall consistency between the two data and models. The most notable outliers with $\chi \approx 2$ correspond to the smooth and gradual turnover around the Ly$\alpha$ break in the low-resolution spectroscopic data and the free-bound jump around the Balmer limit in the photometric data. Both of these sets of outliers can be attributed to \texttt{Prospector} and \texttt{Cue} under-predicting the nebular continuum contributions, where $2 \gamma$ becomes important in the rest-frame UV and free-bound in the rest-frame optical. This is possibly the result of \texttt{Cue} being trained on a limited grid of electron densities with an upper bound of $n_{e} = 10^{4}\ \mathrm{cm}^{-3}$. As inferred in Section~\ref{SectionThreeOne}, the $\mathrm{[O\,\textsc{iii}]}$ electron density for JADES-GS-z7-LAF is roughly an order of magnitude larger than \texttt{Cue}'s upper bound. For this reason, we choose not to report the \texttt{Cue}-inferred nebular properties, since they are biased by the limited training grid. This motivates the expansion of \texttt{Cue}'s grid to even larger electron densities ($n_{e} \gg 10^{4}\ \mathrm{cm}^{-3}$).

\subsection{Spatial-Spectral Correlation for Transmission}
\label{SectionThreeNine}

We qualitatively correlate the blueward-of-Ly$\alpha$ flux from JADES-GS-z7-LAF with the spatial and kinematic locations of foreground galaxies to see if these overlap with one another. As will be discussed in Section~\ref{SectionFour}, the presence of transmitted flux blueward of Ly$\alpha$ suggests an extraordinarily transparent sightline in the direction of JADES-GS-z7-LAF that stretches well beyond this galaxy's proximity zone. For such transmission to occur, ionizing radiation from galaxies needs to dominate over the local cosmic background levels. This sort of spatial-spectral correlation is often performed in the context of high-redshift quasars with observed transmission in the IGM \citep[e.g.,][]{Kashino:2023, Jin:2024, Kakiichi:2025, Kashino:2026}.

\begin{figure*}
    \centering
    \includegraphics[width=1.0\linewidth]{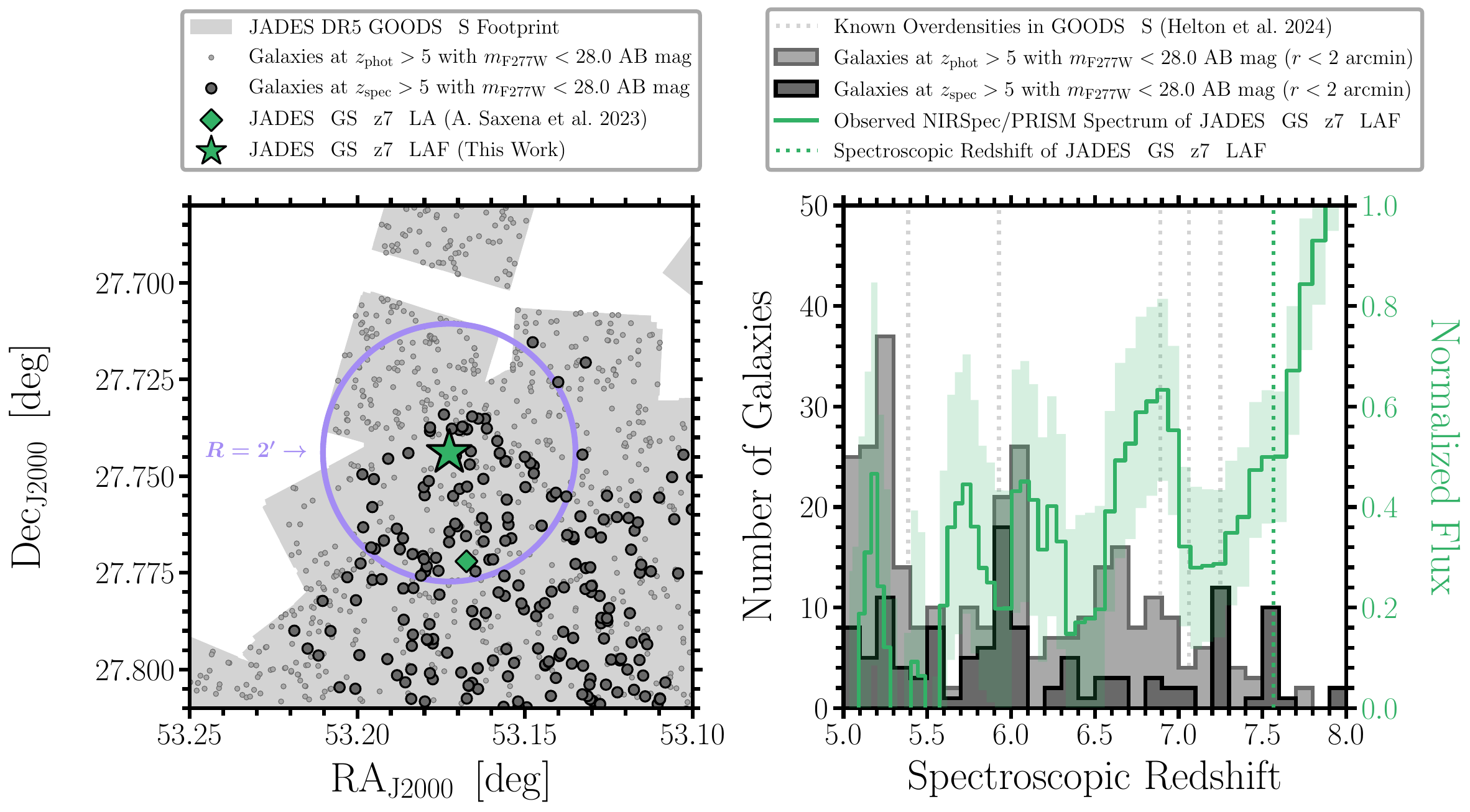}
    \caption{\textbf{Spatial and kinematic distributions of distant galaxies at \boldmath$z > 5$ observed near JADES-GS-z7-LAF.} \textit{Left panel:} We overplot relatively bright galaxies and galaxy candidates with $m_{\mathrm{F277W}} < 28.0\ \mathrm{AB\ mag}$ on the JADES DR5 GOODS-S footprint. The green star indicates the on-sky location of JADES-GS-z7-LAF while the green diamond represents a nearby and extreme LAE, JADES-GS-z7-LA at $z \approx 7.28$ \citep[][]{Saxena:2023}. \textit{Right panel:} We overplot the observed NIRSpec/PRISM spectrum of JADES-GS-z7-LAF on the redshift distributions of galaxies within $2\ \mathrm{arcmin}$ of JADES-GS-z7-LAF. Grey dotted lines indicate the redshifts of known overdensities in GOODS-S from \citet{Helton:2024b}. \textit{Both panels:} Photometrically-selected galaxy candidates at $z_{\mathrm{phot}} > 5$ are shown by light grey points and histograms while spectroscopically-confirmed galaxies at $z_{\mathrm{spec}} > 5$ are shown by dark grey points and histograms. JADES-GS-z7-LAF's blueward-of-Ly$\alpha$ flux is spatially and kinematically coincident with numerous high-redshift galaxy overdensities, which possibly explains how this flux is transmitted. \label{fig:Spatial_Spectral_Correlation_Transmission}}
\end{figure*}

Figure~\ref{fig:Spatial_Spectral_Correlation_Transmission} illustrates our qualitative spatial-spectral correlation for transmission. In the left panel, we plot relatively bright galaxies and galaxy candidates along with the JADES DR5 GOODS-S footprint. The on-sky location of JADES-GS-z7-LAF is indicated by the green star. The green diamond indicates the location of JADES-GS-z7-LA, an extreme LAE at $z \approx 7.28$ that is only $\approx 2\ \mathrm{arcmin}$ away from our source \citep[][]{Saxena:2023}. This LAE has one of the largest ionized bubbles inferred at $z > 5$ \citep[][]{Witstok:2024}, which coincides with a large-scale galaxy overdensity in GOODS-S \citep[][]{Helton:2024b} as indicated by the grey dotted lines in the right panel. In the right panel, we additionally plot the observed NIRSpec/PRISM spectrum of JADES-GS-z7-LAF alongside the redshift distributions of galaxies within $2\ \mathrm{arcmin}$ of JADES-GS-z7-LAF.

We find that JADES-GS-z7-LAF's blueward-of-Ly$\alpha$ flux is spatially and kinematically coincident with numerous high-redshift galaxy overdensities. This can possibly explain how the Ly$\alpha$ forest emission is transmitted. However, not all transmissive windows coincide with overdensities, especially at $z \lesssim 6$. This suggests overdensities and clustering are not necessary requirements for producing large ionized regions, similar to recent conclusions from \citet{YongdaZhu:2026}, who found low Ly$\alpha$ visibility in galaxy overdensities at $z \approx 5-11$. We note that the existing data make it hard to determine whether these are smooth and continuous transmission windows at $z \approx 5.5-7.5$ or a jagged forest of highly transmissive windows surrounded by complete absorption troughs, which motivates deeper and higher-resolution follow-up spectroscopy of JADES-GS-z7-LAF and its surrounding large-scale environment with JWST/NIRSpec.

%% Start of Section Four.
\section{Discussion \& Interpretation}
\label{SectionFour}

\subsection{An Extraordinarily Transparent Sightline \\ \& the Possibility of Escaping Ionizing Radiation}
\label{SectionFourOne}

In their pioneering work, \citet{Gunn:1965} describe the resonant-scattering optical depth that neutral hydrogen in the IGM imposes on Lyman-series photons emitted by distant objects ($z \gg 0$). Any photons emitted blueward of Ly$\alpha$ in the source's rest frame will redshift continuously as it propagates through the expanding, matter-dominated Universe. These photons will eventually redshift into exact resonance with Ly$\alpha$ (i.e., the $2p \rightarrow 1s$ electronic transition of neutral hydrogen), where they will be scattered away from the line of sight by any neutral hydrogen at that redshift.

While it is true that Lyman continuum (LyC; with $\lambda_{\mathrm{rest}} < 912\ \mathrm{\AA}$) and Ly$\alpha$ forest ($\lambda_{\mathrm{rest}} = 912-1216\ \mathrm{\AA}$) photons both arise from neutral hydrogen in its ground state (with $1s$ electrons that couple identically to the underlying radiation field), their absorption cross sections are different from each other by several orders of magnitude. The bound-free photoionization cross-section at the Lyman limit is $\sigma_{\mathrm{LyC}} (912\,\mathrm{\AA}) \approx 6.3 \times 10^{-18}\ \mathrm{cm}^{2}$, then drops quickly as $\propto \nu^{-3}$ (or $\propto \lambda^{3}$) at shorter wavelengths. The bound-bound photoionization cross-section at Ly$\alpha$ is a resonance whose frequency-integrated strength is dependent on the oscillator strength ($f_{\mathrm{Ly}\alpha}$),
\begin{equation}
    \sigma_{\mathrm{Ly}\alpha} (\nu) = \frac{\pi e^2}{m_e c}\,f_{\mathrm{Ly}\alpha}\,\phi(\nu), \qquad f_{\mathrm{Ly}\alpha} = 0.4162,
\end{equation}
where $\phi(\nu)$ is the line profile normalized as $\int \phi(\nu)\,d\nu = 1$ \citep[][]{Gunn:1965}. For a Doppler width that is characteristic of the diffuse IGM\footnote{Assuming pure thermal broadening at $T = 10^{4}\ \mathrm{K}$, we adopt a Doppler width of $b_{\,\mathrm{therm}} = \sqrt{2 k T/ m_{\mathrm{H}}} = 13\ \mathrm{km/s}$ for the value of the bound-bound photoionization cross-section at Ly$\alpha$.}, the line-center cross section is $\sigma_{\mathrm{Ly}\alpha} (1216\,\mathrm{\AA}) \approx 5.8 \times 10^{-14}\ \mathrm{cm}^{2}$. Thus, the absorption cross section of Ly$\alpha$ is roughly four orders of magnitude larger than the LyC. This implies that LyC and Ly$\alpha$ forest photons are each sensitive to opposite ends of the intervening neutral hydrogen column density distribution. The LyC is primarily affected by rare, optically thick clouds; the Ly$\alpha$ forest is affected by more common, optically thin clouds.

For a uniformly distributed IGM \citep[][]{Gunn:1965}, the optical depth is given by
\begin{equation}
    \tau_{\mathrm{GP}} (z) = \frac{\pi e^2 f_{\mathrm{Ly}\alpha} \lambda_{\mathrm{Ly}\alpha}}{m_e c}\,\frac{n_{\mathrm{H\,\textsc{i}}} (z)}{H(z)}, \qquad n_{\mathrm{H}} \propto \Omega_{b} h^{2}.
\end{equation}
Evaluating this expression with our assumed flat $\Lambda$CDM cosmology from Planck18 with $H_{0} = 67.4\ \mathrm{km/s/Mpc}$ and $\Omega_{m} = 0.315$ (\citealt{Planck:2020}), we derive the following for JADES-GS-z7-LAF:
\begin{equation}
    \tau_{\mathrm{GP}} (z = 7.567) \approx 5.2 \times 10^{5}\,x_{\mathrm{H\,\textsc{i}}}, \qquad x_{\mathrm{H\,\textsc{i}}} = \frac{n_{\mathrm{H\,\textsc{i}}}}{n_{\mathrm{H}}}.
\end{equation}
As discussed in \citet{Becker:2001}, this derivation implies that only a small neutral fraction in the IGM is needed to have $\tau_{\mathrm{GP}} \gg 1$, which will produce a complete Gunn-Peterson trough. Thus, the presence of transmitted flux blueward of Ly$\alpha$ in JADES-GS-z7-LAF tells us that the Ly$\alpha$ forest is neutral at the $x_{\mathrm{H\,\textsc{i}}} < 10^{-5}$ level. This means that less than $1$ in $100{,}000$ of the hydrogen atoms in the volume-filling ionization state of the IGM along this particular sightline are neutral, despite JADES-GS-z7-LAF existing in the midst of reionization. However, because the LyC and Ly$\alpha$ forest photons are decoupled from one another due to their vastly different absorption cross sections, we are unsure if this particular sightline is also transparent to ionizing photons. To better understand the escape of ionizing photons from this galaxy, dedicated follow-up observations are needed. Specifically, observations targeting the LyC emission from JADES-GS-z7-LAF, either with NIRCam/F070W or HST/ACS imaging, would provide a direct measurement of this galaxy's ionizing photon escape fraction. This would be the first measurement of its kind at $z > 5$.

\subsection{Comparison with Theoretical Predictions}
\label{SectionFourTwo}

As discussed in Section~\ref{SectionThree}, we report the significant detection of flux blueward of Ly$\alpha$ in JADES-GS-z7-LAF using the available NIRSpec/PRISM spectroscopy and NIRCam/F090W imaging. For the low-resolution spectrum provided in Figure~\ref{fig:Full_PRISM_Spectrum}, there is flux clearly detected above the noise level ($> 1 \sigma$) at $\lambda_{\mathrm{obs}} \approx 0.92-1.04\ \mu\mathrm{m}$, which corresponds to redshifts\footnote{The redshift range quoted here corresponds to the redshift of intervening gas for which a photon detected at that wavelength was passing through when it was in local resonance with Ly$\alpha$.} of $z \approx 6.56-7.57$ and probes a line-of-sight physical size of $d_{\mathrm{LOS}} \approx 43.2\ \mathrm{pMpc}$ (or $d_{\mathrm{LOS}} \approx 349\ \mathrm{cMpc}$ in comoving units). As discussed in Section~\ref{SectionFourOne}, for there to be transmission in the Ly$\alpha$ forest, the IGM must be neutral along the line-of-sight at the $x_{\mathrm{H\,\textsc{i}}} < 10^{-5}$ level. We assess the feasibility of such an extraordinarily transparent sightline occurring in the midst of cosmic reionization ($z > 6$) by analyzing theoretical predictions from the \texttt{THESAN} project.

\begin{figure}
    \centering
    \includegraphics[width=1.0\linewidth]{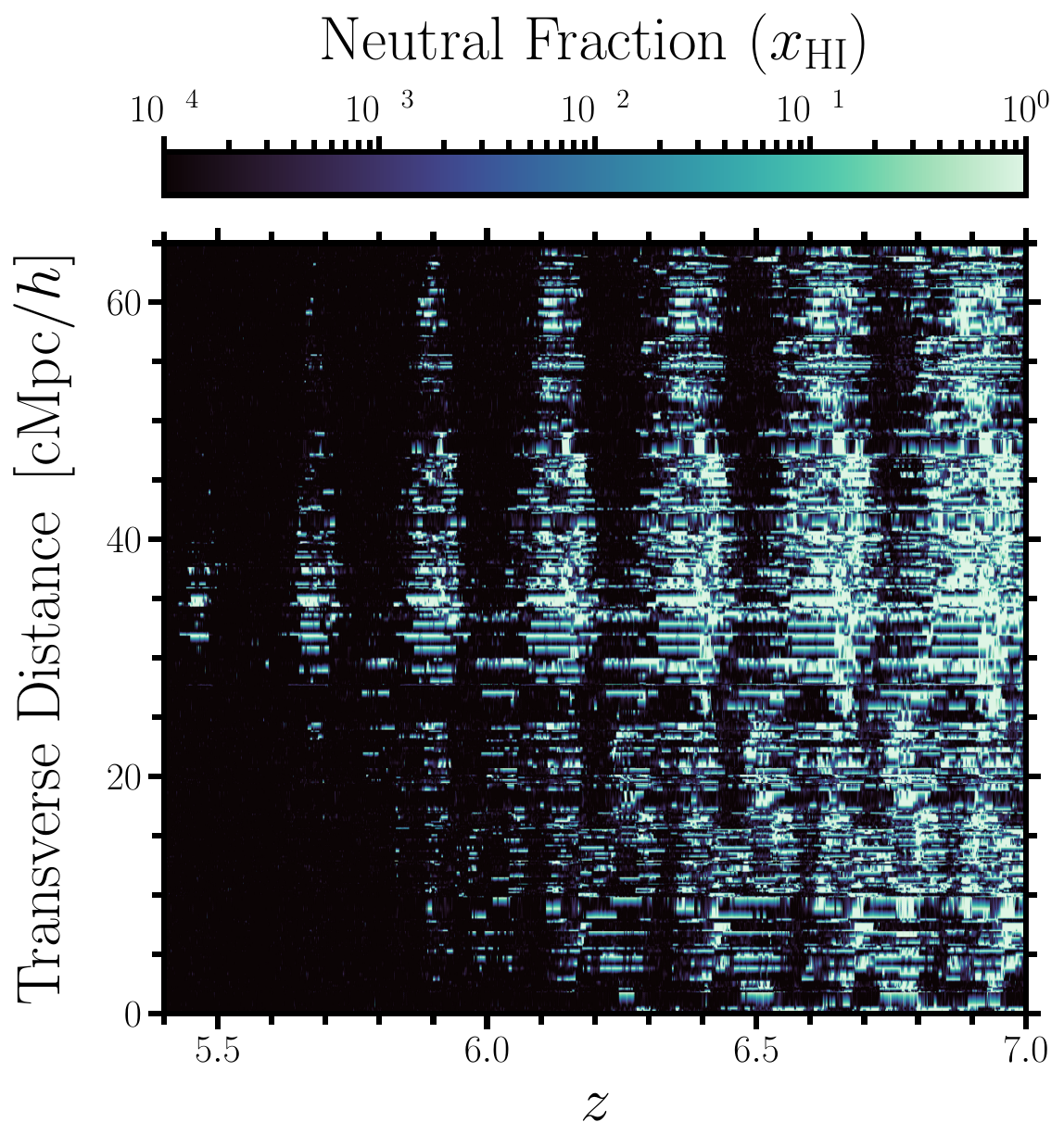}
    \caption{\textbf{Theoretical predictions for IGM neutral fractions from the \texttt{THESAN} project.} We show results from $300$ line-of-sight lightcones provided by the \texttt{THESAN} team, which span redshifts of $z \approx 5.4-7.0$ \citep[][]{Garaldi:2024}. The longest contiguous ionized region in these simulations is $d_{\mathrm{LOS}} \approx 46\ \mathrm{cMpc}$ at $z < 6$ and $d_{\mathrm{LOS}} \approx 15\ \mathrm{cMpc}$ at $z > 6$. For comparison, flux is clearly detected above the noise level in the Ly$\alpha$ forest at $\lambda_{\mathrm{obs}} \approx 0.92-1.04\ \mu\mathrm{m}$, which corresponds to redshifts of $z \approx 6.56-7.57$ and probes a line-of-sight physical size of $d_{\mathrm{LOS}} \approx 43.2\ \mathrm{pMpc}$ (or $d_{\mathrm{LOS}} \approx 349\ \mathrm{cMpc}$). Thus, we find an extraordinarily transparent sightline in the direction of JADES-GS-z7-LAF that is several orders of magnitude larger than the longest fully ionized region in \texttt{THESAN}. However, we note that it is difficult to estimate the length of the ionized region from the existing low-resolution data alone. \label{fig:THESAN_Neutral_Fraction_Lightcones}}
\end{figure}

\texttt{THESAN} is a suite of large volume ($L_{\mathrm{box}} \approx 100\ \mathrm{cMpc}$) cosmological, radiation-magnetohydrodynamical simulations that were specifically designed to simultaneously model the large-scale statistical properties of the IGM during reionization and the resolved properties of the galaxies during this epoch \citep[][]{Kannan:2022, Garaldi:2022, Smith:2022}. We use the publicly available line-of-sight lightcone data products\footnote{\href{https://www.thesan-project.com/thesan/los_lc.html}{https://www.thesan-project.com/thesan/los\_lc.html}} from \texttt{THESAN-1}, since this is the only simulation from the \texttt{THESAN} suite with this data product available. For more information about the lightcones used in this work, we refer the reader to \citet{Garaldi:2024}\footnote{\href{https://www.thesan-project.com/thesan/los.html}{https://www.thesan-project.com/thesan/los.html}}.

There are $300$ line-of-sight lightcones provided by the \texttt{THESAN} team, and each span a total of $27$ simulation snapshots (corresponding to redshifts of $z \approx 5.34-7.06$). The lightcones were assembled using a stepwise-constant approximation so that each one traces a continuous, self-consistent collection of gas properties along random lines of sight through the simulation, despite encompassing multiple periodic replications of the simulation box.

To visualize the structure and history of reionization along these sightlines, we resampled each lightcone's piecewise-constant $x_{\mathrm{H\,\textsc{i}}} (z)$ values onto a finely sampled redshift grid then linearly interpolated between the $300$ sparsely sampled transverse sightlines. Resampling onto the same redshift grid was necessary because each of the lightcones uses a different Voronoi tesselation for the simulation and has a different number of segments, as determined by the native Voronoi structure. Figure~\ref{fig:THESAN_Neutral_Fraction_Lightcones} shows theoretical predictions for IGM neutral fractions from the \texttt{THESAN} project, which we use for comparison.

To quantify how far a Ly$\alpha$ photon could travel through a fully ionized region of the IGM without encountering a neutral patch, we computed the longest contiguous series of segments with $x_{\mathrm{H\,\textsc{i}}} < 10^{-4}$ for each of the $300$ line-of-sight lightcones provided by the \texttt{THESAN} team. We used the native resolutions of the lightcones rather than the resampled grid. Across the full redshift range of the simulated lightcones ($z \approx 5.34-7.06$), the longest contiguous region with $x_{\mathrm{H\,\textsc{i}}} < 10^{-4}$ is $d_{\mathrm{LOS}} \approx 46.0\ \mathrm{cMpc}$ at $z \approx 5.43-5.53$. Limiting ourselves to $z > 6$, the longest fully ionized region is only $d_{\mathrm{LOS}} \approx 15.0\ \mathrm{cMpc}$ at $z \approx 6.12-6.15$. We note that these lightcones from the \texttt{THESAN} team do not extend to the systemic redshift of JADES-GS-z7-LAF (\ResultSpectroscopicRedshift). Despite this shortcoming, we can extrapolate that the longest ionized region at $z > 7$ would be at least a factor of a few smaller.

To summarize, we find an extraordinarily transparent sightline in the direction of JADES-GS-z7-LAF that might be several orders of magnitude larger than the longest fully ionized region from \texttt{THESAN}'s large volume cosmological, radiation-magnetohydrodynamical simulations. This suggests that reionization models are unable to explain relatively large ($d_{\mathrm{LOS}} \gtrsim 100\ \mathrm{cMpc}$) ionized regions above $z > 7$, which might demonstrate a discrepancy between the abundance of large ionized bubbles in the simulations when compared to observations. This potential discrepancy is exacerbated by the fact that we defined ionized regions as those with $x_{\mathrm{H\,\textsc{i}}} < 10^{-4}$, yet $x_{\mathrm{H\,\textsc{i}}} < 10^{-5}$ is needed for there to be transmission in the Ly$\alpha$ forest (see also Section~\ref{SectionFourOne}).

However, we caution the reader that it is enormously difficult to measure the physical and comoving sizes of the transparent sightline from the low-resolution ($R \approx 100$) NIRSpec/PRISM data alone, since the detection of positive flux across $\lambda_{\mathrm{obs}} \approx 0.92-1.04\ \mu\mathrm{m}$ does not necessarily establish continuous transmission at every intervening redshift. This is because unresolved transmission windows separated by saturated absorption troughs can become a smooth, positive continuum signal after convolution with the LSF. To properly compare observations with theoretical predictions, one should generate mock transmitted spectra from the simulated density, temperature, velocity, and neutral-hydrogen fields, multiply by an appropriate intrinsic source spectrum, convolve with the LSF, and apply the same noise and detection statistic as used for the observations. The proposed self-consistent comparison is outside the scope of this work and should be the subject of future work. Deeper medium-resolution ($R \approx 1000$) NIRSpec/G140M data would be invaluable for measuring transmission windows separated by saturated absorption troughs, and therefore could resolve the Ly$\alpha$ forest. Follow-up observations that are $\approx 3 \times$ deeper than existing NIRSpec/G140M data would be able to detect the $\lambda_{\mathrm{obs}} \approx 0.92-1.04\ \mu\mathrm{m}$ flux at a per-pixel significance of $> 1 \sigma$, assuming smooth and continuous transmission windows.

\subsection{Implications for Cosmic Reionization}
\label{SectionFourThree}

Theoretically, \citet{Summerfield:2026} found that the bulk of reionization occurred rapidly after $z \approx 8$ and is primarily driven by ionizing photons from UV-luminous galaxies brighter than $M_{\mathrm{UV}} = -17$. They arrived at this conclusion by analyzing predictions from the \texttt{THESAN-ZOOM} cosmological, radiation-hydrodynamical simulations. Observationally, \citet{Giovinazzo:2026b} found similar results by performing detailed spectrophotometric fitting of JWST data for galaxies at $z > 5$ using a picket-fence model for the escape of ionizing photons \citep[see also][]{Giovinazzo:2026a}. They find that a few strongly leaking sources ($\approx 20\%$ of their sample) are able to drive the reionization process by producing $\approx 90\%$ of all the ionizing photons that reach the IGM. These results indicate that it is necessary to identify strong leaking sources at high redshift to better understand the primary drivers of reionization.

JADES-GS-z7-LAF represents a possible example of these strongly leaking sources at high redshift. Other notable examples of these strongly leaking sources at high redshifts include one galaxy at $z \approx 8.5$ from \citet{Baker:2025} and another galaxy at $z \approx 10.3$ from \citet{MarquesChaves:2026}. However, the ionizing photon leakage for these galaxies has only been inferred indirectly thus far. As mentioned in Section~\ref{SectionOne}, the highest redshift LCE candidate was recently discovered at $z \approx 4.4$ \citep[][]{Goovaerts:2026, Zhu:2026}. Searching for LyC emission from these strongly leaking candidates at $z > 7$ is crucial for directly inferring the ionizing photon escape fractions of these galaxies.

Furthermore, the tentative evidence for strong nebular continuum contributions at rest-frame UV wavelengths (see Section~\ref{SectionThreeSeven}) has implications for the UV luminosity functions (UVLFs) and ionizing photon production efficiencies of distant galaxies, both of which are related to the total ionizing photon budget at early cosmic times. A significant nebular continuum contribution reddens the intrinsic UV slope without needing dust and boosts the observed UV luminosity without needing an increase in the formation of massive stars, which means that stellar population synthesis models may misattribute this excess flux to star formation rather than reprocessed nebular emission. This is important because the calibration to infer $\xi_{\mathrm{ion}}$ typically assumes that the rest-frame UV continuum luminosity traces the ionizing photons from massive stars alone, meaning a strong nebular continuum would underestimate the intrinsic $\xi_{\mathrm{ion}}$ value. Our results suggest that careful modeling of the stellar-plus-nebular continuum is crucial for understanding the sources responsible for cosmic reionization.

%% Start of Section Five.
\section{Summary \& Conclusions}
\label{SectionFive}

Time and time again, JWST continues to challenge our understanding of the early Universe. In this work, we report yet another example of JWST defying our expectations for the earliest galaxies. Using a suite of JWST observations spanning $\approx 5$ days of on-source exposure time and data products from JADES DR5, we report the ``discovery'' of a galaxy at $z > 7$ with flux appearing blueward of the Ly$\alpha$ break at $\lambda_{\mathrm{rest}} = 1216\ \mathrm{\AA}$, which we refer to as JADES-GS-z7-LAF. This feature has only ever been observed among the most luminous quasars currently known at $z \approx 7$. Given the relatively large opacity of the IGM at $z > 7$, and the correspondingly high neutral fractions at these redshifts, this was an unexpected finding that warranted further investigation. We note a companion paper that reports a similar discovery for the same galaxy \citep[][]{Yue:2026}. Our findings can be summarized as follows.

\begin{itemize}
    \itemsep 0pt
    \item JADES-GS-z7-LAF is a high-redshift star-forming galaxy at \ResultSpectroscopicRedshift\ as evidenced by the detection of numerous rest-frame UV and optical emission lines and a prominent Ly$\alpha$ break at $\lambda_{\mathrm{obs}} \approx 1.0-1.1\ \mu\mathrm{m}$ (see Figure~\ref{fig:Full_PRISM_Spectrum} and Section~\ref{SectionThreeOne}). A brief summary of the empirical and inferred physical properties for JADES-GS-z7-LAF are provided in Table~\ref{tab:PhysicalProperties}.
    \item Using the detected hydrogen lines, we determine that JADES-GS-z7-LAF is dust-free, moderately star-forming (\ResultStarFormationRateLin), with strong emission lines (\ResultEmissionLineEquivalentWidth) and a large ionizing photon production efficiency (\ResultsIonizingPhotonProductionEfficiencyLin).
    \item Using the detected oxygen lines (see Figure~\ref{fig:Electron_Density_And_Temperature}), we self-consistently determine that the $\mathrm{[O\,\textsc{iii}]}$-emitting gas of JADES-GS-z7-LAF has a large electron density (\ResultsElectronDensityLinSelfConsistent) and temperature (\ResultsElectronTemperatureLinSelfConsistent), which suggests extreme nebular conditions. Using these values, we further infer a gas-phase oxygen abundance of \ResultsGasPhaseOxygenAbundanceLog\ ($\approx 20\%\ Z_{\odot}$), which confirms that this is a metal-poor galaxy. If we instead assume a density of $n_{e} = 300\ \mathrm{cm}^{-3}$, we would infer an oxygen abundance $\approx 0.6-0.7\ \mathrm{dex}$ smaller than our fiducial value. We conclude that a failure to account for high densities can lead to biased results about chemical enrichment in the early Universe \citep[see also][]{Hsiao:2026}.
    \item Using the detected carbon and nitrogen lines, we determine that JADES-GS-z7-LAF is carbon-deficient (\ResultsCarbonToOxygenRatioLogSolar) but nitrogen-rich (\ResultsNitrogenToOxygenRatioLogSolar) with an elevated nitrogen-to-carbon ratio (\ResultsNitrogenToCarbonRatioLogSolar), suggesting an unusual mode of chemical enrichment and nitrogen production within this metal-poor galaxy.
    \item Using the NIRSpec/PRISM spectrum (Figure~\ref{fig:Full_PRISM_Spectrum}), we significantly detect ($\approx 7 \sigma$) blueward-of-Ly$\alpha$ flux. The spectrum's cross-dispersion profile (Figure~\ref{fig:Cross_Dispersion_Profile}) indicates that the blueward-of-Ly$\alpha$ flux is spatially coincident with the rest-frame UV and optical emission, suggesting a single source of flux rather than chance alignment with an interloper. We independently confirm these conclusions using NIRCam/F090W imaging (Figure~\ref{fig:RGB_Thumbnail_Overlay}), whose flux is spatially coincident with JADES-GS-z7-LAF and emerges $\approx 0.4\ \mathrm{pkpc}$ from its UV centroid.
    \item We perform two-component morphological fitting (Figure~\ref{fig:Morphological_Fitting_Results}) and find that the blueward-of-Ly$\alpha$ flux measured by F090W has morphological properties consistent with those of JADES-GS-z7-LAF's disk component, suggesting a similar physical origin. We note that the F090W and disk centroids are offset from one another by $\Delta \theta = 0.0380 \pm 0.0069^{\prime\prime}$ (or $0.193 \pm 0.035\ \mathrm{pkpc}$), which we can explain with a partial-covering geometry.
    \item We test for instrumental systematics (Section~\ref{SectionThreeFour}) or foreground interlopers (Section~\ref{SectionThreeFive}) as sources for the blueward-of-Ly$\alpha$ flux. We find no evidence for either of these alternatives, although we cannot formally reject the latter with the existing data.
    \item We observe a smooth and gradual turnover in the NIRSpec/PRISM spectrum around the Ly$\alpha$ break (Figure~\ref{fig:Damping_Wing_Fitting_Results}), which is often associated with either damping-wing absorption or emission from the $2 \gamma$ nebular continuum. If the observed turnover is from damping-wing absorption, JADES-GS-z7-LAF would have one of the strongest DLAs ever measured ($N_{\mathrm{H\,\textsc{i}}} \approx 10^{23.3}\ \mathrm{cm}^{-2}$). We find $2 \gamma$ nebular continuum emission is the preferred interpretation and that the nebular continuum contributes $\approx 60-70\%$ of the flux at $\lambda_{\mathrm{rest}} = 1500\ \mathrm{\AA}$. We tentatively detect ($\approx 3.4 \sigma$) a Balmer jump from free-bound nebular continuum at $\lambda_{\mathrm{rest}} = 3645\ \mathrm{\AA}$ (see Figure~\ref{fig:Balmer_Jump_Fit} and Section~\ref{SectionThreeSeven}), consistent with our interpretation of a strong nebular continuum.
    \item We self-consistently determine the properties of the stellar populations, nebular gas, and dust via panchromatic SED fitting (Section~\ref{SectionThreeEight}) and find that JADES-GS-z7-LAF is a relatively massive (\ResultStellarMassLin) star-forming galaxy that experienced a burst $\approx 10-30\ \mathrm{Myr}$ ago and a declining star-formation history ever since the burst.
    \item The blueward-of-Ly$\alpha$ flux implies an extraordinarily transparent sightline ($x_{\mathrm{H\,\textsc{i}}} \ll 0.1\%$) through the IGM at $z \approx 5.5-7.5$, a sightline that is spatially and kinematically coincident with numerous high-redshift galaxy overdensities (Section~\ref{SectionThreeNine}), which possibly explains the emergence of this flux.
    \item We find zero analogous sightlines in the \texttt{THESAN} suite of large volume ($L_{\mathrm{box}} \approx 100\ \mathrm{cMpc}$) cosmological, radiation-magnetohydrodynamical simulations (Figure~\ref{fig:THESAN_Neutral_Fraction_Lightcones} and Section~\ref{SectionFourTwo}), suggesting a possible discrepancy between these observations and theoretical predictions. Although we caution that it is difficult to compare between observations and predictions with the existing spectroscopic data.
\end{itemize}

In the same way that reionization models need to be able to explain relatively large ($L_{\mathrm{neutral}} \gtrsim 100\,\mathrm{cMpc}$) neutral regions below $z < 6$, mimicking late-type reionization histories that have been invoked to explain long troughs in the Ly$\alpha$ forest \citep[e.g.,][]{Zhu:2021}, these same models may also need to be able to explain relatively large ($L_{\mathrm{ionized}} \gtrsim 100\,\mathrm{cMpc}$) ionized regions above $z > 7$, mimicking early-type reionization histories.

As discussed in \citet{Kannan:2022}, different models for reionization produce different bubble size distributions at the same global ionization fraction. When low-mass halos dominate the ionizing photon budget, simulations produce a large amount of small bubbles; when high-mass halos dominate, simulations produce less numerous but physically larger bubbles.

Future telescopes and instruments targeting neutral hydrogen's $21\,\mathrm{cm}$ hyper-fine transition (e.g., the Square Kilometer Array, or SKA) will have enough sensitivity to detect these reionization signatures in the slope and normalization of the $21\,\mathrm{cm}$ power spectrum by placing stringent constraints on the ionized bubble size distribution and its evolution. This will ultimately inform us about the sources that dominate reionization's ionizing photon budget in the early Universe.

With a sample size of one, it is difficult to determine the abundance of such extraordinarily transparent sightlines in the midst of cosmic reionization. Future observations from the Roman Space Telescope's High-Latitude Wide-Area Survey (HLWAS) will identify large samples of luminous and massive galaxies to see if galaxies like JADES-GS-z7-LAF are ubiquitous. If such systems are common, luminous and massive galaxies may contribute more to the ionizing photon budget than suggested by other indirect constraints \citep[e.g.,][]{Atek:2024}, thus reshaping our understanding of the Universe's last major baryonic phase transition -- cosmic reionization.

\newpage

\begin{contribution}

%% This section gives authors the space to recognize author contributions. The text inside this environment is NOT counted towards the total word quanta. At a minimum, manuscripts are expected to include this text:

J.M.H came up with the initial research concept, led the data analyses, and was responsible for writing and submitting the manuscript. J.W. and M.L. both contributed to various aspects of the data analysis, including the damping-wing and nebular continuum analysis presented in Section~\ref{SectionThreeSeven} and the panchromatic SED fitting presented in Section~\ref{SectionThreeEight}. All co-authors contributed to the scientific interpretation of the results and helped to write, review, and edit the manuscript. Many of the co-authors acquired the relevant funding support.

\end{contribution}

%% But authors are expected to provide more specific details, e.g., SC was responsible for writing and submitting the manuscript. WWM came up with the initial research concept and edited the manuscript. OTS obtained the funding and edited the manuscript. EBF provided the formal analysis and validation. He also edited the manuscript. GEH Supervised the undergraduates, wrote the software and administers the project github and Zenodo repositories.
%%
%% Authors can use the Contributor Role Taxonomy (CRediT) at https://credit.niso.org for ideas on how write a good statement tailored to their needs.

\section*{Acknowledgments}

We thank Yongda Zhu for valuable conversations and early guidance that ultimately strengthened this paper. This work is based (in part) on observations acquired by the NASA/ESA/CSA James Webb Space Telescope (JWST). The data were obtained from the Mikulski Archive for Space Telescopes (MAST) at the Space Telescope Science Institute (STScI), which is operated by the Association of Universities for Research in Astronomy, Inc., under NASA contract NAS 5-03127 for JWST. We acknowledge the use of NASA’s Astrophysics Data System and Science Explorer, funded by NASA under Cooperative Agreement 80NSSC21M00561, for this work.

JADES DR5 includes NIRCam data from JWST program IDs 1176, 1180, 1181, 1210, 1264, 1283, 1286, 1287, 1895, 1963, 2079, 2198, 2514, 2516, 2674, 3215, 3577, 3990, 4540, 4762, 5398, 5997, 6434, 6511, and 6541. JADES DR5 includes MIRI data from JWST program IDs 1180, 1181, and 1207. The authors acknowledge the teams of program IDs 1895, 1963, 2079, 2514, 3215, 3577, 3990, 6434, and 6541 for developing each of their observing programs with a zero-exclusive-access period.

We respectfully acknowledge that The Pennsylvania State University campuses are located on the original homelands of the Erie, Haudenosaunee (Seneca, Cayuga, Onondaga, Oneida, Mohawk, and Tuscarora), Lenape (Delaware Nation, Delaware Tribe, Stockbridge-Munsee), Monongahela, Shawnee (Absentee, Eastern, and Oklahoma), Susquehannock, and Wahzhazhe (Osage) Nations. As a land grant institution, we acknowledge and honor the traditional caretakers of these lands and strive to understand and model their responsible stewardship. We also acknowledge the longer history of these lands and our place in that history.

We respectfully acknowledge that the University of Arizona is on the land and territories of Indigenous peoples. Today, Arizona is home to 22 federally recognized tribes, with Tucson being home to the O’odham and the Yaqui. The University of Arizona strives to build sustainable relationships with sovereign Native Nations and Indigenous communities through education offerings, partnerships, and community service.

Generative large language models (LLMs), including Claude Sonnet 5.0 and Opus 5.0, were used during the development of this manuscript. The authors retain full responsibility for the intellectual content of this work, including all aspects of the analysis, results, discussion, scientific interpretation, and conclusions. Every piece of AI-generated content was reviewed, evaluated, and often substantially rewritten by the authors of this work.

J.M.H. acknowledges support from the Evolving Universe Fellowship, which is made possible by a generous donation from Dr. Keiko Miwa Ross. J.M.H., K.N.H., D.J.E., Z.J., B.D.J., B.E.R., and C.N.A.W. acknowledge support from the JWST Near Infrared Camera (NIRCam) \mbox{Science} Team Lead, NAS5-02105, from NASA Goddard Space Flight Center to the University of Arizona. J.M.H., D.J.E., B.D.J., and B.E.R. acknowledge support from JWST Program \#3215. J.M.H., M.L., and K.N.H. acknowledge support from JWST Program \#8544. F.D.E. and I.J. acknowledge support from the Science and Technology Facilities Council (STFC), by the ERC through Advanced Grant 695671 ``QUENCH,'' and funding from the UKRI Frontier Research grant RISEandFALL. I.J. acknowledges support by the Huo Family Foundation through a P.C. Ho PhD Studentship. S.T. acknowledges support by the Royal Society Research Grant G125142. A.J.B. acknowledges funding from the ``FirstGalaxies'' Advanced Grant from the European Research Council (ERC) under the European Union’s Horizon 2020 research and innovation program (grant agreement $\# 789056$). D.J.E. is supported as a Simons Investigator. P.R. acknowledges support from the University of Texas at Austin Cosmic Frontier Center.

%% To help institutions obtain information on the effectiveness of their telescopes the AAS Journals has created a group of keywords for telescope facilities.
%%
%% Following the acknowledgments section, use the following syntax and the \facility{} or \facilities{} macros to list the keywords of facilities used in the research for the paper. Each keyword is check against the master list during copy editing.  Individual instruments can be provided in parentheses, after the keyword, but they are not verified.

\facilities{JWST (NIRSpec, NIRCam, MIRI)}

%% Similar to \facility{}, there is the optional \software command to allow authors a place to specify which programs were used during the creation of the manuscript. Authors should list each code and include either a citation or url to the code inside ()s when available.

\software{\texttt{AstroPy} \citep[][]{Astropy:2013, Astropy:2018}, \texttt{corner} \citep[][]{Foreman-Mackey:2016}, \texttt{Cue} \citep[][]{Li:2024, Li:2025}, \texttt{dynesty} \citep[][]{Speagle:2020}, \texttt{FitsMap} \citep[][]{Hausen:2022}, \texttt{FSPS} \citep[][]{Conroy:2009, Conroy:2010}, \texttt{Matplotlib} \citep[][]{Matplotlib:2007}, \texttt{msafit} \citep[][]{deGraaff:2024}, \texttt{MultiNest} \citep[][]{Feroz:2009}, \texttt{NumPy} \citep[][]{NumPy:2011, NumPy:2020}, \texttt{NUTS} \citep[][]{Hoffman:2011}, \texttt{pandas} \citep[][]{Pandas:2025}, \texttt{Prospector} \citep[][]{Johnson:2021}, \texttt{PyMultiNest} \citep[][]{Perrin:2014}, \texttt{PyNeb} \citep[][]{Luridiana:2015}, \texttt{pysersic} \citep[][]{Pasha:2023}, \texttt{SciPy} \citep[][]{SciPy:2020}, \texttt{seaborn} \citep[][]{Waskom:2020}, \texttt{specutils} \citep[][]{Specutils:2019}}

%% Appendix material should be preceded with a single \appendix command. There should be a \section command for each appendix. Mark appendix subsections with the same markup you use in the main body of the paper.
%%
%% Each Appendix (indicated with \section) will be lettered A, B, C, etc. The equation counter will reset when it encounters the \appendix command and will number appendix equations (A1), (A2), etc. The Figure and Table counter will not reset.

% \appendix

%% For this sample we use BibTeX plus aasjournalv7.bst to generate the bibliography. The sample7.bib file was populated from ADS. To get the citations to show in the compiled file do the following:

\newpage
\bibliography{main}{}

\begin{thebibliography}{}
\expandafter\ifx\csname natexlab\endcsname\relax\def\natexlab#1{#1}\fi
\providecommand{\url}[1]{\href{#1}{#1}}
\providecommand{\dodoi}[1]{doi:~\href{http://doi.org/#1}{\nolinkurl{#1}}}
\providecommand{\doeprint}[1]{\href{http://ascl.net/#1}{\nolinkurl{http://ascl.net/#1}}}
\providecommand{\doarXiv}[1]{\href{https://arxiv.org/abs/#1}{\nolinkurl{https://arxiv.org/abs/#1}}}

% type= article
\bibitem[{T. {Abel} {et~al.}(2002){Abel}, {Bryan}, \& {Norman}}]{Abel:2002}
{Abel}, T., {Bryan}, G.~L., \& {Norman}, M.~L. 2002, \bibinfo{title}{{The Formation of the First Star in the Universe},} Science, 295, 93, \dodoi{10.1126/science.1063991}

% type= article
\bibitem[{S. {Alberts} {et~al.}(2024){Alberts}, {Lyu}, {Shivaei}, {Rieke}, {P{\'e}rez-Gonz{\'a}lez}, {Bonaventura}, {Zhu}, {Helton}, {Ji}, {Morrison}, \& et~al.}]{Alberts:2024}
{Alberts}, S., {Lyu}, J., {Shivaei}, I., {et~al.} 2024, \bibinfo{title}{{SMILES Initial Data Release: Unveiling the Obscured Universe with MIRI Multiband Imaging},} \apj, 976, 224, \dodoi{10.3847/1538-4357/ad7396}

% type= article
\bibitem[{K.~Z. {Arellano-C{\'o}rdova} {et~al.}(2026){Arellano-C{\'o}rdova}, {M{\'e}ndez-Delgado}, {Flury}, {Esteban}, {Kreckel}, {Garc{\'\i}a-Rojas}, {Cullen}, {Carigi}, {Morisset}, {Rosales-Ortega}, \& et~al.}]{ArellanoCordova:2026}
{Arellano-C{\'o}rdova}, K.~Z., {M{\'e}ndez-Delgado}, J.~E., {Flury}, S.~R., {et~al.} 2026, \bibinfo{title}{{A self-consistent direct method for chemical abundances in high-z galaxies with JWST},} \mnras, 547, stag380, \dodoi{10.1093/mnras/stag380}

% type= article
\bibitem[{M. {Asplund} {et~al.}(2021){Asplund}, {Amarsi}, \& {Grevesse}}]{Asplund:2021}
{Asplund}, M., {Amarsi}, A.~M., \& {Grevesse}, N. 2021, \bibinfo{title}{{The chemical make-up of the Sun: A 2020 vision},} \aap, 653, A141, \dodoi{10.1051/0004-6361/202140445}

% type= article
\bibitem[{ {Astropy Collaboration} {et~al.}(2013){Astropy Collaboration}, {Robitaille}, {Tollerud}, {Greenfield}, {Droettboom}, {Bray}, {Aldcroft}, {Davis}, {Ginsburg}, {Price-Whelan}, \& et~al.}]{Astropy:2013}
{Astropy Collaboration}, {Robitaille}, T.~P., {Tollerud}, E.~J., {et~al.} 2013, \bibinfo{title}{{Astropy: A community Python package for astronomy},} \aap, 558, A33, \dodoi{10.1051/0004-6361/201322068}

% type= article
\bibitem[{ {Astropy Collaboration} {et~al.}(2018){Astropy Collaboration}, {Price-Whelan}, {Sip{\H{o}}cz}, {G{\"u}nther}, {Lim}, {Crawford}, {Conseil}, {Shupe}, {Craig}, {Dencheva}, \& et~al.}]{Astropy:2018}
{Astropy Collaboration}, {Price-Whelan}, A.~M., {Sip{\H{o}}cz}, B.~M., {et~al.} 2018, \bibinfo{title}{{The Astropy Project: Building an Open-science Project and Status of the v2.0 Core Package},} \aj, 156, 123, \dodoi{10.3847/1538-3881/aabc4f}

% type=
\bibitem[{ {Astropy-Specutils Development Team}(2019){Astropy-Specutils Development Team}}]{Specutils:2019}
{Astropy-Specutils Development Team}. 2019, {Specutils: Spectroscopic analysis and reduction},, Astrophysics Source Code Library, record ascl:1902.012 \doeprint{1902.012}

% type= article
\bibitem[{H. {Atek} {et~al.}(2024){Atek}, {Labb{\'e}}, {Furtak}, {Chemerynska}, {Fujimoto}, {Setton}, {Miller}, {Oesch}, {Bezanson}, {Price}, \& et~al.}]{Atek:2024}
{Atek}, H., {Labb{\'e}}, I., {Furtak}, L.~J., {et~al.} 2024, \bibinfo{title}{{Most of the photons that reionized the Universe came from dwarf galaxies},} \nat, 626, 975, \dodoi{10.1038/s41586-024-07043-6}

% type= article
\bibitem[{K. {Bach} \& H.-W. {Lee}(2015){Bach} \& {Lee}}]{Bach:2015}
{Bach}, K., \& {Lee}, H.-W. 2015, \bibinfo{title}{{Accurate Ly{\ensuremath{\alpha}} scattering cross-section and red damping wing in the reionization epoch},} \mnras, 446, 264, \dodoi{10.1093/mnras/stu2030}

% type= article
\bibitem[{W.~M. {Baker} {et~al.}(2025){Baker}, {D'Eugenio}, {Maiolino}, {Bunker}, {Simmonds}, {Tacchella}, {Witstok}, {Arribas}, {Carniani}, {Charlot}, \& et~al.}]{Baker:2025}
{Baker}, W.~M., {D'Eugenio}, F., {Maiolino}, R., {et~al.} 2025, \bibinfo{title}{{Zapped then napped? A rapidly quenched remnant leaker candidate with a steep spectroscopic {\ensuremath{\beta}}$_{UV}$ slope at z = 8.5},} \aap, 697, A90, \dodoi{10.1051/0004-6361/202553766}

% type= article
\bibitem[{R. {Barkana} \& A. {Loeb}(2001){Barkana} \& {Loeb}}]{Barkana:2001}
{Barkana}, R., \& {Loeb}, A. 2001, \bibinfo{title}{{In the beginning: the first sources of light and the reionization of the universe},} \physrep, 349, 125, \dodoi{10.1016/S0370-1573(01)00019-9}

% type= article
\bibitem[{R. {Barnett} {et~al.}(2017){Barnett}, {Warren}, {Becker}, {Mortlock}, {Hewett}, {McMahon}, {Simpson}, \& {Venemans}}]{Barnett:2017}
{Barnett}, R., {Warren}, S.~J., {Becker}, G.~D., {et~al.} 2017, \bibinfo{title}{{Observations of the Lyman series forest towards the redshift 7.1 quasar ULAS J1120+0641},} \aap, 601, A16, \dodoi{10.1051/0004-6361/201630258}

% type= article
\bibitem[{G.~D. {Becker} {et~al.}(2015){Becker}, {Bolton}, {Madau}, {Pettini}, {Ryan-Weber}, \& {Venemans}}]{Becker:2015}
{Becker}, G.~D., {Bolton}, J.~S., {Madau}, P., {et~al.} 2015, \bibinfo{title}{{Evidence of patchy hydrogen reionization from an extreme Ly{\ensuremath{\alpha}} trough below redshift six},} \mnras, 447, 3402, \dodoi{10.1093/mnras/stu2646}

% type= article
\bibitem[{R.~H. {Becker} {et~al.}(2001){Becker}, {Fan}, {White}, {Strauss}, {Narayanan}, {Lupton}, {Gunn}, {Annis}, {Bahcall}, {Brinkmann}, \& et~al.}]{Becker:2001}
{Becker}, R.~H., {Fan}, X., {White}, R.~L., {et~al.} 2001, \bibinfo{title}{{Evidence for Reionization at z\raisebox{-0.5ex}\textasciitilde6: Detection of a Gunn-Peterson Trough in a z=6.28 Quasar},} \aj, 122, 2850, \dodoi{10.1086/324231}

% type= article
\bibitem[{S.~V.~W. {Beckwith} {et~al.}(2006){Beckwith}, {Stiavelli}, {Koekemoer}, {Caldwell}, {Ferguson}, {Hook}, {Lucas}, {Bergeron}, {Corbin}, {Jogee}, \& et~al.}]{Beckwith:2006}
{Beckwith}, S. V.~W., {Stiavelli}, M., {Koekemoer}, A.~M., {et~al.} 2006, \bibinfo{title}{{The Hubble Ultra Deep Field},} \aj, 132, 1729, \dodoi{10.1086/507302}

% type= article
\bibitem[{V. {Belokurov} \& A. {Kravtsov}(2023){Belokurov} \& {Kravtsov}}]{Belokurov:2023}
{Belokurov}, V., \& {Kravtsov}, A. 2023, \bibinfo{title}{{Nitrogen enrichment and clustered star formation at the dawn of the Galaxy},} \mnras, 525, 4456, \dodoi{10.1093/mnras/stad2241}

% type= article
\bibitem[{V. {Belokurov} \& A. {Kravtsov}(2024){Belokurov} \& {Kravtsov}}]{Belokurov:2024}
{Belokurov}, V., \& {Kravtsov}, A. 2024, \bibinfo{title}{{In-situ versus accreted Milky Way globular clusters: a new classification method and implications for cluster formation},} \mnras, 528, 3198, \dodoi{10.1093/mnras/stad3920}

% type= misc
\bibitem[{D.~A. Berg(2025)Berg}]{Berg:2025_GISM}
Berg, D.~A. 2025, UV-MIR ISM/Galaxy Studies,, Lecture presented at the 2025 International Summer School on the Interstellar Medium of Galaxies (GISM3) \url{https://ismgalaxies2025.sciencesconf.org/data/pages/GISM3_lecture_slides_D_Berg.pdf}

% type= article
\bibitem[{D.~A. {Berg} {et~al.}(2019){Berg}, {Erb}, {Henry}, {Skillman}, \& {McQuinn}}]{Berg:2019}
{Berg}, D.~A., {Erb}, D.~K., {Henry}, R. B.~C., {Skillman}, E.~D., \& {McQuinn}, K. B.~W. 2019, \bibinfo{title}{{The Chemical Evolution of Carbon, Nitrogen, and Oxygen in Metal-poor Dwarf Galaxies},} \apj, 874, 93, \dodoi{10.3847/1538-4357/ab020a}

% type= article
\bibitem[{S.~E.~I. {Bosman} {et~al.}(2022){Bosman}, {Davies}, {Becker}, {Keating}, {Davies}, {Zhu}, {Eilers}, {D'Odorico}, {Bian}, {Bischetti}, \& et~al.}]{Bosman:2022}
{Bosman}, S. E.~I., {Davies}, F.~B., {Becker}, G.~D., {et~al.} 2022, \bibinfo{title}{{Hydrogen reionization ends by z = 5.3: Lyman-{\ensuremath{\alpha}} optical depth measured by the XQR-30 sample},} \mnras, 514, 55, \dodoi{10.1093/mnras/stac1046}

% type= article
\bibitem[{R.~J. {Bouwens} {et~al.}(2014){Bouwens}, {Illingworth}, {Oesch}, {Labb{\'e}}, {van Dokkum}, {Trenti}, {Franx}, {Smit}, {Gonzalez}, \& {Magee}}]{Bouwens:2014}
{Bouwens}, R.~J., {Illingworth}, G.~D., {Oesch}, P.~A., {et~al.} 2014, \bibinfo{title}{{UV-continuum Slopes of >4000 z \raisebox{-0.5ex}\textasciitilde 4-8 Galaxies from the HUDF/XDF, HUDF09, ERS, CANDELS-South, and CANDELS-North Fields},} \apj, 793, 115, \dodoi{10.1088/0004-637X/793/2/115}

% type= article
\bibitem[{R.~J. {Bouwens} {et~al.}(2021){Bouwens}, {Oesch}, {Stefanon}, {Illingworth}, {Labb{\'e}}, {Reddy}, {Atek}, {Montes}, {Naidu}, {Nanayakkara}, \& et~al.}]{Bouwens:2021}
{Bouwens}, R.~J., {Oesch}, P.~A., {Stefanon}, M., {et~al.} 2021, \bibinfo{title}{{New Determinations of the UV Luminosity Functions from z 9 to 2 Show a Remarkable Consistency with Halo Growth and a Constant Star Formation Efficiency},} \aj, 162, 47, \dodoi{10.3847/1538-3881/abf83e}

% type= article
\bibitem[{V. {Bromm} \& R.~B. {Larson}(2004){Bromm} \& {Larson}}]{Bromm:2004}
{Bromm}, V., \& {Larson}, R.~B. 2004, \bibinfo{title}{{The First Stars},} \araa, 42, 79, \dodoi{10.1146/annurev.astro.42.053102.134034}

% type= article
\bibitem[{D. {Calzetti} {et~al.}(1994){Calzetti}, {Kinney}, \& {Storchi-Bergmann}}]{Calzetti:1994}
{Calzetti}, D., {Kinney}, A.~L., \& {Storchi-Bergmann}, T. 1994, \bibinfo{title}{{Dust Extinction of the Stellar Continua in Starburst Galaxies: The Ultraviolet and Optical Extinction Law},} \apj, 429, 582, \dodoi{10.1086/174346}

% type= article
\bibitem[{A.~J. {Cameron} {et~al.}(2023){Cameron}, {Katz}, {Rey}, \& {Saxena}}]{Cameron:2023}
{Cameron}, A.~J., {Katz}, H., {Rey}, M.~P., \& {Saxena}, A. 2023, \bibinfo{title}{{Nitrogen enhancements 440 Myr after the big bang: supersolar N/O, a tidal disruption event, or a dense stellar cluster in GN-z11?},} \mnras, 523, 3516, \dodoi{10.1093/mnras/stad1579}

% type= article
\bibitem[{A.~J. {Cameron} {et~al.}(2024){Cameron}, {Katz}, {Witten}, {Saxena}, {Laporte}, \& {Bunker}}]{Cameron:2024}
{Cameron}, A.~J., {Katz}, H., {Witten}, C., {et~al.} 2024, \bibinfo{title}{{Nebular dominated galaxies: insights into the stellar initial mass function at high redshift},} \mnras, 534, 523, \dodoi{10.1093/mnras/stae1547}

% type= article
\bibitem[{J.~A. {Cardelli} {et~al.}(1989){Cardelli}, {Clayton}, \& {Mathis}}]{Cardelli:1989}
{Cardelli}, J.~A., {Clayton}, G.~C., \& {Mathis}, J.~S. 1989, \bibinfo{title}{{The Relationship between Infrared, Optical, and Ultraviolet Extinction},} \apj, 345, 245, \dodoi{10.1086/167900}

% type= article
\bibitem[{M. {Chatzikos} {et~al.}(2023){Chatzikos}, {Bianchi}, {Camilloni}, {Chakraborty}, {Gunasekera}, {Guzm{\'a}n}, {Milby}, {Sarkar}, {Shaw}, {van Hoof}, \& et~al.}]{Chatzikos:2023}
{Chatzikos}, M., {Bianchi}, S., {Camilloni}, F., {et~al.} 2023, \bibinfo{title}{{The 2023 Release of Cloudy},} \rmxaa, 59, 327, \dodoi{10.22201/ia.01851101p.2023.59.02.12}

% type= article
\bibitem[{J. {Choi} {et~al.}(2016){Choi}, {Dotter}, {Conroy}, {Cantiello}, {Paxton}, \& {Johnson}}]{Choi:2016}
{Choi}, J., {Dotter}, A., {Conroy}, C., {et~al.} 2016, \bibinfo{title}{{Mesa Isochrones and Stellar Tracks (MIST). I. Solar-scaled Models},} \apj, 823, 102, \dodoi{10.3847/0004-637X/823/2/102}

% type= article
\bibitem[{N.~J. {Cleri} {et~al.}(2026){Cleri}, {Lewis}, {Leja}, {Helton}, {Burnham}, {Curtis}, {de Graaff}, {Hirschmann}, {Katz}, {Maseda}, \& et~al.}]{Cleri:2026}
{Cleri}, N.~J., {Lewis}, Z.~J., {Leja}, J., {et~al.} 2026, \bibinfo{title}{{RUBIES: The Evolution of the Ionization Parameter from 0 < z < 9},} arXiv e-prints, arXiv:2605.30410, \dodoi{10.48550/arXiv.2605.30410}

% type= article
\bibitem[{C. {Conroy} \& J.~E. {Gunn}(2010){Conroy} \& {Gunn}}]{Conroy:2010}
{Conroy}, C., \& {Gunn}, J.~E. 2010, \bibinfo{title}{{The Propagation of Uncertainties in Stellar Population Synthesis Modeling. III. Model Calibration, Comparison, and Evaluation},} \apj, 712, 833, \dodoi{10.1088/0004-637X/712/2/833}

% type= article
\bibitem[{C. {Conroy} {et~al.}(2009){Conroy}, {Gunn}, \& {White}}]{Conroy:2009}
{Conroy}, C., {Gunn}, J.~E., \& {White}, M. 2009, \bibinfo{title}{{The Propagation of Uncertainties in Stellar Population Synthesis Modeling. I. The Relevance of Uncertain Aspects of Stellar Evolution and the Initial Mass Function to the Derived Physical Properties of Galaxies},} \apj, 699, 486, \dodoi{10.1088/0004-637X/699/1/486}

% type= article
\bibitem[{C. {Conroy} {et~al.}(2019){Conroy}, {Naidu}, {Zaritsky}, {Bonaca}, {Cargile}, {Johnson}, \& {Caldwell}}]{Conroy:2019}
{Conroy}, C., {Naidu}, R.~P., {Zaritsky}, D., {et~al.} 2019, \bibinfo{title}{{Resolving the Metallicity Distribution of the Stellar Halo with the H3 Survey},} \apj, 887, 237, \dodoi{10.3847/1538-4357/ab5710}

% type= article
\bibitem[{E. {Curtis-Lake} {et~al.}(2026){Curtis-Lake}, {Cameron}, {Bunker}, {Scholtz}, {Carniani}, {Parlanti}, {D'Eugenio}, {Jakobsen}, {Willmer}, {Arribas}, \& et~al.}]{Curtis-Lake:2026}
{Curtis-Lake}, E., {Cameron}, A.~J., {Bunker}, A.~J., {et~al.} 2026, \bibinfo{title}{{JADES data release 4 ─ Paper I. Sample selection, observing strategy and redshifts of the complete spectroscopic sample},} \mnras, 549, stag836, \dodoi{10.1093/mnras/stag836}

% type= article
\bibitem[{A. {de Graaff} {et~al.}(2024){de Graaff}, {Rix}, {Carniani}, {Suess}, {Charlot}, {Curtis-Lake}, {Arribas}, {Baker}, {Boyett}, {Bunker}, \& et~al.}]{deGraaff:2024}
{de Graaff}, A., {Rix}, H.-W., {Carniani}, S., {et~al.} 2024, \bibinfo{title}{{Ionised gas kinematics and dynamical masses of z {\ensuremath{\gtrsim}} 6 galaxies from JADES/NIRSpec high-resolution spectroscopy},} \aap, 684, A87, \dodoi{10.1051/0004-6361/202347755}

% type= article
\bibitem[{F. {D'Eugenio} {et~al.}(2024){D'Eugenio}, {Maiolino}, {Carniani}, {Chevallard}, {Curtis-Lake}, {Witstok}, {Charlot}, {Baker}, {Arribas}, {Boyett}, \& et~al.}]{DEugenio:2024}
{D'Eugenio}, F., {Maiolino}, R., {Carniani}, S., {et~al.} 2024, \bibinfo{title}{{JADES: Carbon enrichment 350 Myr after the Big Bang},} \aap, 689, A152, \dodoi{10.1051/0004-6361/202348636}

% type= article
\bibitem[{M. {Dijkstra}(2009){Dijkstra}}]{Dijkstra:2009}
{Dijkstra}, M. 2009, \bibinfo{title}{{Continuum Emission by Cooling Clouds},} \apj, 690, 82, \dodoi{10.1088/0004-637X/690/1/82}

% type= article
\bibitem[{A. {Dotter}(2016){Dotter}}]{Dotter:2016}
{Dotter}, A. 2016, \bibinfo{title}{{MESA Isochrones and Stellar Tracks (MIST) 0: Methods for the Construction of Stellar Isochrones},} \apjs, 222, 8, \dodoi{10.3847/0067-0049/222/1/8}

% type= article
\bibitem[{D.~J. {Eisenstein} {et~al.}(2026){Eisenstein}, {Willott}, {Alberts}, {Arribas}, {Bonaventura}, {Bunker}, {Cameron}, {Carniani}, {Charlot}, {Curtis-Lake}, \& et~al.}]{Eisenstein:2026}
{Eisenstein}, D.~J., {Willott}, C., {Alberts}, S., {et~al.} 2026, \bibinfo{title}{{Overview of the JWST Advanced Deep Extragalactic Survey (JADES)},} \apjs, 283, 6, \dodoi{10.3847/1538-4365/ae3163}

% type= article
\bibitem[{R. {Endsley} {et~al.}(2024){Endsley}, {Stark}, {Whitler}, {Topping}, {Johnson}, {Robertson}, {Tacchella}, {Alberts}, {Baker}, {Bhatawdekar}, \& et~al.}]{Endsley:2024}
{Endsley}, R., {Stark}, D.~P., {Whitler}, L., {et~al.} 2024, \bibinfo{title}{{The star-forming and ionizing properties of dwarf z 6-9 galaxies in JADES: insights on bursty star formation and ionized bubble growth},} \mnras, 533, 1111, \dodoi{10.1093/mnras/stae1857}

% type= article
\bibitem[{X. {Fan} {et~al.}(2023){Fan}, {Ba{\~n}ados}, \& {Simcoe}}]{Fan:2023}
{Fan}, X., {Ba{\~n}ados}, E., \& {Simcoe}, R.~A. 2023, \bibinfo{title}{{Quasars and the Intergalactic Medium at Cosmic Dawn},} \araa, 61, 373, \dodoi{10.1146/annurev-astro-052920-102455}

% type= article
\bibitem[{X. {Fan} {et~al.}(2006){Fan}, {Strauss}, {Becker}, {White}, {Gunn}, {Knapp}, {Richards}, {Schneider}, {Brinkmann}, \& {Fukugita}}]{Fan:2006}
{Fan}, X., {Strauss}, M.~A., {Becker}, R.~H., {et~al.} 2006, \bibinfo{title}{{Constraining the Evolution of the Ionizing Background and the Epoch of Reionization with z\raisebox{-0.5ex}\textasciitilde6 Quasars. II. A Sample of 19 Quasars},} \aj, 132, 117, \dodoi{10.1086/504836}

% type= article
\bibitem[{G.~J. {Ferland} {et~al.}(1998){Ferland}, {Korista}, {Verner}, {Ferguson}, {Kingdon}, \& {Verner}}]{Ferland:1998}
{Ferland}, G.~J., {Korista}, K.~T., {Verner}, D.~A., {et~al.} 1998, \bibinfo{title}{{CLOUDY 90: Numerical Simulation of Plasmas and Their Spectra},} \pasp, 110, 761, \dodoi{10.1086/316190}

% type= article
\bibitem[{F. {Feroz} {et~al.}(2009){Feroz}, {Hobson}, \& {Bridges}}]{Feroz:2009}
{Feroz}, F., {Hobson}, M.~P., \& {Bridges}, M. 2009, \bibinfo{title}{{MULTINEST: an efficient and robust Bayesian inference tool for cosmology and particle physics},} \mnras, 398, 1601, \dodoi{10.1111/j.1365-2966.2009.14548.x}

% type= article
\bibitem[{A. {Ferrara} \& A. {Loeb}(2013){Ferrara} \& {Loeb}}]{Ferrara:2013}
{Ferrara}, A., \& {Loeb}, A. 2013, \bibinfo{title}{{Escape fraction of the ionizing radiation from starburst galaxies at high redshifts},} \mnras, 431, 2826, \dodoi{10.1093/mnras/stt381}

% type= article
\bibitem[{S.~L. {Finkelstein} {et~al.}(2012){Finkelstein}, {Papovich}, {Salmon}, {Finlator}, {Dickinson}, {Ferguson}, {Giavalisco}, {Koekemoer}, {Reddy}, {Bassett}, \& et~al.}]{Finkelstein:2012}
{Finkelstein}, S.~L., {Papovich}, C., {Salmon}, B., {et~al.} 2012, \bibinfo{title}{{Candels: The Evolution of Galaxy Rest-frame Ultraviolet Colors from z = 8 to 4},} \apj, 756, 164, \dodoi{10.1088/0004-637X/756/2/164}

% type= article
\bibitem[{S.~L. {Finkelstein} {et~al.}(2019){Finkelstein}, {D'Aloisio}, {Paardekooper}, {Ryan}, {Behroozi}, {Finlator}, {Livermore}, {Upton Sanderbeck}, {Dalla Vecchia}, \& {Khochfar}}]{Finkelstein:2019}
{Finkelstein}, S.~L., {D'Aloisio}, A., {Paardekooper}, J.-P., {et~al.} 2019, \bibinfo{title}{{Conditions for Reionizing the Universe with a Low Galaxy Ionizing Photon Escape Fraction},} \apj, 879, 36, \dodoi{10.3847/1538-4357/ab1ea8}

% type= article
\bibitem[{T.~J. {Fletcher} {et~al.}(2019){Fletcher}, {Tang}, {Robertson}, {Nakajima}, {Ellis}, {Stark}, \& {Inoue}}]{Fletcher:2019}
{Fletcher}, T.~J., {Tang}, M., {Robertson}, B.~E., {et~al.} 2019, \bibinfo{title}{{The Lyman Continuum Escape Survey: Ionizing Radiation from [O III]-strong Sources at a Redshift of 3.1},} \apj, 878, 87, \dodoi{10.3847/1538-4357/ab2045}

% type= article
\bibitem[{D. {Foreman-Mackey}(2016){Foreman-Mackey}}]{Foreman-Mackey:2016}
{Foreman-Mackey}, D. 2016, \bibinfo{title}{{corner.py: Scatterplot matrices in Python},} The Journal of Open Source Software, 1, 24, \dodoi{10.21105/joss.00024}

% type= article
\bibitem[{E. {Garaldi} {et~al.}(2022){Garaldi}, {Kannan}, {Smith}, {Springel}, {Pakmor}, {Vogelsberger}, \& {Hernquist}}]{Garaldi:2022}
{Garaldi}, E., {Kannan}, R., {Smith}, A., {et~al.} 2022, \bibinfo{title}{{The THESAN project: properties of the intergalactic medium and its connection to reionization-era galaxies},} \mnras, 512, 4909, \dodoi{10.1093/mnras/stac257}

% type= article
\bibitem[{E. {Garaldi} {et~al.}(2024){Garaldi}, {Kannan}, {Smith}, {Borrow}, {Vogelsberger}, {Pakmor}, {Springel}, {Hernquist}, {Gal{\'a}rraga-Espinosa}, {Yeh}, \& et~al.}]{Garaldi:2024}
{Garaldi}, E., {Kannan}, R., {Smith}, A., {et~al.} 2024, \bibinfo{title}{{The THESAN project: public data release of radiation-hydrodynamic simulations matching reionization-era JWST observations},} \mnras, 530, 3765, \dodoi{10.1093/mnras/stae839}

% type= article
\bibitem[{M. {Giavalisco} {et~al.}(2004){Giavalisco}, {Ferguson}, {Koekemoer}, {Dickinson}, {Alexander}, {Bauer}, {Bergeron}, {Biagetti}, {Brandt}, {Casertano}, \& et~al.}]{Giavalisco:2004}
{Giavalisco}, M., {Ferguson}, H.~C., {Koekemoer}, A.~M., {et~al.} 2004, \bibinfo{title}{{The Great Observatories Origins Deep Survey: Initial Results from Optical and Near-Infrared Imaging},} \apjl, 600, L93, \dodoi{10.1086/379232}

% type= article
\bibitem[{E. {Giovinazzo} {et~al.}(2026{\natexlab{a}}){Giovinazzo}, {Oesch}, {Verhamme}, {Meyer}, {Witten}, {Chisholm}, {Marques-Chaves}, \& {Simmonds}}]{Giovinazzo:2026b}
{Giovinazzo}, E., {Oesch}, P.~A., {Verhamme}, A., {et~al.} 2026{\natexlab{a}}, \bibinfo{title}{{Reionization driven by the few: the ionizing budget of galaxies at z=5-10 from JWST/NIRSpec},} arXiv e-prints, arXiv:2607.22834, \dodoi{10.48550/arXiv.2607.22834}

% type= article
\bibitem[{E. {Giovinazzo} {et~al.}(2026{\natexlab{b}}){Giovinazzo}, {Oesch}, {Weibel}, {Meyer}, {Witten}, {Bhagwat}, {Brammer}, {Chisholm}, {de Graaff}, {Gottumukkala}, \& et~al.}]{Giovinazzo:2026a}
{Giovinazzo}, E., {Oesch}, P.~A., {Weibel}, A., {et~al.} 2026{\natexlab{b}}, \bibinfo{title}{{Breaking through the cosmic fog: JWST/NIRSpec constraints on ionizing photon escape in reionization-era galaxies},} \aap, 707, A352, \dodoi{10.1051/0004-6361/202556204}

% type= article
\bibitem[{I. {Goovaerts} {et~al.}(2026){Goovaerts}, {Rafelski}, {Beckett}, {Oyarz{\'u}n}, {Citro}, {Hasan}, {Nedkova}, {Hawcroft}, {Koekemoer}, {Revalski}, \& et~al.}]{Goovaerts:2026}
{Goovaerts}, I., {Rafelski}, M., {Beckett}, A., {et~al.} 2026, \bibinfo{title}{{MXDFz4.4: A LyC Emitter 250 Myr after the Epoch of Reionization and a First Test of Ly{\ensuremath{\alpha}} Morphology as a Tracer of LyC Escape at High Redshift},} \apj, 1005, 34, \dodoi{10.3847/1538-4357/ae75b0}

% type= article
\bibitem[{J.~E. {Gunn} \& B.~A. {Peterson}(1965){Gunn} \& {Peterson}}]{Gunn:1965}
{Gunn}, J.~E., \& {Peterson}, B.~A. 1965, \bibinfo{title}{{On the Density of Neutral Hydrogen in Intergalactic Space.},} \apj, 142, 1633, \dodoi{10.1086/148444}

% type= article
\bibitem[{K.~N. {Hainline} {et~al.}(2024){Hainline}, {D'Eugenio}, {Jakobsen}, {Chevallard}, {Carniani}, {Witstok}, {Ji}, {Curtis-Lake}, {Johnson}, {Robertson}, \& et~al.}]{Hainline:2024}
{Hainline}, K.~N., {D'Eugenio}, F., {Jakobsen}, P., {et~al.} 2024, \bibinfo{title}{{Searching for Emission Lines at z > 11: The Role of Damped Ly{\ensuremath{\alpha}} and Hints About the Escape of Ionizing Photons},} \apj, 976, 160, \dodoi{10.3847/1538-4357/ad8447}

% type= article
\bibitem[{Y. {Harikane} {et~al.}(2016){Harikane}, {Ouchi}, {Ono}, {More}, {Saito}, {Lin}, {Coupon}, {Shimasaku}, {Shibuya}, {Price}, \& et~al.}]{Harikane:2016}
{Harikane}, Y., {Ouchi}, M., {Ono}, Y., {et~al.} 2016, \bibinfo{title}{{Evolution of Stellar-to-Halo Mass Ratio at z = 0 - 7 Identified by Clustering Analysis with the Hubble Legacy Imaging and Early Subaru/Hyper Suprime-Cam Survey Data},} \apj, 821, 123, \dodoi{10.3847/0004-637X/821/2/123}

% type= article
\bibitem[{C.~R. {Harris} {et~al.}(2020){Harris}, {Millman}, {van der Walt}, {Gommers}, {Virtanen}, {Cournapeau}, {Wieser}, {Taylor}, {Berg}, {Smith}, \& et~al.}]{NumPy:2020}
{Harris}, C.~R., {Millman}, K.~J., {van der Walt}, S.~J., {et~al.} 2020, \bibinfo{title}{{Array programming with NumPy},} \nat, 585, 357, \dodoi{10.1038/s41586-020-2649-2}

% type= article
\bibitem[{R. {Hausen} \& B.~E. {Robertson}(2022){Hausen} \& {Robertson}}]{Hausen:2022}
{Hausen}, R., \& {Robertson}, B.~E. 2022, \bibinfo{title}{{FitsMap: A simple, lightweight tool for displaying interactive astronomical image and catalog data},} Astronomy and Computing, 39, 100586, \dodoi{10.1016/j.ascom.2022.100586}

% type= article
\bibitem[{T.~M. {Heckman} {et~al.}(2011){Heckman}, {Borthakur}, {Overzier}, {Kauffmann}, {Basu-Zych}, {Leitherer}, {Sembach}, {Martin}, {Rich}, {Schiminovich}, \& et~al.}]{Heckman:2011}
{Heckman}, T.~M., {Borthakur}, S., {Overzier}, R., {et~al.} 2011, \bibinfo{title}{{Extreme Feedback and the Epoch of Reionization: Clues in the Local Universe},} \apj, 730, 5, \dodoi{10.1088/0004-637X/730/1/5}

% type= article
\bibitem[{K.~E. {Heintz} {et~al.}(2024){Heintz}, {Watson}, {Brammer}, {Vejlgaard}, {Hutter}, {Strait}, {Matthee}, {Oesch}, {Jakobsson}, {Tanvir}, \& et~al.}]{Heintz:2024}
{Heintz}, K.~E., {Watson}, D., {Brammer}, G., {et~al.} 2024, \bibinfo{title}{{Strong damped Lyman-{\ensuremath{\alpha}} absorption in young star-forming galaxies at redshifts 9 to 11},} Science, 384, 890, \dodoi{10.1126/science.adj0343}

% type= article
\bibitem[{K.~E. {Heintz} {et~al.}(2025{\natexlab{a}}){Heintz}, {Brammer}, {Watson}, {Oesch}, {Keating}, {Hayes}, {Abdurro'uf}, {Arellano-C{\'o}rdova}, {Carnall}, {Christiansen}, \& et~al.}]{Heintz:2025}
{Heintz}, K.~E., {Brammer}, G.~B., {Watson}, D., {et~al.} 2025{\natexlab{a}}, \bibinfo{title}{{The JWST-PRIMAL archival survey: A JWST/NIRSpec reference sample for the physical properties and Lyman-{\ensuremath{\alpha}} absorption and emission of {\ensuremath{\sim}}600 galaxies at z = 5.0 {\ensuremath{-}} 13.4},} \aap, 693, A60, \dodoi{10.1051/0004-6361/202450243}

% type= article
\bibitem[{K.~E. {Heintz} {et~al.}(2025{\natexlab{b}}){Heintz}, {Pollock}, {Witstok}, {Carniani}, {Hainline}, {D'Eugenio}, {Terp}, {Saxena}, \& {Watson}}]{Heintz:2025_z14}
{Heintz}, K.~E., {Pollock}, C.~L., {Witstok}, J., {et~al.} 2025{\natexlab{b}}, \bibinfo{title}{{Dissecting the Massive Pristine, Neutral Gas Reservoir of a Remarkably Bright Galaxy at z = 14.179},} \apjl, 987, L2, \dodoi{10.3847/2041-8213/ade393}

% type= article
\bibitem[{J.~M. {Helton} {et~al.}(2024{\natexlab{a}}){Helton}, {Sun}, {Woodrum}, {Hainline}, {Willmer}, {Rieke}, {Rieke}, {Alberts}, {Eisenstein}, {Tacchella}, \& et~al.}]{Helton:2024b}
{Helton}, J.~M., {Sun}, F., {Woodrum}, C., {et~al.} 2024{\natexlab{a}}, \bibinfo{title}{{Identification of High-redshift Galaxy Overdensities in GOODS-N and GOODS-S},} \apj, 974, 41, \dodoi{10.3847/1538-4357/ad6867}

% type= article
\bibitem[{J.~M. {Helton} {et~al.}(2024{\natexlab{b}}){Helton}, {Sun}, {Woodrum}, {Hainline}, {Willmer}, {Rieke}, {Rieke}, {Tacchella}, {Robertson}, {Johnson}, \& et~al.}]{Helton:2024a}
{Helton}, J.~M., {Sun}, F., {Woodrum}, C., {et~al.} 2024{\natexlab{b}}, \bibinfo{title}{{The JWST Advanced Deep Extragalactic Survey: Discovery of an Extreme Galaxy Overdensity at z = 5.4 with JWST/NIRCam in GOODS-S},} \apj, 962, 124, \dodoi{10.3847/1538-4357/ad0da7}

% type= article
\bibitem[{J.~M. {Helton} {et~al.}(2026{\natexlab{a}}){Helton}, {Morrison}, {Hainline}, {D'Eugenio}, {Rieke}, {Alberts}, {Carniani}, {Leja}, {Li}, {Rinaldi}, \& et~al.}]{Helton:2026a}
{Helton}, J.~M., {Morrison}, J.~E., {Hainline}, K.~N., {et~al.} 2026{\natexlab{a}}, \bibinfo{title}{{Ionizing Photon Production Efficiencies and Chemical Abundances at Cosmic Dawn Revealed by Ultradeep Rest-frame Optical Spectroscopy of JADES-GS-z14-0},} \apjl, 1004, L30, \dodoi{10.3847/2041-8213/ae7446}

% type= article
\bibitem[{J.~M. {Helton} {et~al.}(2026{\natexlab{b}}){Helton}, {Alberts}, {Rieke}, {Hainline}, {Ji}, {Rieke}, {Johnson}, {Robertson}, {Tacchella}, {Whitler}, \& et~al.}]{Helton:2026b}
{Helton}, J.~M., {Alberts}, S., {Rieke}, G.~H., {et~al.} 2026{\natexlab{b}}, \bibinfo{title}{{The Stellar Populations and Rest-frame Colors of Star-forming Galaxies at z ≍ 8: Exploring the Impact of Filter Choice and Star Formation History Assumption with JADES},} \apj, 1006, 43, \dodoi{10.3847/1538-4357/ae5dcd}

% type= article
\bibitem[{M.~D. {Hoffman} \& A. {Gelman}(2011){Hoffman} \& {Gelman}}]{Hoffman:2011}
{Hoffman}, M.~D., \& {Gelman}, A. 2011, \bibinfo{title}{{The No-U-Turn Sampler: Adaptively Setting Path Lengths in Hamiltonian Monte Carlo},} arXiv e-prints, arXiv:1111.4246, \dodoi{10.48550/arXiv.1111.4246}

% type= article
\bibitem[{T.~Y.-Y. {Hsiao} {et~al.}(2026){Hsiao}, {Berg}, {Finkelstein}, {Gupta}, {Martinez}, {Taylor}, {Akins}, {Chavez Ortiz}, {Chisholm}, {Furtak}, \& et~al.}]{Hsiao:2026}
{Hsiao}, T. Y.-Y., {Berg}, D.~A., {Finkelstein}, S.~L., {et~al.} 2026, \bibinfo{title}{{An Optical Illusion: High Electron Densities Create Extremely Metal-Poor Galaxy Impostors},} arXiv e-prints, arXiv:2608.20339, \dodoi{10.48550/arXiv.2608.20339}

% type= article
\bibitem[{M. {Huberty} {et~al.}(2025){Huberty}, {Scarlata}, {Hayes}, \& {Gazagnes}}]{Huberty:2025}
{Huberty}, M., {Scarlata}, C., {Hayes}, M.~J., \& {Gazagnes}, S. 2025, \bibinfo{title}{{The Pitfalls of Using Ly{\ensuremath{\alpha}} Damping Wings in High-z Galaxy Spectra to Measure the Intergalactic Neutral Hydrogen Fraction},} \apj, 987, 82, \dodoi{10.3847/1538-4357/add5e7}

% type= article
\bibitem[{J.~D. {Hunter}(2007){Hunter}}]{Matplotlib:2007}
{Hunter}, J.~D. 2007, \bibinfo{title}{{Matplotlib: A 2D Graphics Environment},} Computing in Science and Engineering, 9, 90, \dodoi{10.1109/MCSE.2007.55}

% type= article
\bibitem[{Y. {Isobe} {et~al.}(2023){Isobe}, {Ouchi}, {Nakajima}, {Harikane}, {Ono}, {Xu}, {Zhang}, \& {Umeda}}]{Isobe:2023}
{Isobe}, Y., {Ouchi}, M., {Nakajima}, K., {et~al.} 2023, \bibinfo{title}{{Redshift Evolution of Electron Density in the Interstellar Medium at z 0-9 Uncovered with JWST/NIRSpec Spectra and Line-spread Function Determinations},} \apj, 956, 139, \dodoi{10.3847/1538-4357/acf376}

% type= article
\bibitem[{X. {Ji} {et~al.}(2026){Ji}, {Belokurov}, {Maiolino}, {Monty}, {Isobe}, {Kravtsov}, {McClymont}, \& {{\"U}bler}}]{Ji:2026}
{Ji}, X., {Belokurov}, V., {Maiolino}, R., {et~al.} 2026, \bibinfo{title}{{Connecting JWST discovered N/O-enhanced galaxies to globular clusters: evidence from chemical imprints},} \mnras, 545, staf2110, \dodoi{10.1093/mnras/staf2110}

% type= article
\bibitem[{Z. {Ji} {et~al.}(2020){Ji}, {Giavalisco}, {Vanzella}, {Siana}, {Pentericci}, {Jaskot}, {Liu}, {Nonino}, {Ferguson}, {Castellano}, \& et~al.}]{Ji:2020}
{Ji}, Z., {Giavalisco}, M., {Vanzella}, E., {et~al.} 2020, \bibinfo{title}{{HST Imaging of the Ionizing Radiation from a Star-forming Galaxy at z = 3.794},} \apj, 888, 109, \dodoi{10.3847/1538-4357/ab5fdc}

% type= article
\bibitem[{X. {Jin} {et~al.}(2024){Jin}, {Yang}, {Fan}, {Wang}, {Kakiichi}, {Meyer}, {Becker}, {Zou}, {Ba{\~n}ados}, {Champagne}, \& et~al.}]{Jin:2024}
{Jin}, X., {Yang}, J., {Fan}, X., {et~al.} 2024, \bibinfo{title}{{A SPectroscopic Survey of Biased Halos In the Reionization Era (ASPIRE): JWST Supports Earlier Reionization around [O III] Emitters},} \apj, 976, 93, \dodoi{10.3847/1538-4357/ad82de}

% type= article
\bibitem[{B.~D. {Johnson} {et~al.}(2021){Johnson}, {Leja}, {Conroy}, \& {Speagle}}]{Johnson:2021}
{Johnson}, B.~D., {Leja}, J., {Conroy}, C., \& {Speagle}, J.~S. 2021, \bibinfo{title}{{Stellar Population Inference with Prospector},} \apjs, 254, 22, \dodoi{10.3847/1538-4365/abef67}

% type= article
\bibitem[{B.~D. {Johnson} {et~al.}(2026){Johnson}, {Robertson}, {Eisenstein}, {Tacchella}, {Pusk{\'a}s}, {Duan}, {Wu}, {Hainline}, {Rieke}, {Willott}, \& et~al.}]{Johnson:2026}
{Johnson}, B.~D., {Robertson}, B.~E., {Eisenstein}, D.~J., {et~al.} 2026, \bibinfo{title}{{JWST Advanced Deep Extragalactic Survey (JADES) Data Release 5: NIRCam Imaging in GOODS-S and GOODS-N},} arXiv e-prints, arXiv:2601.15954, \dodoi{10.48550/arXiv.2601.15954}

% type= article
\bibitem[{K. {Kakiichi} {et~al.}(2025){Kakiichi}, {Jin}, {Wang}, {Meyer}, {Garaldi}, {Bosman}, {Davies}, {Fan}, {Trebitsch}, {Yang}, \& et~al.}]{Kakiichi:2025}
{Kakiichi}, K., {Jin}, X., {Wang}, F., {et~al.} 2025, \bibinfo{title}{{JWST ASPIRE: How Did Galaxies Complete Reionization? Evidence for Excess IGM Transmission around ${\rm [O\,{\scriptstyle III}]}$ Emitters during Reionization},} arXiv e-prints, arXiv:2503.07074, \dodoi{10.48550/arXiv.2503.07074}

% type= article
\bibitem[{R. {Kannan} {et~al.}(2022){Kannan}, {Garaldi}, {Smith}, {Pakmor}, {Springel}, {Vogelsberger}, \& {Hernquist}}]{Kannan:2022}
{Kannan}, R., {Garaldi}, E., {Smith}, A., {et~al.} 2022, \bibinfo{title}{{Introducing the THESAN project: radiation-magnetohydrodynamic simulations of the epoch of reionization},} \mnras, 511, 4005, \dodoi{10.1093/mnras/stab3710}

% type= article
\bibitem[{D. {Kashino} {et~al.}(2023){Kashino}, {Lilly}, {Matthee}, {Eilers}, {Mackenzie}, {Bordoloi}, \& {Simcoe}}]{Kashino:2023}
{Kashino}, D., {Lilly}, S.~J., {Matthee}, J., {et~al.} 2023, \bibinfo{title}{{EIGER. I. A Large Sample of [O III]-emitting Galaxies at 5.3 < z < 6.9 and Direct Evidence for Local Reionization by Galaxies},} \apj, 950, 66, \dodoi{10.3847/1538-4357/acc588}

% type= article
\bibitem[{D. {Kashino} {et~al.}(2026){Kashino}, {Lilly}, {Matthee}, {Mackenzie}, {Eilers}, {Bordoloi}, {Simcoe}, {Naidu}, {Yue}, \& {Liu}}]{Kashino:2026}
{Kashino}, D., {Lilly}, S.~J., {Matthee}, J., {et~al.} 2026, \bibinfo{title}{{EIGER. VII. The Evolving Relationship between Galaxies and the Intergalactic Medium in the Final Stages of Reionization},} \apj, 997, 280, \dodoi{10.3847/1538-4357/ae2799}

% type= article
\bibitem[{H. {Katz} {et~al.}(2025){Katz}, {Cameron}, {Saxena}, {Barrufet}, {Choustikov}, {Cleri}, {de Graff}, {Ellis}, {Fosbury}, {Heintz}, \& et~al.}]{Katz:2025}
{Katz}, H., {Cameron}, A.~J., {Saxena}, A., {et~al.} 2025, \bibinfo{title}{{21 Balmer Jump Street: The Nebular Continuum at High Redshift and Implications for the Bright Galaxy Problem, UV Continuum Slopes, and Early Stellar Populations},} The Open Journal of Astrophysics, 8, 104, \dodoi{10.33232/001c.142570}

% type= article
\bibitem[{L.~C. {Keating} {et~al.}(2020){Keating}, {Weinberger}, {Kulkarni}, {Haehnelt}, {Chardin}, \& {Aubert}}]{Keating:2020}
{Keating}, L.~C., {Weinberger}, L.~H., {Kulkarni}, G., {et~al.} 2020, \bibinfo{title}{{Long troughs in the Lyman-{\ensuremath{\alpha}} forest below redshift 6 due to islands of neutral hydrogen},} \mnras, 491, 1736, \dodoi{10.1093/mnras/stz3083}

% type= article
\bibitem[{R.~C. {Kennicutt} \& N.~J. {Evans}(2012){Kennicutt} \& {Evans}}]{Kennicutt:2012}
{Kennicutt}, R.~C., \& {Evans}, N.~J. 2012, \bibinfo{title}{{Star Formation in the Milky Way and Nearby Galaxies},} \araa, 50, 531, \dodoi{10.1146/annurev-astro-081811-125610}

% type= article
\bibitem[{I.~G. {Kramarenko} {et~al.}(2026){Kramarenko}, {Rosdahl}, {Blaizot}, {Matthee}, {Katz}, \& {Di Cesare}}]{Kramarenko:2026}
{Kramarenko}, I.~G., {Rosdahl}, J., {Blaizot}, J., {et~al.} 2026, \bibinfo{title}{{H{\ensuremath{\alpha}} as a tracer of star formation in the SPHINX cosmological simulations},} \aap, 707, A184, \dodoi{10.1051/0004-6361/202557114}

% type= article
\bibitem[{G. {Kulkarni} {et~al.}(2019){Kulkarni}, {Keating}, {Haehnelt}, {Bosman}, {Puchwein}, {Chardin}, \& {Aubert}}]{Kulkarni:2019}
{Kulkarni}, G., {Keating}, L.~C., {Haehnelt}, M.~G., {et~al.} 2019, \bibinfo{title}{{Large Ly {\ensuremath{\alpha}} opacity fluctuations and low CMB {\ensuremath{\tau}} in models of late reionization with large islands of neutral hydrogen extending to z < 5.5},} \mnras, 485, L24, \dodoi{10.1093/mnrasl/slz025}

% type= article
\bibitem[{M.~M. {Lee} {et~al.}(2025){Lee}, {Magdis}, {Brammer}, {Liu}, {Magnelli}, {Gillman}, {Gullberg}, {Ito}, {Sillassen}, {Valentino}, \& et~al.}]{Lee:2025}
{Lee}, M.~M., {Magdis}, G., {Brammer}, G., {et~al.} 2025, \bibinfo{title}{{ECOGAL I. Project design and the first catalogue},} arXiv e-prints, arXiv:2511.20751, \dodoi{10.48550/arXiv.2511.20751}

% type= article
\bibitem[{Y. {Li} {et~al.}(2025){Li}, {Leja}, {Johnson}, {Tacchella}, {Davies}, {Belli}, {Park}, \& {Emami}}]{Li:2025}
{Li}, Y., {Leja}, J., {Johnson}, B.~D., {et~al.} 2025, \bibinfo{title}{{Cue: A Fast and Flexible Photoionization Emulator for Modeling Nebular Emission Powered by Almost Any Ionizing Source},} \apj, 986, 9, \dodoi{10.3847/1538-4357/adcab4}

% type= article
\bibitem[{Y. {Li} {et~al.}(2024){Li}, {Leja}, {Johnson}, {Tacchella}, \& {Naidu}}]{Li:2024}
{Li}, Y., {Leja}, J., {Johnson}, B.~D., {Tacchella}, S., \& {Naidu}, R.~P. 2024, \bibinfo{title}{{No Top-heavy Stellar Initial Mass Function Needed: The Ionizing Radiation of GS9422 Can Be Powered by a Mixture of an Active Galactic Nucleus and Stars},} \apjl, 969, L5, \dodoi{10.3847/2041-8213/ad5280}

% type= article
\bibitem[{V. {Luridiana} {et~al.}(2015){Luridiana}, {Morisset}, \& {Shaw}}]{Luridiana:2015}
{Luridiana}, V., {Morisset}, C., \& {Shaw}, R.~A. 2015, \bibinfo{title}{{PyNeb: a new tool for analyzing emission lines. I. Code description and validation of results},} \aap, 573, A42, \dodoi{10.1051/0004-6361/201323152}

% type= article
\bibitem[{R. {Marques-Chaves} {et~al.}(2022){Marques-Chaves}, {Schaerer}, {{\'A}lvarez-M{\'a}rquez}, {Verhamme}, {Ceverino}, {Chisholm}, {Colina}, {Dessauges-Zavadsky}, {P{\'e}rez-Fournon}, {Saldana-Lopez}, \& et~al.}]{Marques-Chaves:2022}
{Marques-Chaves}, R., {Schaerer}, D., {{\'A}lvarez-M{\'a}rquez}, J., {et~al.} 2022, \bibinfo{title}{{An extreme blue nugget, UV-bright starburst at z = 3.613 with 90 per cent of Lyman continuum photon escape},} \mnras, 517, 2972, \dodoi{10.1093/mnras/stac2893}

% type= article
\bibitem[{R. {Marques-Chaves} {et~al.}(2026){Marques-Chaves}, {{\'A}lvarez-M{\'a}rquez}, {Colina}, {Kendrew}, {Abdurro'uf}, {Blanco-Prieto}, {Boogaard}, {Castellano}, {Caputi}, {Crespo-G{\'o}mez}, \& et~al.}]{MarquesChaves:2026}
{Marques-Chaves}, R., {{\'A}lvarez-M{\'a}rquez}, J., {Colina}, L., {et~al.} 2026, \bibinfo{title}{{PRISMS: U37126, a very blue ISM-naked starburst at z = 10.255 with a nearly 100\% Lyman continuum escape fraction},} \aap, 711, A301, \dodoi{10.1051/0004-6361/202659281}

% type= article
\bibitem[{C.~A. {Mason} {et~al.}(2026){Mason}, {Chen}, {Stark}, {Yi Lu}, {Topping}, \& {Tang}}]{Mason:2026}
{Mason}, C.~A., {Chen}, Z., {Stark}, D.~P., {et~al.} 2026, \bibinfo{title}{{Constraints on the z {\ensuremath{\sim}} 6{\ensuremath{-}}13 intergalactic medium from JWST spectroscopy of Lyman-alpha damping wings in galaxies},} \aap, 705, A114, \dodoi{10.1051/0004-6361/202553820}

% type= article
\bibitem[{C.~A. {Mason} \& M. {Gronke}(2020){Mason} \& {Gronke}}]{Mason:2020}
{Mason}, C.~A., \& {Gronke}, M. 2020, \bibinfo{title}{{Measuring the properties of reionized bubbles with resolved Ly{\ensuremath{\alpha}} spectra},} \mnras, 499, 1395, \dodoi{10.1093/mnras/staa2910}

% type= article
\bibitem[{C.~A. {Mason} {et~al.}(2018){Mason}, {Treu}, {Dijkstra}, {Mesinger}, {Trenti}, {Pentericci}, {de Barros}, \& {Vanzella}}]{Mason:2018}
{Mason}, C.~A., {Treu}, T., {Dijkstra}, M., {et~al.} 2018, \bibinfo{title}{{The Universe Is Reionizing at z {\ensuremath{\sim}} 7: Bayesian Inference of the IGM Neutral Fraction Using Ly{\ensuremath{\alpha}} Emission from Galaxies},} \apj, 856, 2, \dodoi{10.3847/1538-4357/aab0a7}

% type= article
\bibitem[{I.~D. {McGreer} {et~al.}(2015){McGreer}, {Mesinger}, \& {D'Odorico}}]{McGreer:2015}
{McGreer}, I.~D., {Mesinger}, A., \& {D'Odorico}, V. 2015, \bibinfo{title}{{Model-independent evidence in favour of an end to reionization by z {\ensuremath{\approx}} 6},} \mnras, 447, 499, \dodoi{10.1093/mnras/stu2449}

% type= article
\bibitem[{R.~A. {Meyer} {et~al.}(2025){Meyer}, {Roberts-Borsani}, {Oesch}, \& {Ellis}}]{Meyer:2025}
{Meyer}, R.~A., {Roberts-Borsani}, G., {Oesch}, P.~A., \& {Ellis}, R.~S. 2025, \bibinfo{title}{{Probing patchy reionization with JWST: IGM opacity constraints from the Lyman {\ensuremath{\alpha}} forest of galaxies in legacy extragalactic fields},} \mnras, 542, 1952, \dodoi{10.1093/mnras/staf1312}

% type= article
\bibitem[{R.~A. {Meyer} {et~al.}(2024){Meyer}, {Oesch}, {Giovinazzo}, {Weibel}, {Brammer}, {Matthee}, {Naidu}, {Bouwens}, {Chisholm}, {Covelo-Paz}, \& et~al.}]{Meyer:2024}
{Meyer}, R.~A., {Oesch}, P.~A., {Giovinazzo}, E., {et~al.} 2024, \bibinfo{title}{{JWST FRESCO: a comprehensive census of H {\ensuremath{\beta}} + [O III] emitters at 6.8 < z < 9.0 in the GOODS fields},} \mnras, 535, 1067, \dodoi{10.1093/mnras/stae2353}

% type= article
\bibitem[{J.~B. {Mu{\~n}oz} {et~al.}(2024){Mu{\~n}oz}, {Mirocha}, {Chisholm}, {Furlanetto}, \& {Mason}}]{Munoz:2024}
{Mu{\~n}oz}, J.~B., {Mirocha}, J., {Chisholm}, J., {Furlanetto}, S.~R., \& {Mason}, C. 2024, \bibinfo{title}{{Reionization after JWST: a photon budget crisis?},} \mnras, 535, L37, \dodoi{10.1093/mnrasl/slae086}

% type= article
\bibitem[{R.~P. {Naidu} {et~al.}(2020){Naidu}, {Tacchella}, {Mason}, {Bose}, {Oesch}, \& {Conroy}}]{Naidu:2020}
{Naidu}, R.~P., {Tacchella}, S., {Mason}, C.~A., {et~al.} 2020, \bibinfo{title}{{Rapid Reionization by the Oligarchs: The Case for Massive, UV-bright, Star-forming Galaxies with High Escape Fractions},} \apj, 892, 109, \dodoi{10.3847/1538-4357/ab7cc9}

% type= article
\bibitem[{R.~P. {Naidu} {et~al.}(2026){Naidu}, {Oesch}, {Brammer}, {Weibel}, {Li}, {Matthee}, {Chisolm}, {Pollock}, {Heintz}, {Johnson}, \& et~al.}]{Naidu:2026}
{Naidu}, R.~P., {Oesch}, P.~A., {Brammer}, G., {et~al.} 2026, \bibinfo{title}{{A Cosmic Miracle: A Remarkably Luminous Galaxy at zspec = 14.44 Confirmed with JWST},} The Open Journal of Astrophysics, 9, 56033, \dodoi{10.33232/001c.156033}

% type= article
\bibitem[{L. {Napolitano} {et~al.}(2026){Napolitano}, {Pentericci}, {Dickinson}, {Arrabal Haro}, {Taylor}, {Calabr{\`o}}, {Bhagwat}, {Santini}, {Arevalo-Gonzalez}, {Begley}, \& et~al.}]{Napolitano:2026}
{Napolitano}, L., {Pentericci}, L., {Dickinson}, M., {et~al.} 2026, \bibinfo{title}{{Ly{\ensuremath{\alpha}} visibility from z = 4.5 to 11 in the UDS field: Evidence for a high neutral hydrogen fraction and small ionized bubbles at z {\ensuremath{\sim}} 7},} \aap, 708, A102, \dodoi{10.1051/0004-6361/202556912}

% type= article
\bibitem[{P.~A. {Oesch} {et~al.}(2023){Oesch}, {Brammer}, {Naidu}, {Bouwens}, {Chisholm}, {Illingworth}, {Matthee}, {Nelson}, {Qin}, {Reddy}, \& et~al.}]{Oesch:2023}
{Oesch}, P.~A., {Brammer}, G., {Naidu}, R.~P., {et~al.} 2023, \bibinfo{title}{{The JWST FRESCO survey: legacy NIRCam/grism spectroscopy and imaging in the two GOODS fields},} \mnras, 525, 2864, \dodoi{10.1093/mnras/stad2411}

% type= article
\bibitem[{J.~B. {Oke} \& J.~E. {Gunn}(1983){Oke} \& {Gunn}}]{Oke:1983}
{Oke}, J.~B., \& {Gunn}, J.~E. 1983, \bibinfo{title}{{Secondary standard stars for absolute spectrophotometry.},} \apj, 266, 713, \dodoi{10.1086/160817}

% type= article
\bibitem[{M. {Orte-Garc{\'\i}a} {et~al.}(2026){Orte-Garc{\'\i}a}, {Esteban}, {Garc{\'\i}a-Rojas}, {M{\'e}ndez-Delgado}, {Arellano-C{\'o}rdova}, {Lugo-Aranda}, {Toribio San Cipriano}, {Rosales-Ortega}, {Mart{\'\i}nez-Hern{\'a}ndez}, \& {Reyes-Rodr{\'\i}guez}}]{OrteGarcia:2026}
{Orte-Garc{\'\i}a}, M., {Esteban}, C., {Garc{\'\i}a-Rojas}, J., {et~al.} 2026, \bibinfo{title}{{The DESIRED electron temperature relations in star-forming regions of the local Universe},} \aap, 711, A165, \dodoi{10.1051/0004-6361/202659162}

% type= book
\bibitem[{D.~E. {Osterbrock} \& G.~J. {Ferland}(2006){Osterbrock} \& {Ferland}}]{Osterbrock:2006}
{Osterbrock}, D.~E., \& {Ferland}, G.~J. 2006, {Astrophysics of gaseous nebulae and active galactic nuclei}

% type= article
\bibitem[{M. {Ouchi} {et~al.}(2009){Ouchi}, {Mobasher}, {Shimasaku}, {Ferguson}, {Fall}, {Ono}, {Kashikawa}, {Morokuma}, {Nakajima}, {Okamura}, \& et~al.}]{Ouchi:2009}
{Ouchi}, M., {Mobasher}, B., {Shimasaku}, K., {et~al.} 2009, \bibinfo{title}{{Large Area Survey for z = 7 Galaxies in SDF and GOODS-N: Implications for Galaxy Formation and Cosmic Reionization},} \apj, 706, 1136, \dodoi{10.1088/0004-637X/706/2/1136}

% type= article
\bibitem[{C. {Papovich} {et~al.}(2026){Papovich}, {Cole}, {Hu}, {Finkelstein}, {Shen}, {Arrabal Haro}, {Amor{\'\i}n}, {Backhaus}, {Bagley}, {Bhatawdekar}, \& et~al.}]{Papovich:2026}
{Papovich}, C., {Cole}, J.~W., {Hu}, W., {et~al.} 2026, \bibinfo{title}{{Galaxies in the Epoch of Reionization Are All Bark and No Bite{\textemdash}Plenty of Ionizing Photons, Low Escape Fractions},} \apj, 1000, 111, \dodoi{10.3847/1538-4357/ae3b25}

% type= article
\bibitem[{I. {Pasha} \& T.~B. {Miller}(2023){Pasha} \& {Miller}}]{Pasha:2023}
{Pasha}, I., \& {Miller}, T.~B. 2023, \bibinfo{title}{{pysersic: A Python package for determining galaxy structural properties via Bayesian inference, accelerated with jax},} The Journal of Open Source Software, 8, 5703, \dodoi{10.21105/joss.05703}

% type= inproceedings
\bibitem[{M.~D. {Perrin} {et~al.}(2014){Perrin}, {Sivaramakrishnan}, {Lajoie}, {Elliott}, {Pueyo}, {Ravindranath}, \& {Albert}}]{Perrin:2014}
{Perrin}, M.~D., {Sivaramakrishnan}, A., {Lajoie}, C.-P., {et~al.} 2014, \bibinfo{title}{{Updated point spread function simulations for JWST with WebbPSF},} in Society of Photo-Optical Instrumentation Engineers (SPIE) Conference Series, Vol. 9143, Space Telescopes and Instrumentation 2014: Optical, Infrared, and Millimeter Wave, ed. J.~M. {Oschmann}, Jr., M.~{Clampin}, G.~G. {Fazio}, \& H.~A. {MacEwen}, 91433X, \dodoi{10.1117/12.2056689}

% type= inproceedings
\bibitem[{M.~D. {Perrin} {et~al.}(2012){Perrin}, {Soummer}, {Elliott}, {Lallo}, \& {Sivaramakrishnan}}]{Perrin:2012}
{Perrin}, M.~D., {Soummer}, R., {Elliott}, E.~M., {Lallo}, M.~D., \& {Sivaramakrishnan}, A. 2012, \bibinfo{title}{{Simulating point spread functions for the James Webb Space Telescope with WebbPSF},} in Society of Photo-Optical Instrumentation Engineers (SPIE) Conference Series, Vol. 8442, Space Telescopes and Instrumentation 2012: Optical, Infrared, and Millimeter Wave, ed. M.~C. {Clampin}, G.~G. {Fazio}, H.~A. {MacEwen}, \& J.~M. {Oschmann}, Jr., 84423D, \dodoi{10.1117/12.925230}

% type= article
\bibitem[{ {Planck Collaboration} {et~al.}(2020){Planck Collaboration}, {Aghanim}, {Akrami}, {Ashdown}, {Aumont}, {Baccigalupi}, {Ballardini}, {Banday}, {Barreiro}, {Bartolo}, \& et~al.}]{Planck:2020}
{Planck Collaboration}, {Aghanim}, N., {Akrami}, Y., {et~al.} 2020, \bibinfo{title}{{Planck 2018 results. VI. Cosmological parameters},} \aap, 641, A6, \dodoi{10.1051/0004-6361/201833910}

% type= article
\bibitem[{G.~H. {Rieke} {et~al.}(2024){Rieke}, {Alberts}, {Shivaei}, {Lyu}, {Willmer}, {P{\'e}rez-Gonz{\'a}lez}, \& {Williams}}]{Rieke:2024}
{Rieke}, G.~H., {Alberts}, S., {Shivaei}, I., {et~al.} 2024, \bibinfo{title}{{SMILES: A Prototype JWST Multiband Mid-infrared Survey},} \apj, 975, 83, \dodoi{10.3847/1538-4357/ad6cd2}

% type= article
\bibitem[{B.~E. {Robertson}(2022){Robertson}}]{Robertson:2022}
{Robertson}, B.~E. 2022, \bibinfo{title}{{Galaxy Formation and Reionization: Key Unknowns and Expected Breakthroughs by the James Webb Space Telescope},} \araa, 60, 121, \dodoi{10.1146/annurev-astro-120221-044656}

% type= article
\bibitem[{B.~E. {Robertson} {et~al.}(2015){Robertson}, {Ellis}, {Furlanetto}, \& {Dunlop}}]{Robertson:2015}
{Robertson}, B.~E., {Ellis}, R.~S., {Furlanetto}, S.~R., \& {Dunlop}, J.~S. 2015, \bibinfo{title}{{Cosmic Reionization and Early Star-forming Galaxies: A Joint Analysis of New Constraints from Planck and the Hubble Space Telescope},} \apjl, 802, L19, \dodoi{10.1088/2041-8205/802/2/L19}

% type= article
\bibitem[{B.~E. {Robertson} {et~al.}(2013){Robertson}, {Furlanetto}, {Schneider}, {Charlot}, {Ellis}, {Stark}, {McLure}, {Dunlop}, {Koekemoer}, {Schenker}, \& et~al.}]{Robertson:2013}
{Robertson}, B.~E., {Furlanetto}, S.~R., {Schneider}, E., {et~al.} 2013, \bibinfo{title}{{New Constraints on Cosmic Reionization from the 2012 Hubble Ultra Deep Field Campaign},} \apj, 768, 71, \dodoi{10.1088/0004-637X/768/1/71}

% type= article
\bibitem[{B.~E. {Robertson} {et~al.}(2026){Robertson}, {Johnson}, {Tacchella}, {Eisenstein}, {Hainline}, {Alberts}, {Arribas}, {Baker}, {Bunker}, {Cameron}, \& et~al.}]{Robertson:2026}
{Robertson}, B.~E., {Johnson}, B.~D., {Tacchella}, S., {et~al.} 2026, \bibinfo{title}{{JWST Advanced Deep Extragalactic Survey (JADES) Data Release 5: Photometric Catalog},} arXiv e-prints, arXiv:2601.15956, \dodoi{10.48550/arXiv.2601.15956}

% type= article
\bibitem[{R.~L. {Sanders} {et~al.}(2024){Sanders}, {Shapley}, {Topping}, {Reddy}, \& {Brammer}}]{Sanders:2024}
{Sanders}, R.~L., {Shapley}, A.~E., {Topping}, M.~W., {Reddy}, N.~A., \& {Brammer}, G.~B. 2024, \bibinfo{title}{{Direct T $_{e}$-based Metallicities of z = 2─9 Galaxies with JWST/NIRSpec: Empirical Metallicity Calibrations Applicable from Reionization to Cosmic Noon},} \apj, 962, 24, \dodoi{10.3847/1538-4357/ad15fc}

% type= article
\bibitem[{R.~L. {Sanders} {et~al.}(2026){Sanders}, {Shapley}, {Topping}, {Reddy}, {Berg}, {Khostovan}, {Bouwens}, {Brammer}, {Carnall}, {Cullen}, \& et~al.}]{Sanders:2026}
{Sanders}, R.~L., {Shapley}, A.~E., {Topping}, M.~W., {et~al.} 2026, \bibinfo{title}{{The AURORA Survey: High-redshift Empirical Metallicity Calibrations from Electron Temperature Measurements at z = 2─10},} \apj, 1003, 228, \dodoi{10.3847/1538-4357/ae66e2}

% type= article
\bibitem[{A. {Saxena} {et~al.}(2023){Saxena}, {Robertson}, {Bunker}, {Endsley}, {Cameron}, {Charlot}, {Simmonds}, {Tacchella}, {Witstok}, {Willott}, \& et~al.}]{Saxena:2023}
{Saxena}, A., {Robertson}, B.~E., {Bunker}, A.~J., {et~al.} 2023, \bibinfo{title}{{JADES: Discovery of extremely high equivalent width Lyman-{\ensuremath{\alpha}} emission from a faint galaxy within an ionized bubble at z = 7.3},} \aap, 678, A68, \dodoi{10.1051/0004-6361/202346245}

% type= article
\bibitem[{A. {Saxena} {et~al.}(2024){Saxena}, {Bunker}, {Jones}, {Stark}, {Cameron}, {Witstok}, {Arribas}, {Baker}, {Baum}, {Bhatawdekar}, \& et~al.}]{Saxena:2024}
{Saxena}, A., {Bunker}, A.~J., {Jones}, G.~C., {et~al.} 2024, \bibinfo{title}{{JADES: The production and escape of ionizing photons from faint Lyman-alpha emitters in the epoch of reionization},} \aap, 684, A84, \dodoi{10.1051/0004-6361/202347132}

% type= article
\bibitem[{A. {Saxena} {et~al.}(2026){Saxena}, {Cameron}, {Katz}, {Bunker}, {Chevallard}, {D'Eugenio}, {Arribas}, {Bhatawdekar}, {Boyett}, {Cargile}, \& et~al.}]{Saxena:2026}
{Saxena}, A., {Cameron}, A.~J., {Katz}, H., {et~al.} 2026, \bibinfo{title}{{Hitting the slopes: a spectroscopic view of UV continuum slopes of galaxies reveals a reddening at z > 9.5},} \mnras, 548, stag808, \dodoi{10.1093/mnras/stag808}

% type= article
\bibitem[{D. {Schaerer} {et~al.}(2025){Schaerer}, {Guibert}, {Marques-Chaves}, \& {Martins}}]{Schaerer:2025}
{Schaerer}, D., {Guibert}, J., {Marques-Chaves}, R., \& {Martins}, F. 2025, \bibinfo{title}{{Observable and ionizing properties of star-forming galaxies with very massive stars and different initial mass functions},} \aap, 693, A271, \dodoi{10.1051/0004-6361/202451454}

% type= article
\bibitem[{J. {Scholtz} {et~al.}(2026){Scholtz}, {Carniani}, {Parlanti}, {D'Eugenio}, {Curtis-Lake}, {Jakobsen}, {Bunker}, {Cameron}, {Arribas}, {Baker}, \& et~al.}]{Scholtz:2026}
{Scholtz}, J., {Carniani}, S., {Parlanti}, E., {et~al.} 2026, \bibinfo{title}{{JADES Data Release 4 ─ Paper II. Data reduction, analysis, and emission-line fluxes of the complete spectroscopic sample},} \mnras, 549, stag939, \dodoi{10.1093/mnras/stag939}

% type= article
\bibitem[{A.~E. {Shapley} {et~al.}(2023){Shapley}, {Sanders}, {Reddy}, {Topping}, \& {Brammer}}]{Shapley:2023}
{Shapley}, A.~E., {Sanders}, R.~L., {Reddy}, N.~A., {Topping}, M.~W., \& {Brammer}, G.~B. 2023, \bibinfo{title}{{JWST/NIRSpec Balmer-line Measurements of Star Formation and Dust Attenuation at z 3-6},} \apj, 954, 157, \dodoi{10.3847/1538-4357/acea5a}

% type= article
\bibitem[{A. {Smith} {et~al.}(2022){Smith}, {Kannan}, {Garaldi}, {Vogelsberger}, {Pakmor}, {Springel}, \& {Hernquist}}]{Smith:2022}
{Smith}, A., {Kannan}, R., {Garaldi}, E., {et~al.} 2022, \bibinfo{title}{{The THESAN project: Lyman-{\ensuremath{\alpha}} emission and transmission during the Epoch of Reionization},} \mnras, 512, 3243, \dodoi{10.1093/mnras/stac713}

% type= article
\bibitem[{J.~S. {Speagle}(2020){Speagle}}]{Speagle:2020}
{Speagle}, J.~S. 2020, \bibinfo{title}{{DYNESTY: a dynamic nested sampling package for estimating Bayesian posteriors and evidences},} \mnras, 493, 3132, \dodoi{10.1093/mnras/staa278}

% type= article
\bibitem[{L. {Spitzer} \& J.~L. {Greenstein}(1951){Spitzer} \& {Greenstein}}]{Spitzer:1951}
{Spitzer}, Jr., L., \& {Greenstein}, J.~L. 1951, \bibinfo{title}{{Continuous Emission from Planetary Nebulae},} \apj, 114, 407, \dodoi{10.1086/145480}

% type= article
\bibitem[{C.~C. {Steidel} {et~al.}(2018){Steidel}, {Bogosavljevi{\'c}}, {Shapley}, {Reddy}, {Rudie}, {Pettini}, {Trainor}, \& {Strom}}]{Steidel:2018}
{Steidel}, C.~C., {Bogosavljevi{\'c}}, M., {Shapley}, A.~E., {et~al.} 2018, \bibinfo{title}{{The Keck Lyman Continuum Spectroscopic Survey (KLCS): The Emergent Ionizing Spectrum of Galaxies at z {\ensuremath{\sim}} 3},} \apj, 869, 123, \dodoi{10.3847/1538-4357/aaed28}

% type= article
\bibitem[{A.~L. {Strom} {et~al.}(2018){Strom}, {Steidel}, {Rudie}, {Trainor}, \& {Pettini}}]{Strom:2018}
{Strom}, A.~L., {Steidel}, C.~C., {Rudie}, G.~C., {Trainor}, R.~F., \& {Pettini}, M. 2018, \bibinfo{title}{{Measuring the Physical Conditions in High-redshift Star-forming Galaxies: Insights from KBSS-MOSFIRE},} \apj, 868, 117, \dodoi{10.3847/1538-4357/aae1a5}

% type= article
\bibitem[{Z. {Summerfield} {et~al.}(2026){Summerfield}, {McClymont}, {Tacchella}, {Smith}, {Kannan}, {Garaldi}, {Puchwein}, {Shen}, {Borrow}, {Danhaive}, \& et~al.}]{Summerfield:2026}
{Summerfield}, Z., {McClymont}, W., {Tacchella}, S., {et~al.} 2026, \bibinfo{title}{{From THESAN-ZOOM to JWST: Predicting ionizing photon escape and the rise of UV-bright reionization sources},} arXiv e-prints, arXiv:2606.14671, \dodoi{10.48550/arXiv.2606.14671}

% type= article
\bibitem[{S. {Tacchella} {et~al.}(2018){Tacchella}, {Bose}, {Conroy}, {Eisenstein}, \& {Johnson}}]{Tacchella:2018}
{Tacchella}, S., {Bose}, S., {Conroy}, C., {Eisenstein}, D.~J., \& {Johnson}, B.~D. 2018, \bibinfo{title}{{A Redshift-independent Efficiency Model: Star Formation and Stellar Masses in Dark Matter Halos at z {\ensuremath{\gtrsim}} 4},} \apj, 868, 92, \dodoi{10.3847/1538-4357/aae8e0}

% type= article
\bibitem[{A. {Tasitsiomi}(2006){Tasitsiomi}}]{Tasitsiomi:2006}
{Tasitsiomi}, A. 2006, \bibinfo{title}{{Ly{\ensuremath{\alpha}} Radiative Transfer in Cosmological Simulations and Application to a z \raisebox{-0.5ex}\textasciitilde= 8 Ly{\ensuremath{\alpha}} Emitter},} \apj, 645, 792, \dodoi{10.1086/504460}

% type= article
\bibitem[{S.~S. {Tayal} \& O. {Zatsarinny}(2017){Tayal} \& {Zatsarinny}}]{Tayal:2017}
{Tayal}, S.~S., \& {Zatsarinny}, O. 2017, \bibinfo{title}{{Transition and Electron Impact Excitation Collision Rates for O III},} \apj, 850, 147, \dodoi{10.3847/1538-4357/aa9613}

% type=
\bibitem[{ {The pandas development Team}(2025){The pandas development Team}}]{Pandas:2025}
{The pandas development Team}. 2025, {pandas-dev/pandas: Pandas}, v2.3.1 Zenodo, \dodoi{10.5281/zenodo.3509134}

% type= article
\bibitem[{M.~W. {Topping} {et~al.}(2024){Topping}, {Stark}, {Endsley}, {Whitler}, {Hainline}, {Johnson}, {Robertson}, {Tacchella}, {Chen}, {Alberts}, \& et~al.}]{Topping:2024}
{Topping}, M.~W., {Stark}, D.~P., {Endsley}, R., {et~al.} 2024, \bibinfo{title}{{The UV continuum slopes of early star-forming galaxies in JADES},} \mnras, 529, 4087, \dodoi{10.1093/mnras/stae800}

% type= article
\bibitem[{M.~W. {Topping} {et~al.}(2025){Topping}, {Sanders}, {Shapley}, {Pahl}, {Reddy}, {Stark}, {Berg}, {Clarke}, {Cullen}, {Dunlop}, \& et~al.}]{Topping:2025}
{Topping}, M.~W., {Sanders}, R.~L., {Shapley}, A.~E., {et~al.} 2025, \bibinfo{title}{{The AURORA survey: the evolution of multiphase electron densities at high redshift},} \mnras, 541, 1707, \dodoi{10.1093/mnras/staf903}

% type= article
\bibitem[{C. {Turner} {et~al.}(2025){Turner}, {Tacchella}, {D'Eugenio}, {Carniani}, {Curti}, {Glazebrook}, {Johnson}, {Lim}, {Looser}, {Maiolino}, \& et~al.}]{Turner:2025}
{Turner}, C., {Tacchella}, S., {D'Eugenio}, F., {et~al.} 2025, \bibinfo{title}{{Age-dating early quiescent galaxies: high star formation efficiency, but consistent with direct, higher-redshift observations},} \mnras, 537, 1826, \dodoi{10.1093/mnras/staf128}

% type= article
\bibitem[{H. {Umeda} {et~al.}(2026){Umeda}, {Ouchi}, {Kageura}, {Harikane}, {Nakane}, {Thai}, \& {Nakajima}}]{Umeda:2026}
{Umeda}, H., {Ouchi}, M., {Kageura}, Y., {et~al.} 2026, \bibinfo{title}{{Probing the Cosmic Reionization History with JWST: Gunn─Peterson and Ly{\ensuremath{\alpha}} Damping Wing Absorption at 4.5 < z < 13},} \apj, 997, 86, \dodoi{10.3847/1538-4357/ae232b}

% type= article
\bibitem[{H. {Umeda} {et~al.}(2024){Umeda}, {Ouchi}, {Nakajima}, {Harikane}, {Ono}, {Xu}, {Isobe}, \& {Zhang}}]{Umeda:2024}
{Umeda}, H., {Ouchi}, M., {Nakajima}, K., {et~al.} 2024, \bibinfo{title}{{JWST Measurements of Neutral Hydrogen Fractions and Ionized Bubble Sizes at z = 7─12 Obtained with Ly{\ensuremath{\alpha}} Damping Wing Absorptions in 27 Bright Continuum Galaxies},} \apj, 971, 124, \dodoi{10.3847/1538-4357/ad554e}

% type= article
\bibitem[{S. {Van Der Walt} {et~al.}(2011){Van Der Walt}, {Colbert}, \& {Varoquaux}}]{NumPy:2011}
{Van Der Walt}, S., {Colbert}, S.~C., \& {Varoquaux}, G. 2011, \bibinfo{title}{{The NumPy Array: A Structure for Efficient Numerical Computation},} Computing in Science and Engineering, 13, 22, \dodoi{10.1109/MCSE.2011.37}

% type= article
\bibitem[{P. {Virtanen} {et~al.}(2020){Virtanen}, {Gommers}, {Oliphant}, {Haberland}, {Reddy}, {Cournapeau}, {Burovski}, {Peterson}, {Weckesser}, {Bright}, \& et~al.}]{SciPy:2020}
{Virtanen}, P., {Gommers}, R., {Oliphant}, T.~E., {et~al.} 2020, \bibinfo{title}{{SciPy 1.0: fundamental algorithms for scientific computing in Python},} Nature Methods, 17, 261, \dodoi{10.1038/s41592-019-0686-2}

% type=
\bibitem[{M. {Waskom} {et~al.}(2020){Waskom}, {Botvinnik}, {Gelbart}, {Ostblom}, {Hobson}, {Lukauskas}, {Gemperline}, {Augspurger}, {Halchenko}, {Warmenhoven}, \& et~al.}]{Waskom:2020}
{Waskom}, M., {Botvinnik}, O., {Gelbart}, M., {et~al.} 2020, {seaborn: Statistical data visualization},, Astrophysics Source Code Library, record ascl:2012.015 \doeprint{2012.015}

% type= article
\bibitem[{J. {Witstok} {et~al.}(2024){Witstok}, {Smit}, {Saxena}, {Jones}, {Helton}, {Sun}, {Maiolino}, {Kumari}, {Stark}, {Bunker}, \& et~al.}]{Witstok:2024}
{Witstok}, J., {Smit}, R., {Saxena}, A., {et~al.} 2024, \bibinfo{title}{{Inside the bubble: exploring the environments of reionisation-era Lyman-{\ensuremath{\alpha}} emitting galaxies with JADES and FRESCO},} \aap, 682, A40, \dodoi{10.1051/0004-6361/202347176}

% type= article
\bibitem[{J. {Witstok} {et~al.}(2025{\natexlab{a}}){Witstok}, {Maiolino}, {Smit}, {Jones}, {Bunker}, {Helton}, {Johnson}, {Tacchella}, {Saxena}, {Arribas}, \& et~al.}]{Witstok:2025a}
{Witstok}, J., {Maiolino}, R., {Smit}, R., {et~al.} 2025{\natexlab{a}}, \bibinfo{title}{{JADES: primaeval Lyman {\ensuremath{\alpha}} emitting galaxies reveal early sites of reionization out to redshift z \raisebox{-0.5ex}\textasciitilde 9},} \mnras, 536, 27, \dodoi{10.1093/mnras/stae2535}

% type= article
\bibitem[{J. {Witstok} {et~al.}(2025{\natexlab{b}}){Witstok}, {Jakobsen}, {Maiolino}, {Helton}, {Johnson}, {Robertson}, {Tacchella}, {Cameron}, {Smit}, {Bunker}, \& et~al.}]{Witstok:2025b}
{Witstok}, J., {Jakobsen}, P., {Maiolino}, R., {et~al.} 2025{\natexlab{b}}, \bibinfo{title}{{Witnessing the onset of reionization through Lyman-{\ensuremath{\alpha}} emission at redshift 13},} \nat, 639, 897, \dodoi{10.1038/s41586-025-08779-5}

% type= article
\bibitem[{H. {Xu} {et~al.}(2016){Xu}, {Wise}, {Norman}, {Ahn}, \& {O'Shea}}]{Xu:2016}
{Xu}, H., {Wise}, J.~H., {Norman}, M.~L., {Ahn}, K., \& {O'Shea}, B.~W. 2016, \bibinfo{title}{{Galaxy Properties and UV Escape Fractions during the Epoch of Reionization: Results from the Renaissance Simulations},} \apj, 833, 84, \dodoi{10.3847/1538-4357/833/1/84}

% type= article
\bibitem[{H. {Yan} {et~al.}(2012){Yan}, {Finkelstein}, {Huang}, {Ryan}, {Ferguson}, {Koekemoer}, {Grogin}, {Dickinson}, {Newman}, {Somerville}, \& et~al.}]{Yan:2012}
{Yan}, H., {Finkelstein}, S.~L., {Huang}, K.-H., {et~al.} 2012, \bibinfo{title}{{Luminous and High Stellar Mass Candidate Galaxies at z {\ensuremath{\approx}} 8 Discovered in the Cosmic Assembly Near-Infrared Deep Extragalactic Legacy Survey},} \apj, 761, 177, \dodoi{10.1088/0004-637X/761/2/177}

% type= article
\bibitem[{M. {Yue} {et~al.}(2026){Yue}, {Zhu}, {Ding}, {Fan}, \& {Luo}}]{Yue:2026}
{Yue}, M., {Zhu}, Y., {Ding}, J., {Fan}, X., \& {Luo}, Y. 2026, \bibinfo{title}{{An Ionized Superstructure at Cosmic Dawn Revealed by a Foundation Model for Astrophysical Research},} arXiv e-prints, arXiv:2609.36656, \dodoi{10.48550/arXiv.2609.36656}

% type= article
\bibitem[{S. {Zhu} {et~al.}(2026){Zhu}, {Zheng}, {Bian}, {Yuan}, {Jiang}, {Zhang}, {Lin}, \& {Guo}}]{Zhu:2026}
{Zhu}, S., {Zheng}, Z.-Y., {Bian}, F., {et~al.} 2026, \bibinfo{title}{{LCEz4-M1: A Lyman Continuum Emitter Candidate at z = 4.444 in the MUSE Hubble Ultra Deep Field},} \apjl, 1005, L3, \dodoi{10.3847/2041-8213/ae75e1}

% type= article
\bibitem[{Y. {Zhu} {et~al.}(2021){Zhu}, {Becker}, {Bosman}, {Keating}, {Christenson}, {Ba{\~n}ados}, {Bian}, {Davies}, {D'Odorico}, {Eilers}, \& et~al.}]{Zhu:2021}
{Zhu}, Y., {Becker}, G.~D., {Bosman}, S. E.~I., {et~al.} 2021, \bibinfo{title}{{Chasing the Tail of Cosmic Reionization with Dark Gap Statistics in the Ly{\ensuremath{\alpha}} Forest over 5 < z < 6},} \apj, 923, 223, \dodoi{10.3847/1538-4357/ac26c2}

% type= article
\bibitem[{Y. {Zhu} {et~al.}(2022){Zhu}, {Becker}, {Bosman}, {Keating}, {D'Odorico}, {Davies}, {Christenson}, {Ba{\~n}ados}, {Bian}, {Bischetti}, \& et~al.}]{Zhu:2022}
{Zhu}, Y., {Becker}, G.~D., {Bosman}, S. E.~I., {et~al.} 2022, \bibinfo{title}{{Long Dark Gaps in the Ly{\ensuremath{\beta}} Forest at z < 6: Evidence of Ultra-late Reionization from XQR-30 Spectra},} \apj, 932, 76, \dodoi{10.3847/1538-4357/ac6e60}

% type= article
\bibitem[{Y. {Zhu} {et~al.}(2026){Zhu}, {Fan}, {Keating}, {Becker}, {Egami}, {Lin}, {Sun}, {Cain}, {Rieke}, {Bunker}, \& et~al.}]{YongdaZhu:2026}
{Zhu}, Y., {Fan}, X., {Keating}, L.~C., {et~al.} 2026, \bibinfo{title}{{Low Ly$α$ Visibility in Galaxy Overdensities: Reionization Topology and Neutral-Fraction Ceilings from DIVER over $4.8<z<11$},} arXiv e-prints, arXiv:2608.19311, \dodoi{10.48550/arXiv.2608.19311}

\end{thebibliography}
\bibliographystyle{aasjournalv7}

%% This command is needed to show the entire author and affiliation list when the collaboration and author truncation commands are used. It has to go at the end of the manuscript.
%%
\allauthors
%%
%% Include this line if you are using the \added, \replaced, \deleted commands to see a summary list of all changes at the end of the article.
%%
%% \listofchanges

\end{document}